\documentclass[11pt]{article}
\usepackage[T1]{fontenc}
\usepackage[utf8]{inputenc}
\usepackage{lmodern,amsmath,amssymb,amsthm,mathtools,mathrsfs}
\usepackage[margin=1in]{geometry}
\usepackage{microtype,enumitem}
\usepackage{array}
\usepackage[hidelinks]{hyperref}
\usepackage{bookmark}
\numberwithin{equation}{section}
\newtheorem{theorem}{Theorem}[section]
\newtheorem{proposition}[theorem]{Proposition}
\newtheorem{lemma}[theorem]{Lemma}
\newtheorem{corollary}[theorem]{Corollary}
\newtheorem{proofclaim}{Claim}[theorem]

\theoremstyle{definition}
\newtheorem{definition}[theorem]{Definition}
\newtheorem{remark}[theorem]{Remark}
\setlist{topsep=5pt,itemsep=3pt}
\allowdisplaybreaks[2]
\newcommand{\R}{\mathbb R}
\newcommand{\C}{\mathbb C}
\newcommand{\Hh}{\mathcal H}
\newcommand{\F}{\mathcal F}
\newcommand{\A}{\mathcal A}
\newcommand{\D}{\mathcal D}
\newcommand{\op}{\mathrm{op}}
\newcommand{\ind}{\mathbf 1}
\newcommand{\dd}{\,d}
\newcommand{\sres}{\xrightarrow{\mathrm{s.r.}}}
\newcommand{\Ad}{\operatorname{Ad}}
\newcommand{\supp}{\operatorname{supp}}

\hypersetup{pdftitle={On JT boundary algebras},pdfauthor={Feng Xu}}
\title{On JT boundary algebras}
\author{Feng Xu\\[0.4em]
 Department of Mathematics\\
University of California at Riverside\\
E-mail: {\tt xufeng@math.ucr.edu}}
\date{}
\begin{document}
\maketitle
\begin{abstract}
We construct the canonical boundary algebras of Jackiw--Teitelboim
gravity coupled to free bosonic matter. For a nonempty finite or
countable family of species with lowest weights greater than one half,
the two boundary algebras are mutual commutants and nonhyperfinite
type $\mathrm{II}_\infty$ factors. Their nonzero finite corners are
nonhyperfinite type $\mathrm{II}_1$ factors. The construction uses
bounded heat-regularized fields and a Hilbert algebra, which gives
a faithful normal semifinite trace without a path-integral argument.
The weight restriction is sharp: elementary point fields remain
unbounded after positive heat regularization when $0<d\leq1/2$.
For one massless scalar, specified bounded observables and normal
states have a high energy limit described by a hyperfinite continuous
core. We explain why this limit does not preserve the
nonhyperfiniteness of the exact algebra.
\end{abstract}

\section{Introduction}\label{sec:intro}
Jackiw--Teitelboim (JT) gravity coupled to matter is a model in which
one can ask precisely what algebra is generated by gravitational
observables. Canonical quantization leaves a length coordinate and a
matter state. Each of the two boundaries has a Hamiltonian and dressed
matter fields. The problem is to turn these fields, initially
distributions, into an algebra of bounded operators and to determine
its trace and type.

A key result of this paper is that the boundary algebras are
nonhyperfinite: every nonzero finite corner is a nonhyperfinite
type $\mathrm{II}_1$ factor, already for one massless scalar.
This distinguishes the exact JT algebras from the familiar hyperfinite
local algebras of quantum field theory and their modular crossed
products used in gravity \cite{W,Takesaki}. The distinction goes beyond
the change from type $\mathrm{III}$ to type $\mathrm{II}$: the exact
boundary algebra cannot be approximated by finite-dimensional
subalgebras. For one massless scalar, we also obtain a hyperfinite
high-energy limit, showing that this limit loses an algebraic feature
of the full gravitational theory. Similar nonhyperfiniteness results
hold for super JT gravity, as established in the author's forthcoming
paper \cite{XuSuperJT}.

Penington and Witten \cite[Sections 3.2--3.3]{PW} described the
boundary algebras and their trace through Euclidean amplitudes.
Kolchmeyer \cite[Sections 3--5]{K} developed the joint energy
representation and the factor and commutant arguments.
For free bosonic matter we prove heat bounds on the full Hilbert
space, left--right commutation, and the trace theorem by a Hilbert
algebra. Our factoriality proof follows
\cite[equations (5.12)--(5.19)]{K}.

The sharp boundedness threshold is $d>1/2$.
For $0<d<1/2$ the alternative boundary condition still gives a
positive-energy scalar theory on $\mathrm{AdS}_2$; many
low-momentum bosons cause the unboundedness.

Recall that a factor is a von Neumann algebra with scalar center.
A type $\mathrm{II}_1$ factor has a normalized faithful normal trace
and no minimal projections; a type $\mathrm{II}_\infty$ factor has
the corresponding semifinite trace, infinite on the identity.
A finite corner is $pMp$ for a nonzero finite-trace projection $p$.
Hyperfiniteness means that the algebra can be approximated by
finite-dimensional subalgebras. It is stronger than the existence
of matrix approximations to selected correlation functions.

\subsection*{Main results}
Let $\mathcal I$ be a nonempty finite or countable set of species,
with lowest weights $d_a>1/2$, and put
\[
 \mathcal F=\Gamma_s\!\left(\bigoplus_{a\in\mathcal I}
           L^2(\mathbb R_+,dp)\right),\qquad
 \mathcal H=L^2(\mathbb R_x;\mathcal F).
\]
Here $\Gamma_s$ denotes symmetric Fock space. Let $\mathcal P$ index
all nonvacuum lowest-weight copies in $\mathcal F$, including
multiplicities. Each copy has weight $\lambda_i$ and a normalized
finite Wick primary $O_{i,R}$, dressed at the right boundary.
The construction is given in Sections~\ref{sec:model}--\ref{sec:primaries}.
For the canonical Hamiltonians $H_R,H_L$ and reflection $J$, define
\[
 M_R=\{e^{-tH_R},\,T_{i;a,b}:t,a,b>0,\ i\in\mathcal P\}'',
 \qquad
 T_{i;a,b}=e^{-aH_R}O_{i,R}e^{-bH_R},\qquad M_L=JM_RJ.
\]
The sandwiches mean the bounded extensions proved below, not
products of already defined point operators. Fix $\phi_b>0$ and write
\begin{equation}\label{eq:density}
 E(s)=\frac{s^2}{4\phi_b},\qquad
 d\mu(s)=\rho(s)\,ds,\qquad
 \rho(s)=\frac{s\sinh(2\pi s)}{\pi^2},\quad s>0.
\end{equation}
\begin{theorem}\label{thm:main}
For the free bosonic matter theories described above, the following hold.
\begin{enumerate}[label=(\arabic*)]
\item Every two-sided positive heat regularization of a finite Wick
primary has a canonical bounded extension. For a primary of dimension
$\lambda_i$, its symmetric heat sandwich has norm
$O(\varepsilon^{-\lambda_i})$ as $\varepsilon\downarrow0$.
\item The Hamiltonians strongly commute, the bounded opposite-boundary
operators commute, and
$M_R'=M_L$, $M_L'=M_R$.
\item There is a faithful normal semifinite trace $\tau$ on $M_R$ such that
\begin{equation}\label{eq:maintrace}
 \tau(f(H_R))=\int_0^\infty f(E(s))\,d\mu(s)
 \quad\text{for every nonnegative Borel function }f.
\end{equation}
The canonical Hilbert space is $L^2(M_R,\tau)$ and $J$ acts as adjunction.
\item Both boundary algebras are factors of type $\mathrm{II}_\infty$.
For every nonzero projection $p\in M_R$ with $\tau(p)<\infty$,
the corner $pM_Rp$ is a factor of type $\mathrm{II}_1$.
The trace is unique up to a positive scalar multiple.
\item For $\beta>0$, let $\xi_\beta$ be the vacuum-sector vector with
wavefunction $e^{-\beta E(s)}$ in $L^2((0,\infty),d\mu)$. For every $x\in(M_R)_+$,
\begin{equation}\label{eq:mainlimit}
 \tau(x)=\lim_{\beta\downarrow0}
              \langle\xi_\beta,x\xi_\beta\rangle,
\end{equation}
where the limit may be infinite.
\item Both boundary factors are noninjective.
Every nonzero finite-trace corner of either factor is a
nonhyperfinite type $\mathrm{II}_1$ factor.
\end{enumerate}
\end{theorem}

For the threshold statement, let $q$ multiply by $p$ on $L^2(dp)$,
$Q=d\Gamma(q)$, and $P=d\Gamma(-i(p\partial_p+1/2))$.
Set $g_d(p)=p^{d-1/2}/\sqrt{\Gamma(2d)}$ and
$\Phi_d(0)=a^*(g_d)+a(g_d)$ as a form. The unitary
$(U\psi)(x)=e^{ixP/2}\psi(x)$ changes to the right gauge.
\begin{theorem}[Weight threshold]\label{thm:weight-threshold}
For one elementary species of weight $d>0$ and every $t>0$, the
canonical symmetric point sandwich
\[
 T_{d,t}=e^{-tL}\bigl(2^de^{-dx}\Phi_d(0)\bigr)e^{-tL},
 \qquad L=-\tfrac12\partial_x^2+(Q+2)e^{-x},
\]
defines an essentially self-adjoint operator on the algebraic
sum of the full particle sectors. Its closure is bounded if and
only if $d>1/2$. Conjugation by the canonical unitary $U$ gives
the same assertion for $e^{-\varepsilon H_R}O_Re^{-\varepsilon H_R}$,
where $t=\varepsilon/(2\phi_b)$. For $0<d<1/2$, the underlying
scalar with the alternative boundary condition is nevertheless
a unitary positive-energy theory.
\end{theorem}

\subsection*{High energy limit}
For one massless scalar ($d=1$), put $M=U^*M_RU$ and
$L=2\phi_bU^*H_RU$. The scaling
\begin{equation}\label{eq:introscale}
 \begin{split}
 (S_k\psi)(x)&=e^{ikx}\psi(x+\log k),\\
 K_k&=\frac{S_k^*LS_k-k^2/2}{k}
     =D+(Q+2)e^{-y}+\frac{D^2}{2k},\qquad D=-i\partial_y
 \end{split}
\end{equation}
has the self-adjoint limit $K_\infty=GDG^*$, where
$G(y)=e^{i(Q+2)e^{-y}}$.
Theorems~\ref{thm:boundedconv} and~\ref{thm:core} prove convergence
of specified unitary observables to an algebra
\begin{equation}\label{eq:introcore}
 M_\infty\cong\mathcal A\rtimes_{\sigma^\varphi}\mathbb R,
\end{equation}
where $\mathcal A$ is the negative half-line current algebra and
$\sigma^\varphi$ is its vacuum modular group.
This continuous core \cite{PW,W} is hyperfinite. Proposition~\ref{prop:statelimit}
identifies the limiting normal states.

In a fixed interval of rescaled energy, the compressed field norms
diverge while their normalized tracial second moments converge to a
positive finite value (Theorem~\ref{thm:normdivergence} and
Proposition~\ref{prop:varianceconvergence}). Normalizing these fields
to contractions therefore sends their tracial second moments to zero.

\subsection*{Notation}\label{guide:notation}
Inner products are antilinear in the first variable. We use
$D(A)$ for an operator domain, $\mathbf1_I(A)$ for a spectral
projection, and $\sres$ for strong resolvent convergence.
The main notation is as follows; auxiliary symbols are defined
where they are used.
\begin{center}
\begin{tabular}{@{}ll@{}}
$\mathcal F,\Omega,N$ & matter Fock space, vacuum, particle number\\
$Q,P,K,H_m$ & matter translation, dilation, opposite translation,
and energy\\
$H_R,H_L;\ A_R,A_L$ & boundary Hamiltonians;
$A_{R,L}=2\phi_bH_{R,L}$\\
$U,\ L=U^*A_RU$ & right gauge and its Hamiltonian\\
$J,\ M_R,M_L$ & reflection and boundary algebras\\
$\lambda_i,\ O_{i,R},T_{i;a,b}$ & primary weight, dressed form,
bounded heat letter\\
$\mu,\ G_d,\ F_d$ & spectral measure, Gamma factor, insertion kernel\\
$\mathfrak A_0,\Lambda,\tau$ & heat-word algebra, vacuum map, trace\\
$M=U^*M_RU,\ e,\ b=c+c^*$ & right-gauge algebra, energy projection,
compressed field
\end{tabular}
\end{center}
The canonical operators are defined in Section~\ref{sec:model},
the kernels in Section~\ref{sec:spectral}, and the scaling notation
in Section~\ref{sec:scale}.

\subsection*{Proof outline}\label{guide:proof}
The heat bound rests on
$\|a(h)\psi\|\le\|p^{-1/2}h\|_2\|Q^{1/2}\psi\|$.
For $h=e^{-bp}g_d$ the weighted norm is finite exactly when
$d>1/2$; its length dependence cancels the dressing factor.
Weighted semigroup bounds transfer this estimate to JT heat.
Coherent states prove unboundedness at and below the threshold.

Separation of the Hamiltonians and their endpoint conditions give
commutation with opposite resolvents. The mixed-product estimate
then identifies bounded products with Wick expansions; reality of
the opposite-boundary contractions proves locality.

A central operator acts as $c(r)$ on the vacuum summand. The
nonvanishing right and left vacuum columns force its action on each
primary summand to be both $c(r)$ and $c(s)$, hence constant.

For heat words the norm limit
$\Lambda(a)=\lim_{\beta\downarrow0}a\xi_\beta$ exists.
Reflection and locality give
$J\Lambda(a)=\Lambda(a^*)$ and
$\Lambda(ab)=a\Lambda(b)=(Jb^*J)\Lambda(a)$.
The resulting Hilbert algebra supplies the trace and mutual
commutants; the spectral measure determines the type.

In a finite energy corner write $0\ne b=c+c^*$.
The full-Fock estimate and Parseval give a corner unitary $u$ with
$\sum_{j=1}^n u^jcc^*u^{-j}\le R^2e$ for every $n$.
A conditional expectation onto the corner would send $c$ to
$a$ satisfying $n\tau_e(aa^*)\le R^2$. Thus $a=0$ and
$b=a+a^*=0$, a contradiction.

\section{Preliminaries and the canonical operators}\label{sec:model}

We use the canonical Hamiltonians of \cite{PW,K}. The following
conventions fix their domains and the normalization of the matter fields.

Write $\mathbb R_+=(0,\infty)$, $\mathbb N_0=\{0,1,\ldots\}$,
and let $M_f$ denote multiplication by $f$. Unless stated otherwise,
$C_c^\infty$ means compactly supported smooth functions on the open
set; the closed-half-line convention is given in
Section~\ref{sub:endpoint}.

For a von Neumann algebra $N$, write $N'$ for its commutant,
$Z(N)$ for its center and $N_+$ for its positive cone.
The opposite algebra $N^{\op}$ has reversed multiplication;
$\bar\otimes$ and $\otimes_{\min}$ denote the spatial von Neumann
and minimal $C^*$ tensor products. Strong-$*$ convergence means
strong convergence of operators and adjoints. Constants $C$ may
change from line to line; subscripts indicate their dependencies.

\subsection{The matter representation}\label{sub:matter}

In the canonical quantization of \cite[Sections 2.1--2.3]{PW},
the left, right and matter charges satisfy the diagonal
$\widetilde{\mathrm{SL}}(2,\mathbb R)$ constraint. Group averaging
gives the physical inner product. Fixing the endpoints at a common
global time leaves the renormalized geodesic length $x\in\mathbb R$
and a matter state. Thus physical wavefunctions are matter-valued
functions of $x$. Solving the constraint gives the boundary
Hamiltonians; attaching matter insertions to the endpoints gives
the fields \cite[equations (2.28), (2.35)--(2.39) and Section 3.2]{PW}.

A free bosonic species has lowest global energy $d$ and symmetric
Fock space. We assume $d>1/2$ until Section~\ref{sec:smallweight}.

Let \(\mathcal I\) be a nonempty finite or countable set, and choose a
real number \(d_a>1/2\) for each \(a\in\mathcal I\). Define

\begin{equation}\label{eq:19.1}
\mathfrak h_a=L^2(\mathbb R_+,dp),\qquad
\mathfrak h=\bigoplus_{a\in\mathcal I}\mathfrak h_a,
\qquad \mathcal F=\Gamma_s(\mathfrak h),\qquad
\mathcal H=L^2(\mathbb R_x;\mathcal F).
\end{equation}

Here $\Gamma_s(\mathfrak h)=\bigoplus_{n\ge0}\mathfrak h^{\otimes_s n}$,
with vacuum $\Omega$ in the zeroth summand, and
$\Gamma_s(T)=\bigoplus_{n\ge0}T^{\otimes_s n}$.
The second quantization $d\Gamma(A)$ sums $A$ over the particle
slots and vanishes on $\Omega$; $N=d\Gamma(1)$.
Let $\mathcal H_{\mathrm{fin}}$ be the algebraic sum of the full
particle sectors. Creation $a^*(f)$ is linear in $f$, annihilation
$a(f)$ is antilinear, and $\Phi(f)=a^*(f)+a(f)$.
For distributional $f$ these symbols denote forms on the tests below.
In momentum coordinates the positive parabolic generator is
multiplication by $p$, and dilations act by
$(u_tf)(p)=e^{t/2}f(e^tp)$.
The opposite generator is fixed by the $\mathfrak{sl}_2$ relations
and Casimir $d_a(d_a-1)$. Put $\alpha_a=d_a-1/2$ and define

\begin{equation}\label{eq:19.2}
q_a=p,\qquad P_a=-i(p\partial_p+1/2),\qquad
\kappa_a=-\partial_p p\partial_p+\alpha_a^2/p.
\end{equation}

For example, direct differentiation on compact interior tests gives $\frac12(q_a\kappa_a+\kappa_a q_a)-P_a^2=d_a(d_a-1)\,1$. This checks the coefficient of $p^{-1}$ in \eqref{eq:19.2}.
Multiplication \(q_a\) has its maximal domain; \(P_a\) is the dilation
Stone generator. The endpoint at $p=0$ also requires a choice of
self-adjoint realization. For $d_a>1/2$, the lowest-weight behavior
is $p^{d_a-1/2}$ and is selected by the nonnegative Friedrichs form.
Accordingly, \(\kappa_a\) is the Friedrichs operator of the closure of

\begin{equation}\label{eq:19.3}
\mathfrak k_a[f]=\int_0^\infty
\left(p|f'(p)|^2+\frac{\alpha_a^2}{p}|f(p)|^2\right)dp,
\quad f\in C_c^\infty(0,\infty).
\end{equation}

The compact generator of global matter time is the form sum
$h_a=(q_a+\kappa_a)/2$. Its eigenvalues should be the lowest energy
$d_a$ and its integer-spaced descendants. Solving this eigenvalue
equation produces a ground state proportional to $p^{d_a-1/2}e^{-p}$
and Laguerre polynomials for the higher modes. Thus the following
basis records the global-energy interpretation of the momentum
representation and will also make reflection explicit.
For a single weight $d$, let $L_n^\alpha$ denote the generalized
Laguerre polynomial of degree $n$. The normalized Laguerre basis is
\begin{equation}\label{eq:2.1}
f_n^d(p)=2^d\sqrt{\frac{n!}{\Gamma(n+2d)}}
p^{d-1/2}e^{-p}L_n^{2d-1}(2p),\quad n\geq0.
\end{equation}

Changing variables $r=2p$ in Laguerre orthogonality shows that
$f_n=f_n^d$ form a complete orthonormal basis.
For $d=d_a$ they have eigenvalues $n+d_a$ under $h_a$.
Each is $p^{\alpha_a}$ times a smooth function, lies in the
Friedrichs form domain, and satisfies the boundary condition of
Section~\ref{sub:endpoint}. Its differential image is
$(n+d_a)f_n$. Completeness identifies the self-adjoint realization.

For $t=q,\kappa,h$, take $t=\bigoplus_a t_a$ on its maximal
direct-sum domain, and put $p_0=\bigoplus_aP_a$. The total charges are
$Q=d\Gamma(q)$, $P=d\Gamma(p_0)$, $K=d\Gamma(\kappa)$ and
$H_m=d\Gamma(h)$. Finite-particle vectors with finitely many species
and Laguerre modes form a common invariant analytic core. Indeed,
the ladder coefficients below grow linearly, so a word of length
$r$ on any such vector has norm at most $C^{r+1}r!$.
Direct-sum and particle truncation give the core on the full space.

We next construct the reflection which exchanges the two boundaries.
On the matter space it must exchange the two parabolic generators.
A half-period of the compact motion does this. Removing the phase
of the lowest-energy mode makes its eigenvalues $(-1)^n$, so that
the resulting operator is a real involution. In momentum coordinates
this operator is the Hankel transform.
Let $J_\nu$ denote the Bessel function of the first kind. On each
species the Hankel involution \(s_a\) has kernel
\(J_{2d_a-1}(2\sqrt{pv})\). Equivalently
\(s_af_n^{d_a}=(-1)^nf_n^{d_a}\), which defines it on all of
\(L^2(dp)\). Equivalently $s_a=e^{i\pi(h_a-d_a)}$.
The Laguerre recurrence, or the Lie relations on the analytic core,
gives
\(s_aq_as_a=\kappa_a,\qquad s_aP_as_a=-P_a.\)
For any weight $\lambda>1/2$, the symbols
$q_\lambda$, $p_{0,\lambda}$, $\kappa_\lambda$, $h_\lambda$ and
$s_\lambda$ denote these same one-particle operators with $d_a$
replaced by $\lambda$. Set \(s=\bigoplus_a s_a\) and let \(C\)
be conjugation in the real occupation basis of \(\mathcal F\).
Boundary reflection is antiunitary: besides exchanging the matter
generators it conjugates amplitudes and reverses the length momentum,
while keeping the length coordinate fixed. Complex conjugation
reverses the sign of the dilation generator; the Hankel involution
reverses it a second time. Their combination therefore exchanges
$Q$ and $K$ while fixing $P$, exactly as required by the two
Hamiltonians in Section~\ref{sub:hamiltonians}. Apply this operation
to every particle, with the identity on the vacuum, and define
\begin{equation}\label{eq:reflection-fock}
 \Gamma_s(s)=\bigoplus_{N=0}^\infty s^{\otimes_s N},\qquad
 \Theta=\Gamma_s(s)C,\qquad (J\psi)(x)=\Theta\psi(x),
\end{equation}
where the zeroth tensor power is the identity on \(\mathbb C\Omega\).
This defines \(J\) on the whole of \(\mathcal H\). More explicitly, if
\(\mathbf m=(m_{a,n})\) is a finite occupation multi-index,
$|\mathbf m\rangle$ is the normalized occupation vector with
$m_{a,n}$ particles in the mode $f_n^{d_a}$, and
\(\psi(x)=\sum_{\mathbf m}\psi_{\mathbf m}(x)|\mathbf m\rangle\), then
\begin{equation}\label{eq:reflection-occupation}
 (J\psi)(x)=\sum_{\mathbf m}
 (-1)^{\sum_{a,n}n m_{a,n}}
 \overline{\psi_{\mathbf m}(x)}|\mathbf m\rangle.
\end{equation}
The series converges in \(L^2(\mathbb R_x;\mathcal F)\), with the same
norm as \(\psi\). The formula gives \(J^2=1\), so \(J\) is an
antiunitary involution on all particle sectors, including the vacuum.
It conjugates the length-dependent coefficients and leaves \(x\) fixed.
In particular, writing $k=-i\partial_x$ for the momentum conjugate
to length, we have \(JQJ=K\), \(JPJ=P\) and \(JkJ=-k\).
Thus \eqref{eq:reflection-fock} is fixed by the boundary-exchange
symmetry and the real field convention. We define it before the
trace; Section~\ref{sub:normaltrace} will prove that this same $J$
is the modular conjugation of the tracial representation.

On finite Laguerre vectors the raising and lowering coefficients are
$\sqrt{(n+1)(n+2d)}$ and $\sqrt{n(n+2d-1)}$. In particular
$h_a\geq d_a$ and $H_m\geq\frac12N$. Second quantization of the dilation unitaries gives
$e^{itP}=\Gamma_s(\bigoplus_a u_t)$.
On the common analytic core,
\begin{equation}\label{eq:lie}
 [P,Q]=-iQ,\qquad [P,K]=iK,\qquad [Q,K]=2iP,
 \qquad Q+K=2H_m.
\end{equation}
The last equality also holds as an equality of closed forms.

The zero-particle space is $\mathbb C\Omega$, where $\Omega$ is the
Fock vacuum. On a finite occupation sector $Q$ is multiplication by
the sum of the particle momenta.

Dilation covariance fixes the elementary primary distribution as
$g_d(p)=p^{d-1/2}/\sqrt{\Gamma(2d)}$ and
$\Phi_d(0)=a^*(g_d)+a(g_d)$.
Euler's integral gives
$\int_0^\infty e^{-\beta p}|g_d(p)|^2\,dp=\beta^{-2d}$, and
\begin{equation}\label{eq:2.3}
 e^{itP}Qe^{-itP}=e^tQ,\qquad
 e^{itP}\Phi_d(0)e^{-itP}=e^{td}\Phi_d(0).
\end{equation}
The field identity holds as a form on the tests below.
The boost $B$ is the self-adjoint Lie generator whose restriction
to the analytic core is $(Q-K)/2$; thus $Q=H_m+B$.

\subsection{The canonical Hamiltonians}\label{sub:hamiltonians}

In the length convention $x=\ell$, the canonical kinetic momenta
are $k-P/2$ and $k+P/2$, the matter potentials are $Qe^{-x/2}$
and $Ke^{-x/2}$, and the pure-gravity potential is $2e^{-x}$
\cite[equations (4.22)--(4.23)]{K};
see also \cite[equation (2.39)]{PW}.

A length-dependent matter dilation removes $P/2$ from the right
kinetic term and changes $Qe^{-x/2}$ to $Qe^{-x}$.
Put $k=-i\partial_x$ and define \((U\psi)(x)=e^{ixP/2}\psi(x).\)

Direct differentiation and \eqref{eq:2.3} give \(UkU^*=k-P/2,\qquad UQU^*=e^{x/2}Q.\)

Accordingly, introduce the simpler right-gauge expression
$L=\tfrac12 k^2+(Q+2)e^{-x}$, with its self-adjoint realization
specified below. It follows that

\begin{equation}\label{eq:3.1}
ULU^*=\frac12(k-P/2)^2+Qe^{-x/2}+2e^{-x}.
\end{equation}

The dressing of a weight-$d$ field is
$(2e^{-x/2})^d\Phi_d(0)$ in the canonical gauge
\cite[equations (4.24)--(4.25)]{K}. Dilation covariance gives

\begin{equation}\label{eq:3.2}
U\bigl(2^de^{-dx}\Phi_d(0)\bigr)U^*
=(2e^{-x/2})^d\Phi_d(0).
\end{equation}

Here $Q=H_m+B$ and $K=H_m-B$. We define $L$ by the
nonnegative closed quadratic form

\begin{equation}\label{eq:3.3}
\mathfrak l[\psi]=\frac12\int\|\partial_x\psi(x)\|^2dx
+\int e^{-x}\|(Q+2)^{1/2}\psi(x)\|^2dx.
\end{equation}

The intersection of the two form domains, with the sum of their graph
norms, is complete. The form is densely defined and nonnegative. It
therefore defines a unique positive self-adjoint operator. Truncating
particle number, momenta and \(x\), followed by smooth approximation,
shows that the test space $\mathcal D$ of finite-species,
finite-particle functions, jointly smooth and compactly supported in
length and in strictly positive particle momenta, with the required
species-wise symmetries, is a form core. Its elements also belong to
every integer power domain of \(L\), since the differential expression
preserves smoothness, compact supports and finite particle number.

The matching length factors in \eqref{eq:3.2} and \eqref{eq:3.3}
will be essential in Section~\ref{sub:schur}: the field's growth
cancels the length dependence of the matter heat estimate.

Let $\phi_b>0$ be the renormalized boundary dilaton parameter,
which fixes boundary time. Returning to the canonical gauge, set

\begin{equation}\label{eq:19.4}
A_R=ULU^*,\qquad A_L=JA_RJ,\qquad
H_R=A_R/(2\phi_b),\quad H_L=A_L/(2\phi_b),\quad \phi_b>0.
\end{equation}

The operator $L$ has the following domain: $\psi$ belongs to the form
domain of \eqref{eq:3.3} and there is $\eta\in\mathcal H$ such that
$\mathfrak l(\varphi,\psi)=\langle\varphi,\eta\rangle$ for all
form-domain $\varphi$; then $L\psi=\eta$. We set
$D(H_R)=UD(L)$ and $D(H_L)=JD(H_R)$. 

Thus $A_R=2\phi_bH_R$ and $A_L=2\phi_bH_L$ have the closed forms
associated with
\begin{equation}\label{eq:canonicalpair}
 A_R=\tfrac12(k-P/2)^2+Qe^{-x/2}+2e^{-x},\qquad
 A_L=\tfrac12(k+P/2)^2+Ke^{-x/2}+2e^{-x}.
\end{equation}
For the left form, conjugation by $U$ gives the form of
$\tfrac12k^2+(K+2)e^{-x}$. Compact length truncation and the
Friedrichs form cores in each finite occupation sector show that
$\mathcal D$ is a form core for both canonical forms.

\subsection{Field distributions}\label{sub:fields}

Parabolic-time derivatives multiply the creation distribution
by $ip$. The corresponding right and left jets are

The elementary right jets are the forms with
one-particle distributions

\begin{equation}\label{eq:19.5}
g_{a,R,k}(p)=\frac{i^k p^{d_a+k-1/2}}{\sqrt{\Gamma(2d_a)}},
\quad k\in\mathbb N_0,
\qquad g_{a,L,k}=(-i)^k\kappa_a^k s_ag_{a,R,0}.
\end{equation}

Their common initial field test space is compact smooth length functions
tensored with finite Laguerre occupation vectors of finitely many
species. The left distribution \(s_ag_{a,R,0}\) acts by
\(\sqrt{\Gamma(2d_a)}\lim_{p\downarrow0}p^{-\alpha_a}f(p)\); its higher
jets act by applying \(\kappa_a^k\) to the test function, with the
phases in \eqref{eq:19.5}. These formulas define distributions on this test
space, not vectors in \(\mathfrak h\).

For a genuine one-particle vector $f$, creation and annihilation obey
$[a(f),a^*(g)]=\langle f,g\rangle$; annihilation is antilinear in $f$.
For a distribution vector these symbols denote sesquilinear forms on
the test space just specified. We write
$\Phi_{a,R,k}=a^*(g_{a,R,k})+a(g_{a,R,k})$ and define the left fields
by reflection, $\Phi_{a,L,k}=\Theta\Phi_{a,R,k}\Theta$. When a single
species is under discussion the index $a$ is omitted. For scalar time
tests we use the Fourier convention
$\widehat f(v)=\int_{\mathbb R}f(t)e^{itv}\,dt$.
An ordinary time-smearing of an
elementary field has wavefunction $g_d\widehat f\in L^2(dp)$ for
$f\in C_c^\infty(\mathbb R)$ and every $d>0$. In particular it obeys
the usual $(N+1)^{1/2}$ bound. The heat-regularized point form studied
below requires a different estimate.

\subsection{The massless scalar}\label{sub:massless}

For \(d=1\), the global one-particle energies are \(1,2,3,\ldots\). This
is the massless scalar with Dirichlet boundary conditions on the
conformal strip \(0<\sigma<\pi\): its modes are proportional to
\(k^{-1/2}\sin(k\sigma)e^{-ikt}\). The boundary normal derivative has
mode coefficients proportional to \((-1)^k\sqrt{k}\) at the right
endpoint. Formula \eqref{eq:2.1} maps these modes to the present momentum
realization. In fact, pairing \(g_1\) with \(f_n\) gives
\(2(-1)^n\sqrt{n+1}\), fixing the otherwise arbitrary field
normalization.

This gives a concrete example with infinitely many Fock sectors.

\section{Lowest-weight summands and primary fields}\label{sec:primaries}

The lowest-weight decomposition supplies the joint spectral
representation and its vacuum columns.

\subsection{Radial and angular decomposition}\label{sub:decomposition}

Use total momentum as a radial variable and momentum fractions
as angular variables.

Fix a nonempty finite occupation list \((a_1,\ldots,a_n)\), with
$n\ge1$, and write
\(d_j=d_{a_j}\), \(D_0=\sum_jd_j\), \(p=\sum_jp_j\), \(u_j=p_j/p\).
Symmetrize only among particles of the same species. The sector is
equivalently the appropriate subspace of \(L^2(d\mathbf p)\), where
\(d\mathbf p=p^{n-1}dp\,d\mathbf u\). Here
$\Delta_{n-1}=\{\mathbf u:u_j>0,\ \sum_j u_j=1\}$ and
$d\mathbf u=du_1\cdots du_{n-1}$ with
$u_n=1-\sum_{j<n}u_j$; for $n=1$ this is the unit point measure.
On this simplex use the finite measure

\begin{equation}\label{eq:19.7}
d\omega(\mathbf u)=\prod_{j=1}^n u_j^{2d_j-1}\,d\mathbf u.
\end{equation}

The orthogonal degree-\(m\) polynomial space means polynomials of degree
at most \(m\) perpendicular to those of degree at most \(m-1\),
restricted to the required permutation invariants. Choose a real
orthonormal basis \(P_{m,\ell}\) of each nonzero such space. These
spaces are finite dimensional and their orthogonal sum is dense.
Polynomial density follows first on the compact simplex and then in
\(L^2(d\omega)\); averaging over the finite permutation group proves
density in the invariant subspace. For \(n=1\) there is only the
constant angular space, with \(m=0\).

With $\partial_j=\partial/\partial u_j$, the angular part of
the opposite parabolic generator is

\begin{equation}\label{eq:19.8}
\mathcal J=-\sum_{i,j<n}u_i(\delta_{ij}-u_j)\partial_i\partial_j
-\sum_{i<n}(2d_i-2D_0u_i)\partial_i.
\end{equation}

It is symmetric on polynomials for \eqref{eq:19.7}. In its divergence form the
flux across \(u_j=0\) contains \(u_j^{2d_j}\), so the boundary
contribution is zero. Its top-degree action is multiplication by
\(m(m+2D_0-1)\). Because it preserves degree and is symmetric,
subtracting this multiple from its action on an orthogonal degree-\(m\)
polynomial gives a lower-degree polynomial orthogonal to all
lower-degree polynomials. That difference is zero. Thus

\begin{equation}\label{eq:19.9}
\mathcal JP_{m,\ell}=m(m+2D_0-1)P_{m,\ell}.
\end{equation}

Conjugating \(\sum_j\kappa_{a_j}\) by \(\prod_jp_j^{d_j-1/2}\) gives
\(-\sum_j(p_j\partial_{p_j}^2+2d_j\partial_{p_j})\). In radial/angular
variables this becomes
\(-p\partial_p^2-2D_0\partial_p+p^{-1}\mathcal J\). The unitary radial
identification for a normalized angular vector is

\begin{equation}\label{eq:19.10}
f(p)\longmapsto p^{-(n-1)/2}f(p)
\prod_j u_j^{d_j-1/2}P_{m,\ell}(\mathbf u).
\end{equation}

Substitution, using \eqref{eq:19.9}, gives

\begin{equation}\label{eq:19.11}
Q=p,\qquad P=-i(p\partial_p+1/2),\qquad
K=-\partial_p p\partial_p+\frac{(\lambda-1/2)^2}{p},
\quad\lambda=D_0+m>1/2.
\end{equation}

Radial Laguerre vectors in \eqref{eq:19.10} are
$e^{-\sum p_j}\prod_jp_j^{d_j-1/2}$ times polynomials and belong
to the product-Laguerre core. Conversely, product polynomials
decompose into angular and radial polynomials. The two bases
therefore span the same analytic core, identifying the
self-adjoint closures and the Friedrichs realization in \eqref{eq:19.11}.

There is exactly one vacuum summand. Every nonvacuum summand has weight
\(\lambda>1/2\). The list of such summands, including multiplicities, is
countable, also for countably many species. We denote the index set
of these copies by $\mathcal P$. An index $i\in\mathcal P$ records
the occupation list, the angular degree $m$, and its multiplicity
label $\ell$; write $P_i=P_{m,\ell}$ and $\lambda_i=D_0+m$ for this
copy. The notation $D^+_\lambda$ denotes the positive-energy
lowest-weight representation of weight $\lambda$.

\subsection{Normalized Wick primaries}\label{sub:primaries}

For the primary copy $i$ with $P_i=P_{m,\ell}$, let
$\widetilde P_i(\mathbf p)=p^mP_i(\mathbf p/p)$, a homogeneous
polynomial of degree $m$, and write
$\widetilde P_i(\mathbf p)=\sum_{|\mathbf k|=m}c_{i,\mathbf k}
\prod_jp_j^{k_j}$, where $\mathbf k=(k_1,\ldots,k_n)$ and
$|\mathbf k|=\sum_j k_j$. Its coefficients $c_{i,\mathbf k}$ are real.
Replace \(\prod_jp_j^{k_j}\) by the Wick product of the \(k_j\)-th right
jets of species \(a_j\). All monomials have \(\sum_jk_j=m\), so this is
a finite Hermitian Wick polynomial of dimension \(\lambda=D_0+m\). It is
invariant under permutations of identical species.

To match the elementary weight-$\lambda$ convention, set

\begin{equation}\label{eq:19.12}
N_{P_i}=\left(
\frac{\prod_{j=1}^n\Gamma(2d_j)}
{\Gamma(2\lambda)\prod_a n_a!}\right)^{1/2}.
\end{equation}

The normalized undressed right primary is, explicitly,
\begin{equation}\label{eq:primary-definition}
 X_{i,R}:=N_{P_i}\sum_{|\mathbf k|=m}c_{i,\mathbf k}
       :\!\prod_{j=1}^n\Phi_{a_j,R,k_j}\!:\, .
\end{equation}
The colons denote vacuum Wick ordering, with all creation operators
placed to the left of all annihilation operators. This is a form on
the common field tests of Section~\ref{sub:fields}.

Acting on the vacuum, the unnormalized Wick product has coefficient
\(\prod_a\sqrt{n_a!}/\sqrt{\prod_j\Gamma(2d_j)}\). Equation \eqref{eq:19.10}
therefore makes its normalized radial creation distribution
\(i^m g_\lambda\). Absorb the common phase \(i^m\) in the radial basis,
leaving the field Hermitian. The polynomial orthonormalization is itself
explicit: every matrix entry follows from

\begin{equation}\label{eq:19.13}
\int_{\Delta_{n-1}}\prod_j u_j^{2d_j+k_j-1}\,d\mathbf u
=\frac{\prod_j\Gamma(2d_j+k_j)}
{\Gamma(2D_0+\sum_jk_j)}.
\end{equation}

Primarity follows from

\begin{equation}\label{eq:19.14}
\kappa_a(p^kg_{a,R,0})
=-k(k+2d_a-1)p^{k-1}g_{a,R,0},
\quad
[K,\Phi_{a,R,k}]=-ik(k+2d_a-1)\Phi_{a,R,k-1}.
\end{equation}

The homogeneous polynomial \(\widetilde P_i\) is annihilated by
\(\sum_j(p_j\partial_{p_j}^2+2d_j\partial_{p_j})\): this follows by applying the
radial/angular formula preceding \eqref{eq:19.10} to \(p^mP_i\) and using \eqref{eq:19.9}.
Consequently the finite sum of commutators \eqref{eq:19.14} cancels. Second
quantization preserves Wick ordering, since its generators annihilate
the vacuum. Thus the normalized field satisfies

\begin{equation}\label{eq:19.15}
[K,X_{i,R}]=0,\qquad [P,X_{i,R}]=-i\lambda_iX_{i,R}
\end{equation}

as form identities on the stated tests.

On an angular component of degree \(m\), \(\Gamma_s(s)\) is
\((-1)^m s_\lambda\). This can be read off from the product-Laguerre
degrees: total degree equals angular degree plus radial degree.
Let \(E_{m,\ell}\) denote the real radial embedding \eqref{eq:19.10}
before the phase choice, put \(E'_{m,\ell}=i^mE_{m,\ell}\), and let
$C_{\rm rad}f=\overline f$ be radial complex conjugation. Since
\(CE_{m,\ell}=E_{m,\ell}C_{\rm rad}\), antilinearity gives
\[
 \Theta E'_{m,\ell}f
 =(-i)^m(-1)^mE_{m,\ell}s_\lambda\overline f
 =E'_{m,\ell}s_\lambda\overline f.
\]
Thus \(\Theta\) preserves each chosen primary copy and is
\(s_\lambda\) followed by radial conjugation in that copy. The
identity holds separately for each real multiplicity vector, including
equivalent copies of the same lowest-weight representation. We put
$X_{i,L}=\Theta X_{i,R}\Theta$.

A primary of weight $\lambda_i$ has the same boundary redshift as
an elementary field of that weight. We therefore apply the dressing
explained in Section~\ref{sub:hamiltonians}, and use reflection for
the opposite boundary. Define the dressed forms by
\begin{equation}\label{eq:dressed}
 O_{i,R}(x)=2^{\lambda_i}e^{-\lambda_i x/2}X_{i,R},
 \qquad O_{i,L}=JO_{i,R}J.
\end{equation}
In the $U^*$ gauge, $O_{i,R}$ is
$V_i(x)=2^{\lambda_i}e^{-\lambda_i x}X_{i,R}$.
All products of fields used below are finite Wick expansions.

\subsection{Relation with Kolchmeyer's notation}\label{sub:kolchmeyer-notation}
In \cite[Sections 3--4]{K}, the matter space $\mathcal H_0$ is our
$\mathcal F$, the global Hamiltonian $H$ is $H_m$, and
$B=(Q-K)/2$, $P=P$, $\ell=x$, $\Delta_i=\lambda_i$.
Thus $H+B=Q$ and $H-B=K$, and his canonical Hamiltonians
(4.22)--(4.23) are \eqref{eq:canonicalpair}.
His $O_i(\pi)$ and $\mathcal O_{i,R}$ are our $X_{i,R}$ and
$O_{i,R}$. Our parabolic label $0$ is not his left angular endpoint.

For $L_0=H_m$ and $L_{\pm1}=B\pm iP$, the normalized descendants are
\begin{equation}\label{eq:kolchmeyer-descendants}
 |i,n\rangle=\frac{L_{-1}^n|i,0\rangle}{\sqrt{n!(2\lambda_i)_n}},
 \qquad L_0|i,n\rangle=(\lambda_i+n)|i,n\rangle.
\end{equation}
Here $(z)_n=\Gamma(z+n)/\Gamma(z)$; $n$ is a descendant level,
not particle number. The normalization follows by induction from
$[L_1,L_{-1}]=2L_0$.
The identity
$U(-n,2d,2p)=(-1)^n n!L_n^{2d-1}(2p)$
identifies his Appendix C Laguerre functions with $(-1)^nf_n^d$.
The transforms below give his joint energy representation.
We use square-integrable wave packets throughout; fixed-energy
kets are distributions.

\section{Bounded heat regularizations}\label{sec:bounds}

We prove that positive two-sided heat regularizations of the
fields are bounded when $d_a>1/2$ and use them to generate the
boundary algebras.

We first prove the estimates for one elementary species of weight
$d>1/2$. The same estimates apply to a jet of species $a$ with
dimension $d_a+k$, with its normalization factor retained. All
annihilation estimates use the total positive momentum $Q$ of the full
Fock space. For the chosen species, abbreviate the right-gauge field by
$V(x)=2^de^{-dx}\Phi_d(0)$ and its canonical dressed form by $O_R=UVU^*$.

\subsection{Annihilation and matter heat estimates}\label{sub:annihilation}

\begin{lemma}\label{lem:annihilation} Let $h\in\mathfrak h$ satisfy
$q^{-1/2}h\in\mathfrak h$. Then $D(Q^{1/2})\subset D(a(h))$, and
for every $\psi\in D(Q^{1/2})$ one has

\begin{equation}\label{eq:6.1}
\|a(h)\psi\|\leq\|q^{-1/2}h\|\,\|Q^{1/2}\psi\|.
\end{equation}

\end{lemma}

\begin{proof}
On the $m$-particle sector use $z=(a,p)$, with counting measure
in $a$ and Lebesgue measure in $p$. Weighted Cauchy--Schwarz gives
\[
 \|(a(h)\psi)_{m-1}\|^2
 \leq \|q^{-1/2}h\|^2\,m\int p_1|\psi_m(z_1,\ldots,z_m)|^2\,d\mathbf z.
\]
By symmetry, the last factor is
$\int(\sum_{j=1}^m p_j)|\psi_m|^2\,d\mathbf z
=\|Q^{1/2}\psi_m\|^2$.
Summing the orthogonal output sectors proves \eqref{eq:6.1} for
finite-particle vectors. Particle truncations converge in the
$Q^{1/2}$ graph norm; the estimate and closedness of $a(h)$
give the assertion on $D(Q^{1/2})$.
\end{proof}

For an integer cutoff $N\ge1$, choose
\(g_{d,N}=\mathbf1_{[1/N,N]}g_d\). For $b>0$, the pull-through identity
gives

\[
a(g_{d,N})e^{-bQ}
=e^{-bQ/2}a(e^{-bp/2}g_{d,N})e^{-bQ/2}.
\]

For \(d>1/2\),

\[
\left\|\frac{e^{-bp/2}g_{d,N}}{\sqrt p}\right\|_2^2
\leq\frac{\Gamma(2d-1)}{\Gamma(2d)}b^{-(2d-1)}
=\frac{b^{-(2d-1)}}{2d-1},
\]

while \(\|Q^{1/2}e^{-bQ/2}\|=(eb)^{-1/2}\). Consequently,

\begin{equation}\label{eq:6.2}
\|a(g_{d,N})e^{-bQ}\|
\leq c_d b^{-d},\qquad
c_d=\frac1{\sqrt{e(2d-1)}}.
\end{equation}

The weighted \(L^2\) tails converge to zero. Thus these operators
converge in norm for every fixed \(b>0\) when the momentum cutoff is
removed. Taking adjoints proves the corresponding estimate for
\(e^{-aQ}a^*(g_d)\). Therefore

\begin{equation}\label{eq:6.3}
\boxed{\|e^{-aQ}\Phi_d(0)e^{-bQ}\|
\leq c_d(a^{-d}+b^{-d}),\qquad a,b>0.}
\end{equation}

Here \eqref{eq:6.3} defines the bounded extension of the sandwich; it does not
assert that \(\Phi_d(0)\) is an operator.

\subsection{Weighted semigroup estimates}\label{sub:heat-estimates}

Put $\mathsf W=(Q+2)e^{-x}$, a positive multiplication operator.
Then $L=\tfrac12k^2+\mathsf W$ is associated with \eqref{eq:3.3}.

\paragraph{Weighted energy identity}
For $\psi\in\mathcal H$, write $u_s=e^{-sL}\psi$. For $R>0$ and
integers $j\geq0$, define the bounded multiplication operator
$w_R=\mathsf W(1+\mathsf W/R)^{-1}$ and the scalar function
\(F_{j,R}(s)=\langle u_s,w_R^ju_s\rangle.\)
Here $w_R^0=1$. In a fixed momentum fiber, differentiation in $x$
gives
\[
 \partial_x^2(w_R^j)
 =\frac{j(j-\mathsf W/R)}{(1+\mathsf W/R)^2}w_R^j
 \leq j^2w_R^j.
\]
Multiplication by $w_R^j$ preserves the form domain: it is bounded,
has bounded first length derivative, and commutes with $\mathsf W$.
Since $u_s\in D(L)$ for $s>0$, the form identity and integration by
parts therefore give
\begin{equation}\label{eq:weighted-energy}
\begin{aligned}
 F_{j,R}'(s)
 &=-\int\langle\partial_xu_s,w_R^j\partial_xu_s\rangle\,dx
   +\frac12\int\langle u_s,\partial_x^2(w_R^j)u_s\rangle\,dx
   -2\|\mathsf W^{1/2}w_R^{j/2}u_s\|^2\\
 &\leq \frac{j^2}{2}F_{j,R}(s)-2F_{j+1,R}(s).
\end{aligned}
\end{equation}
The integrals and inner products include all momentum variables
and particle sectors. The calculation follows first with compact
length cutoffs; the form norm removes them. The inequality uses
$\mathsf W\geq w_R$, and its constants are independent of $R$,
the momenta and particle number.

For an integer $m\geq0$, set $G_{m,R}(s)=\sum_{j=0}^m\frac{(2s)^j}{j!}F_{j,R}(s)$. In its derivative, the term obtained by differentiating $(2s)^j$
cancels the negative term with index $j-1$ in
\eqref{eq:weighted-energy}. Consequently
$G_{m,R}'(s)\leq(m^2/2)G_{m,R}(s)$ and
$G_{m,R}(0)=\|\psi\|^2$. Gronwall's inequality, followed by
monotone convergence as $R\to\infty$, proves
\begin{equation}\label{eq:6.4}
 \|\mathsf W^{m/2}e^{-tL}\|
 \leq\frac{\sqrt{m!}}{(2t)^{m/2}}e^{m^2t/4},
 \qquad t>0,\quad m=0,1,2,\ldots.
\end{equation}
In particular the range of $e^{-tL}$ lies in every integer power
domain of $\mathsf W$.

To factor a heat sandwich through inverse powers of the potential,
we need to multiply heat evolution by the corresponding positive
powers. The preceding estimate makes these corrected heat operators
bounded. For an integer $r\geq1$, define
\begin{equation}\label{eq:6.5}
 \begin{split}
 B_r(t)&=(1+t\mathsf W)^r e^{-tL},\\
 b_r(t)&=\sum_{j=0}^r\binom rj
          \frac{\sqrt{(2j)!}}{2^j}e^{j^2t}.
 \end{split}
\end{equation}
The binomial formula and \eqref{eq:6.4} give
$\|B_r(t)\|\leq b_r(t)$. For fixed $r$, the function $b_r$ is
bounded on every bounded interval of nonnegative times.

We also record the elementary semigroup domination needed below.
For $q\geq0$, let
$L_q=-\tfrac12\partial_x^2+(q+2)e^{-x}$ be the scalar momentum
fiber, and set $p_t(v)=(2\pi t)^{-1/2}e^{-v^2/(2t)}$.
Its heat semigroup has a measurable positive kernel
$s_t(q;x,y)$ satisfying
\begin{equation}\label{eq:6.6}
 0\leq s_t(q;x,y)\leq p_t(x-y).
\end{equation}
Indeed, the product formula for nonnegative closed-form sums
\cite[equations (1.3)--(1.4)]{TK} approximates $e^{-tL_q}$ by
products of Gaussian convolution and multiplication by functions
between zero and one. Every product is positivity preserving and
dominated by free heat evolution. The strong limit has the same
domination, which supplies the kernel and its bound. Truncating the
potential and using monotone convergence of closed forms
\cite[Theorems 3.1 and 4.2]{SMC} gives the same conclusion. The
operator-valued kernel of $e^{-tL}$ is $S_t(x,y)=s_t(Q;x,y)$,
defined by the spectral calculus of $Q$.

\subsection{Resolvent regularization and removal of cutoffs}\label{sub:schur}

Set $V_N(x)=2^de^{-dx}\Phi(g_{d,N})$, fix heat times $t,u>0$,
and choose an integer $r>d$. Define
$R_r(t)=(1+t\mathsf W)^{-r}$. The spectral theorem
and Euler's Gamma integral give the norm-convergent integral
\begin{equation}\label{eq:resolvent-laplace}
 R_r(t)=\frac1{\Gamma(r)}\int_0^\infty
       s^{r-1}e^{-s}e^{-st\mathsf W}\,ds.
\end{equation}
We first regularize $V_N$ by these resolvent powers. At each fixed
length $x$, put $a=st e^{-x}$ and $b=vu e^{-x}$ for $s,v>0$.
The matter estimate \eqref{eq:6.3} gives
\begin{equation}\label{eq:7.1}
\begin{aligned}
 &\|e^{-st\mathsf W(x)}V_N(x)e^{-vu\mathsf W(x)}\|\\
 &\qquad=2^de^{-dx}e^{-2(a+b)}
            \|e^{-aQ}\Phi(g_{d,N})e^{-bQ}\|\\
 &\qquad\leq 2^dc_d\bigl[(st)^{-d}+(vu)^{-d}\bigr].
\end{aligned}
\end{equation}
Thus the length dependence cancels before any integration. For
$0\leq\alpha<r$, write
$\gamma_r(\alpha)=\Gamma(r-\alpha)/\Gamma(r)$. Integrating
\eqref{eq:7.1} against the two Gamma densities defines the bounded
decomposable operator
$Z_N^{t,u}=R_r(t)V_NR_r(u)$ and proves
\begin{equation}\label{eq:7.2}
 \|Z_N^{t,u}\|
 \leq 2^dc_d\gamma_r(d)(t^{-d}+u^{-d}).
\end{equation}
This notation denotes the bounded extension of the initial form.
The double integral converges in operator norm, uniformly in the
length variable and in $N$.

The range assertion following \eqref{eq:6.4} now gives the
factorization
\begin{equation}\label{eq:7.3}
 T_N:=e^{-tL}V_Ne^{-uL}
       =B_r(t)^*Z_N^{t,u}B_r(u).
\end{equation}
It is first an identity of forms on finite-particle tests and then
an identity of bounded extensions. Indeed \eqref{eq:7.2} extends
$V_N$ continuously to
$D((1+t\mathsf W)^r)\times D((1+u\mathsf W)^r)$, and both heat
operators take every vector into these domains. Consequently
\begin{equation}\label{eq:7.4}
 \begin{split}
 \sup_N\|T_N\|&\leq C_d(t,u),\\
 C_d(t,u)&=2^dc_d\gamma_r(d)b_r(t)b_r(u)(t^{-d}+u^{-d}).
 \end{split}
\end{equation}
One may fix $r=\lfloor d\rfloor+1$ throughout. In particular
$C_d(t,t)=O(t^{-d})$ as $t\downarrow0$. The estimate holds on the
entire Fock space, since the only matter estimate used was
Lemma~\ref{lem:annihilation}.

For fixed $x$ and positive $s,v$, the cutoff matter sandwich in
\eqref{eq:7.1} converges in operator norm by Section~\ref{sub:annihilation}.
The integrable majorant in that equation is independent of $N$.
Dominated convergence in the two Gamma integrals therefore gives a
bounded measurable operator field $Z^{t,u}(x)$ with
\begin{equation}\label{eq:7.4a}
 \|Z_N^{t,u}(x)-Z^{t,u}(x)\|\longrightarrow0,
 \qquad
 \sup_{N,x}\|Z_N^{t,u}(x)\|
 \leq2^dc_d\gamma_r(d)(t^{-d}+u^{-d}).
\end{equation}
For $\psi\in L^2(\mathbb R;\mathcal F)$, another dominated
convergence argument gives
\begin{equation}\label{eq:7.4b}
 \|(Z_N^{t,u}-Z^{t,u})\psi\|^2
 =\int_{\mathbb R}
   \|(Z_N^{t,u}(x)-Z^{t,u}(x))\psi(x)\|^2\,dx
 \longrightarrow0.
\end{equation}
The same proof applies to the adjoints. Multiplying by the fixed
bounded factors in \eqref{eq:7.3} proves strong-$*$ convergence to
$T_{t,u}=B_r(t)^*Z^{t,u}B_r(u)$. This is the canonical bounded
extension denoted by $e^{-tL}Ve^{-uL}$. It agrees with the original
heat-regularized form, and is independent of the choice of $r>d$,
because every such factorization is the limit of the same cutoff
forms. The proof also applies to any momentum cutoff of modulus at
most one converging pointwise to one, with conjugate cutoffs for
adjoints. This proves regulator independence in the stated
strong-$*$ sense.

In particular, for $t=\varepsilon/(2\phi_b)$,
\begin{equation}\label{eq:1.1}
 \|e^{-\varepsilon H_R}O_Re^{-\varepsilon H_R}\|
 \leq 2^{d+1}c_d\gamma_r(d)b_r(t)^2t^{-d},
 \qquad r\in\mathbb N,\quad r>d.
\end{equation}

\subsection{Graph-continuous forms and time smearing}\label{sub:distributions}

The limits inherit the heat compatibility identities, for \(t'\geq t\),
\(u'\geq u\),

\begin{equation}\label{eq:7.5}
T_{t',u'}=e^{-(t'-t)L}T_{t,u}e^{-(u'-u)L}.
\end{equation}

They follow for the cutoff forms and pass to the strong limits because
multiplication by a fixed bounded operator is strongly continuous.

Equation \eqref{eq:7.5} defines a point form on the union of bounded spectral
subspaces of \(L\): for vectors supported in \([0,R]\), put $q_V(\psi,\varphi)=\langle e^{tL}\psi,T_{t,t}e^{tL}\varphi\rangle$. It is independent of \(t>0\). Moreover it extends continuously to
\(D((1+L)^r)\times D((1+L)^r)\) for any \(r>d\). Here is a proof which
also identifies it with the original form. Define the dyadic spectral
projections $\Pi_0=\ind_{[0,1)}(L)$ and
$\Pi_j=\ind_{[2^{j-1},2^j)}(L)$ for integers $j\ge1$. Using \eqref{eq:7.4} with \(t\asymp(1+2^j+2^k)^{-1}\)
gives, uniformly in the cutoff, $\|\Pi_jV_N\Pi_k\|\leq C(1+2^j+2^k)^d$. The block series after multiplication on both sides by \((1+L)^{-r}\) is
absolutely convergent in operator norm, since

\[
\sum_{j,k\geq0}2^{d\max(j,k)-r(j+k)}<\infty\qquad(r>d).
\]

To identify the de-regularized form with $V_N$ on $\mathcal D$, insert \(e^{-vL}\)
on both test vectors and let \(v\downarrow0\). The semigroup preserves
particle number and momentum support. Its Gaussian kernel bound gives
convergence in every fixed exponential length weight for compactly
supported tests. The resulting cutoff-field pairings therefore converge
to those of \(V_N\). The graph-domain form has the same limit, because
\(e^{-vL}\) converges in every power graph norm on \(\mathcal D\).

On each pair of blocks the cutoff limit is \eqref{eq:7.5}.
The summable bound gives a bounded limiting weighted form.
On $\mathcal D$ it equals the original momentum-integral form
by convergence of cutoff pairings in graph norm.

Transfer this form by \(U\) and rescale the Hamiltonian to \(H_R\). On
the invariant smooth-energy domain
$\bigcap_{m\ge0}D((1+H_R)^m)$, let $O_R(0)$ denote the form at
boundary time zero and, for boundary time $v\in\mathbb R$, define \(O_R(v)=e^{ivH_R}O_R(0)e^{-ivH_R}\)

as a form. Its pairings are continuous, hence define a distribution.
For a scalar time test $f$, set $O_R(f)=\int f(v)O_R(v)\,dv$ in form
pairings. For $\epsilon>0$, define its bounded point sandwich by
\[
 T_\epsilon:=e^{-\epsilon H_R}O_R(0)e^{-\epsilon H_R}
       =U T_{t,t}U^*,\qquad t=\epsilon/(2\phi_b).
\]
For an approximate identity \(f_\delta\) with integral one, uniformly
bounded \(L^1\) norms and mass concentrating at zero,

\begin{equation}\label{eq:7.6}
e^{-\epsilon H_R}O_R(f_\delta)e^{-\epsilon H_R}
=\int f_\delta(v)e^{ivH_R}T_\epsilon e^{-ivH_R}dv.
\end{equation}

The right side is a strong operator integral on each vector. Strong
continuity of the unitary group implies convergence strongly to
\(T_\epsilon\), independently of the approximate identity. Its norm is
at most \(\|f_\delta\|_1\|T_\epsilon\|\). For nonnegative unit-mass
mollifiers the constant is exactly \(\|T_\epsilon\|\). Taking adjoints
gives strong-star convergence when the adjoint mollifiers are treated in
the same way.

Thus the regularized time smearings converge strongly together with adjoints to the bounded point sandwich.

\subsection{Wick polynomials and the boundary algebras}\label{sub:wick}

We next treat finite homogeneous Wick polynomials. For a monomial involving species $a_j$ and derivative orders $k_j$, put $D_j=d_{a_j}+k_j$.

Let $m\geq1$, let $h_j\in\mathfrak h$ satisfy
$q^{-1/2}h_j\in\mathfrak h$ for $1\leq j\leq m$, and let
$\psi\in D(Q^{m/2})$. Then

\begin{equation}\label{eq:8.1}
\|a(h_1)\cdots a(h_m)\psi\|
\leq\prod_{j=1}^m\|h_j/\sqrt p\|_2\,\|Q^{m/2}\psi\|.
\end{equation}

\begin{proof}[Proof of \eqref{eq:8.1}]
On the $n$-particle sector, apply Cauchy--Schwarz in the $m$
annihilated variables with weight $p_1\cdots p_m$.
Writing $c_j=\|q^{-1/2}h_j\|$ gives
\[
 \|a(h_1)\cdots a(h_m)\psi_n\|^2
 \leq\left(\prod_j c_j^2\right)\frac{n!}{(n-m)!}
                \int p_1\cdots p_m|\psi_n|^2\,d\mathbf z.
\]
Symmetry replaces the factorial and selected momenta by the sum
of $p_{i_1}\cdots p_{i_m}$ over ordered distinct indices.
That sum is at most $(\sum_jp_j)^m$, since the omitted terms
are nonnegative. Sum the orthogonal sectors to obtain the bound.
For a general vector in $D(Q^{m/2})$, particle truncation converges
in every lower power graph norm. Apply the same bound to each
suffix of the product and use closedness of the annihilators
successively. This identifies the limit with the stated operator
composition and proves its domain assertion.
\end{proof}

In applying \eqref{eq:8.1} to distributional jets, first insert
momentum cutoffs. The heat-damped weighted one-particle norms then
control their removal, just as in Section~\ref{sub:annihilation}.

Put \(D_A=\sum_{j=1}^mD_j\). Pulling half the heat
through all annihilations, using \eqref{eq:8.1}, and maximizing
\(q^{m/2}e^{-tq/2}\) gives

\begin{equation}\label{eq:8.2}
\left\|\prod_{j=1}^m a(g_{a_j,R,k_j})e^{-tQ}\right\|
\leq A_{\mathbf k}\,t^{-D_A},
\end{equation}

where

\[
A_{\mathbf k}=\left(\frac m e\right)^{m/2}
\prod_{j=1}^m\sqrt{\frac{\Gamma(2D_j-1)}{\Gamma(2d_{a_j})}}.
\]

For \(m=0\) set \(A_{\varnothing}=1\), \(D_A=0\). All Gamma integrals
converge because $d_{a_j}>1/2$.

Let $C$ and $A$ be finite lists of creation and annihilation jets,
respectively, and define
$X_{C,A}=\prod_{j\in C}a^*(g_{a_j,R,k_j})
           \prod_{j\in A}a(g_{a_j,R,k_j})$.
Write $D_C=\sum_{j\in C}(d_{a_j}+k_j)$ and
$D_A=\sum_{j\in A}(d_{a_j}+k_j)$. Let $A_C$ and $A_A$ be the constants
$A_{\mathbf k}$ above, evaluated on the corresponding lists, with the
same empty-list convention. This normal-ordered term satisfies

\begin{equation}\label{eq:8.3}
\|e^{-aQ}X_{C,A}e^{-bQ}\|
\leq A_C A_A a^{-D_C}b^{-D_A}.
\end{equation}

Write a homogeneous Wick polynomial as
\(X=\sum_\nu c_\nu X_{C_\nu,A_\nu}\), where \(D_C+D_A=D\) for every
term. Its dressed form after conjugation is \(2^De^{-Dx}X\).
Choose an integer $r>D$. At a fixed length $x$, apply
\eqref{eq:8.3} with $a=st e^{-x}$ and $b=vu e^{-x}$.
The factor $e^{-Dx}$ cancels $e^{(D_C+D_A)x}$ exactly.
Integrating against the Gamma densities in
\eqref{eq:resolvent-laplace}, and then applying the two factors
$B_r(t)^*$ and $B_r(u)$, proves the bound
\begin{equation}\label{eq:8.4}
\begin{aligned}
 C_X(t,u)=2^D b_r(t)b_r(u)\sum_\nu
 &|c_\nu|A_{C_\nu}A_{A_\nu}
   \gamma_r(D_{C_\nu})\gamma_r(D_{A_\nu})\\
 &\times t^{-D_{C_\nu}}u^{-D_{A_\nu}}.
\end{aligned}
\end{equation}
Here $b_r$ and $\gamma_r$ are defined in
Sections~\ref{sub:heat-estimates} and~\ref{sub:schur}; in particular
$\gamma_r(0)=1$. Thus empty creation or annihilation lists require
no separate estimate. Since every summand has total dimension $D$,
$C_X(t,t)=O(t^{-D})$ as $t\downarrow0$.

For each fixed $x$ and positive Gamma-integration variables, the
heat-damped matter monomials converge in operator norm as their
momentum cutoffs are removed. Their bounds are integrable because
$r>D$. The proof of \eqref{eq:7.4a}--\eqref{eq:7.4b}, applied to
the finite normal-ordered expansion, therefore gives strong-$*$
convergence of the resolvent-regularized polynomial. The bounded
factorization \eqref{eq:7.3} gives strong-$*$ convergence of its
heat sandwich. The form and time-smearing arguments of
Section~\ref{sub:distributions} then apply with $d$ replaced by $D$.

Polynomial Hamiltonian factors cause no new difficulty. For
\(n,m\geq0\), split the positive heat times in half and use $\|H_R^ne^{-\epsilon H_R/2}\| =\left(\frac{2n}{e\epsilon}\right)^n$, with value one when \(n=0\). Thus every asymmetric regularized letter is bounded. Its adjoint is the corresponding letter for
\(X^*\) with the left and right times and powers exchanged. Finite
linear combinations and finite products form a genuine \(*\)-algebra of
bounded operators. Its double commutant is consequently well-defined.

For a jet of species $a$ and order $k$, the one-particle integral used
in this proof is
\begin{equation}\label{eq:19.25}
 \|q^{-1/2}e^{-Aq}p^kg_{a,R,0}\|^2
 =\frac{\Gamma(2(d_a+k)-1)}{\Gamma(2d_a)(2A)^{2(d_a+k)-1}}.
\end{equation}
The iterated annihilation bound controls all spectator species and
all particle numbers. Since a fixed polynomial contains only finitely
many jets, the proof takes no sum over the family of all primary
fields.

For $i\in\mathcal P$ and $a,b>0$, denote the bounded
strong-$*$ cutoff limit by
\begin{equation}\label{eq:bounded-letter-definition}
 T_{i;a,b}=e^{-aH_R}O_{i,R}e^{-bH_R}\in B(\mathcal H).
\end{equation}
For Hermitian $O_{i,R}$, $T_{i;a,b}^*=T_{i;b,a}$.

\begin{definition}\label{def:algebras}
Let $\mathfrak A_0$ be the nonunital $*$-algebra of finite linear
combinations of finite products of positive heat operators and the
bounded letters $T_{i;a,b}$. Polynomial Hamiltonian factors are allowed
when positive heat intervals are retained. We define
\begin{equation}\label{eq:16.1}
 M_R=\mathfrak A_0'',\qquad M_L=JM_RJ.
\end{equation}
\end{definition}
The closure agrees with the algebra generated by the bounded spectral
functions of $H_R$ and the symmetric letters. Indeed, a positive heat
has the same spectral projections as $H_R$, and an asymmetric letter
factors through the symmetric one at the smaller time. Moreover
$e^{-tH_R}\to1$ strongly. The Hamiltonian is affiliated with $M_R$;
it is not a bounded element of $M_R$.

\section{Strong commutation and endpoint domains}\label{sec:commutation}

We identify the separated self-adjoint Hamiltonians and prove
commutation with opposite-boundary fields.

\subsection{Separation of the Hamiltonians}\label{sub:separation}

Work first in the $U^*$ gauge and set
$L_R=U^*A_RU=L$ and $L_L=U^*A_LU$. On a nonvacuum
representation of weight \(\lambda\), the pair becomes

\begin{equation}\label{eq:9.8}
L_R=\tfrac12 k^2+(p+2)e^{-x},\qquad
L_L=\tfrac12(k+P)^2+K+2e^{-x}.
\end{equation}

Now set \(y=x-\log((p+2)/2)\) and
\((W\psi)(y,p)=\psi(y+\log((p+2)/2),p)\). This is unitary with Jacobian
one. Let

\[
L_0=-\tfrac12\partial_y^2+2e^{-y},\qquad
B_\lambda=-\tfrac12\partial_p p(p+2)\partial_p
-\tfrac18+\frac{(\lambda-1/2)^2}{p}.
\]

Then on compact smooth tests

\begin{equation}\label{eq:9.9}
WL_RW^*=L_0\otimes1,\qquad
WL_LW^*=B_\lambda+\frac2{p+2}L_0.
\end{equation}

Here is the cancellation responsible for \eqref{eq:9.9}. Write
\(D=-i\partial_p\), \(a=(p+2)^{-1}\) and \(h_1=2/(p+2)\). Under \(W\),
\(P\) becomes \(P-pa k\) and \(D\) becomes \(D-ak\). Consequently
\(k+P\) becomes \(P+h_1k\). The coefficient of \(kD\) in
\(\tfrac12(P+h_1k)^2+(D-ak)p(D-ak)\) is \(ph_1-2pa=0\). Its remaining
coefficient of \(k\) is \(-i(ph_1'/2+h_1/2)+i(pa)'=0\). The coefficient
of \(k^2\) is \(h_1^2/2+pa^2=(p+2)^{-1}\). The \(p\)-only terms are
\(B_\lambda\), and the potential is \(4e^{-y}/(p+2)\).

There is also an equality of closed forms here. Indeed
\(B_\lambda=\tfrac12P^2+K\) on compact tests, so its Friedrichs form is
positive. Let $K_\nu$ denote the modified Bessel function of the
second kind. The normalized Kontorovich--Lebedev transform in $y$ is
\[
 (\mathcal K f)(s)=\int_{\mathbb R}2K_{2is}(4e^{-y/2})f(y)\,dy,
 \qquad \mathcal K:L^2(dy)\longrightarrow L^2(d\mu(s)),
\]
initially on compact smooth tests and then by unitary extension; its
normalization is verified in Section~\ref{sub:kl}. It satisfies
$\mathcal K L_0\mathcal K^*=M_{s^2/2}$. After this transform, the
second operator in \eqref{eq:9.9} is

\begin{equation}\label{eq:9.10}
\int_0^\infty{}^\oplus B_{\lambda,s}\,d\mu(s),\qquad
B_{\lambda,s}=B_\lambda+\frac{s^2}{p+2}.
\end{equation}

The added potential is bounded for each fixed \(s\), with a uniform
bound on bounded \(s\) intervals. Thus \(D(B_{\lambda,s})=D(B_\lambda)\)
on every fiber. Finite sums of compact-energy functions times a form
core for \(B_\lambda\) form a core for the direct-integral form:
truncate \(s\), then approximate simple vector-valued functions in the
form norm. Replacing the \(y\) functions by a form-core approximation
for \(L_0\) gives the same closed form as the closure of \eqref{eq:9.8}. This
proves that \eqref{eq:9.10} is the previously chosen physical \(L_L\).

\subsection{The Friedrichs endpoint condition}\label{sub:endpoint}

The boundary condition at zero must be inherited from the
original positive representation, not chosen from the separated
differential expression alone.

For \(\lambda>1/2\), put \(\alpha=\lambda-1/2\) and let

\begin{equation}\label{eq:19.16}
B_{\lambda,s}=-\tfrac12\partial_p p(p+2)\partial_p
-\tfrac18+\frac{\alpha^2}{p}+\frac{s^2}{p+2},\qquad s\geq0.
\end{equation}

This is the Friedrichs form sum
$\tfrac12P^2+\kappa_\lambda+s^2/(p+2)$.
The maximal differential domain consists of $f\in L^2(dp)$ with
$f,p(p+2)f'$ locally absolutely continuous and
$B_{\lambda,s}f\in L^2(dp)$.

Put $p=2\sinh^2u$ and use the unitary square-root Jacobian from
$L^2(dp)$ to $L^2(du)$. Direct calculation gives
\begin{equation}\label{eq:9.11}
 8B_{\lambda,s}\simeq-\partial_u^2+
 \frac{(2\lambda-1)^2-1/4}{\sinh^2u}
 +\frac{4s^2+1/4}{\cosh^2u}.
\end{equation} At zero
its two indicial behaviors are \(u^{1/2\pm(2\lambda-1)}\), or
\(p^{\pm\alpha}\) before the square-root Jacobian. Infinity is limit
point: in the \(u\) variable the potential decays exponentially, so at a
nonreal energy exactly one of the two Volterra solutions is square
integrable at infinity. At zero the endpoint is limit point when
\(\lambda\geq1\), and limit circle when \(1/2<\lambda<1\), since both
\(p^{\alpha}\) and \(p^{-\alpha}\) then belong to \(L^2(0,1;dp)\).

In the latter case the Friedrichs condition is

\begin{equation}\label{eq:19.17}
\boxed{\quad b_-(f):=\lim_{p\downarrow0}p^\alpha f(p)=0.\quad}
\end{equation}

Here is the local calculation fixing its meaning. Solutions of
\((B_{\lambda,s}+1)u=0\) have expansions
\(u_\pm(p)=p^{\pm\alpha}(1+O(p))\). The exponent difference \(2\alpha\)
lies strictly between zero and one, so no resonant logarithm occurs. For
\(f\) in the maximal domain, variation of parameters with
\(g=(B_{\lambda,s}+1)f\in L^2\) gives

\begin{equation}\label{eq:19.18}
f=b_-u_-+b_+u_++R,
\qquad R=o(p^{1/2}),\quad pR'=o(p^{1/2}).
\end{equation}

Indeed the two remainder terms are constant multiples of
\(u_+(p)\int_0^p u_-(t)g(t)dt\) and \(u_-(p)\int_0^p u_+(t)g(t)dt\).
Cauchy-Schwarz bounds each by \(Cp^{1/2}\|g\|_{L^2(0,p)}\);
differentiation gives the second assertion. Thus \eqref{eq:19.17} is exactly the
coefficient of the nonprincipal solution. For \(b_-=0\), the second
boundary coefficient is \(b_+=\lim p^{-\alpha}f(p)\).

The $p^{-\alpha}$ term has infinite form energy, whereas a cutoff
of $p^\alpha$ has finite energy. Thus $b_-=0$ is the Friedrichs
condition by \cite[Theorem 4.7 and Remark 4.8(v)]{GLN};
semiboundedness follows from the positive form sum. This also
identifies the maximal and minimal boundary domains.

Define the weighted regular test space

\begin{equation}\label{eq:19.19}
\mathcal C_\lambda=
\{p^{\lambda-1/2}F(p):F\in C_c^\infty([0,\infty))\}.
\end{equation}

Here \(C_c^\infty([0,\infty))\) means restrictions of smooth functions
on a neighborhood of the closed half-line, with compact support on that
half-line. This is an operator core for every $B_{\lambda,s}$. For \(\lambda\geq1\), interior compact smooth
functions already form a core and are contained in \eqref{eq:19.19}. For
\(1/2<\lambda<1\), the minimal domain has both boundary coefficients
zero; the Friedrichs domain has only \(b_-=0\). Hence the quotient of
the latter by the former is one dimensional. If \(\chi=1\) near zero is
smooth and compactly supported, \(\chi p^\alpha\) has \(b_+=1\) and lies
in the Friedrichs operator domain. Directly, near zero,

\begin{equation}\label{eq:19.20}
B_{\lambda,s}p^\alpha=
\left(-\frac{\lambda^2}{2}+\frac{s^2}{p+2}\right)p^\alpha\in L^2(0,1).
\end{equation}

Hence $C_c^\infty(0,\infty)+\mathbb C\chi p^\alpha$ is graph dense
in the Friedrichs domain and contained in \eqref{eq:19.19}.
Every vector in \eqref{eq:19.19} satisfies the boundary condition
and has square-integrable differential image, proving the core
assertion. The bounded perturbation $s^2/(p+2)$ preserves the domain,
with uniformly equivalent graph norms for bounded $s$.

Interior compact tests remain a form core throughout. To
specify the cutoff rate, choose a smooth function \(\theta\) equal to
zero near zero and to one on \([2,\infty)\). The discarded part of a
vector equal to \(p^\alpha\) near zero is
\(v_\varepsilon(p)=p^\alpha[1-\theta(p/\varepsilon)]\) for sufficiently
small \(\varepsilon\). The substitution \(p=\varepsilon t\) gives

\begin{equation}\label{eq:19.20a}
\begin{aligned}
\int_0^\infty\left(p|v_\varepsilon'|^2+
       \frac{\alpha^2}{p}|v_\varepsilon|^2\right)dp
 &=C_{\alpha,\theta}\varepsilon^{2\alpha},\\
\|v_\varepsilon\|_2^2
 &=D_{\alpha,\theta}\varepsilon^{2\alpha+1},\qquad
\|Pv_\varepsilon\|_2^2=O(\varepsilon^{2\alpha+1}).
\end{aligned}
\end{equation}

Thus the form-norm error is $O(\varepsilon^\alpha)$.
Interior tests are a form core, but in the limit-circle range
they are not an operator core.

\subsection{Operator cores and the opposite Hamiltonian}\label{sub:ward}

We now identify the separated operators with the prescribed
self-adjoint realizations. Using the transform $\mathcal K$ of
Section~\ref{sub:separation}, put $\mathcal V=(\mathcal K\otimes1)W$. The length and spectral
representations are related by

\begin{equation}\label{eq:19.21}
\begin{aligned}
WL_RW^*&=L_0\otimes1,&
WL_LW^*&=B_\lambda+\frac{2}{p+2}L_0,\\
\mathcal V L_R\mathcal V^*&=M_{s^2/2},&
\mathcal V L_L\mathcal V^*
 &=\int_0^\infty{}^\oplus B_{\lambda,s}\,d\mu(s).
\end{aligned}
\end{equation}

The first line acts in the \(y,p\) representation, and the second in
\(L^2(d\mu(s);L^2(dp))\). Both lines use exactly the physical Friedrichs
forms. Equality of their closed forms follows by truncating \(s\), using
interior compact form cores and the bounded perturbation bound in
Section~\ref{sub:endpoint}. In the second line the first Hamiltonian is scalar
multiplication and the second is decomposable over the same \(s\)
variable. Their spectral projections commute. Thus \(H_R\) and \(H_L\)
strongly commute for every summand \eqref{eq:19.11}.

The operator core needed for the field commutator can now be taken, on
each finite occupation sector, to consist of finite sums of

\begin{equation}\label{eq:19.22}
\psi(x,\mathbf p)=\prod_jp_j^{d_j-1/2}F(x,\mathbf p),
\qquad F\in C_c^\infty(\mathbb R\times[0,\infty)^n),
\end{equation}

with the required permutation symmetry. These vectors belong to the
physical operator domain in the gauge of \eqref{eq:9.8}: the singular terms of
every \(\kappa_{a_j}\) cancel on \(p_j^{d_j-1/2}\), leaving that same
weight times a smooth compactly supported function. They lie in the
closed form domain by the cutoff calculation just given. Integration by
parts against the interior form core identifies their differential image
with their operator image.

The right-field pairings on \eqref{eq:19.22} agree with the previously
constructed forms. Indeed momentum cutoffs away from zero commute with
\(k,Q\) in the right gauge and converge in every fixed \(L\)-power graph
norm on these vectors. The graph-continuous extension of Section~\ref{sub:distributions}
therefore identifies their limits with the absolutely convergent
weighted momentum pairings.

To prove graph density, first take finitely many angular summands in
\eqref{eq:19.11} and truncate \(s\) in \eqref{eq:19.21}. Approximate the radial vectors by
\eqref{eq:19.19} and the inverse spectral transforms in the graph norm of \(L_0\)
by compact smooth \(y\) functions. The estimate \(2/(p+2)\leq1\)
controls this second approximation in the full graph norm. For \(n=1\),
changing back from \(y\) to \(x\) takes these vectors into \eqref{eq:19.22}. For
\(n\geq2\), \(\lambda\geq D_0>1\), so the radial tests may be supported
away from total momentum zero. Then \eqref{eq:19.10}, with the homogeneous
polynomial \(\widetilde P_i\), is a product \(\prod p_j^{d_j-1/2}\) times
a smooth compactly supported function up to each individual face; all
inverse powers of total momentum occur away from zero. It also belongs
to \eqref{eq:19.22}. Finite angular, particle and species truncations are graph
dense. This proves that \eqref{eq:19.22} is an operator core for \(L_L\) on the
full Fock space.

There is no boundary anomaly in the Ward calculation on this core. If
\(u=p^\alpha F\) and \(v=p^\alpha G\) are regular weighted tests at one
face, the Green boundary term of \(\kappa\) is

\begin{equation}\label{eq:19.23}
p(\overline u v'-\overline{u'}v)
=p^{2\alpha+1}(\overline F G'-\overline{F'}G)
=p^{2d}(\overline F G'-\overline{F'}G)\longrightarrow0.
\end{equation}

Pairing a jet \(p^{\alpha+k}\) with a weighted test gives the same
conclusion: the extra term for \(k>0\) is \(O(p^{2\alpha+k})\), and for
\(k=0\) it cancels. The first-order dilation boundary term is
\(p\overline uv=O(p^{2d})\). All tend to zero since \(\alpha>0\). Apply
this one coordinate at a time; the remaining weights are integrable, so
Fubini is legitimate for each finite Wick term.

Right heat evolution preserves momentum support and the regular weighted
behavior in \eqref{eq:19.22}. To verify the required regularity,
use the scalar fibers $L_q$ from Section~\ref{sub:heat-estimates}.
Their forms have common domain and depend holomorphically on $q$
for $\operatorname{Re}q>-2$. The associated semigroups are
holomorphic in this parameter, by the theory of holomorphic closed
forms \cite[Chapters VII and IX]{KatoPert}. For real $q\geq0$, they
are contractions throughout $|z-q|\leq1$; Cauchy's estimate hence
gives $\|\partial_q^j e^{-tL_q}\|\leq j!$ for every integer $j\geq0$.
Since $q=\sum p_j$ on a particle sector, this controls all momentum
derivatives, including at an individual zero-momentum face.
The weighted energy argument also applies on these parameter disks,
with the positive potential $(\operatorname{Re}z+2)e^{-x}$;
the imaginary potential contributes zero to the real energy identity.
Cauchy's estimates therefore hold in the resulting weighted norms.
Together with the equation for $L_q$ and positive-time semigroup
regularity, this controls length derivatives as well. The Gaussian
domination \eqref{eq:6.6} gives the required length tails on compact
tests. Equivalently one may insert compact length cutoffs in the
form identity and let them tend to one; the weighted $L^2$ bounds
make the resulting boundary terms vanish. These estimates justify
the integrations by parts after heat evolution, with
\eqref{eq:19.23} at each momentum face.
In the right gauge the identities \eqref{eq:19.15} imply
$[k+P,V_i]=[K,V_i]=0$, hence $[L_L,V_i]=0$ as forms. Since the two
Hamiltonians strongly commute, the preceding integration by parts
gives, for a bounded right sandwich $T$ and weighted core vectors,
\begin{equation}\label{eq:9.14}
 \langle L_L\psi,T\varphi\rangle
 =\langle\psi,TL_L\varphi\rangle.
\end{equation}

If
\(T\) is the bounded right heat sandwich, the core identity
\(\langle L_L\psi,T\varphi\rangle=\langle\psi,TL_L\varphi\rangle\) first
implies \(T\varphi\in D(L_L^*)=D(L_L)\) for core \(\varphi\).
Approximation of any \(\varphi\in D(L_L)\) in graph norm and closedness
give \(L_LT\varphi=TL_L\varphi\) on the full domain. Substituting
\(\varphi=(L_L-z)^{-1}\eta\) proves resolvent commutation. Reflection
gives the corresponding left statement. Thus every bounded letter
commutes with the spectral projections of the opposite Hamiltonian.

Since $H_L=H_R$ on the vacuum, the orthogonal sum proves strong
commutation on $\mathcal H$.

\section{Commutation of the boundary algebras}\label{sec:locality}

Reality of the opposite-boundary contractions gives locality.
We justify the Wick expansion of bounded mixed products by a
Gaussian majorant on finite-particle tests.

\subsection{A positive-matrix semigroup}\label{sub:semigroup}

For \(\lambda>1/2\), use $q=q_\lambda$,
$P=p_{0,\lambda}$, $\kappa=\kappa_\lambda$, and
$h=h_\lambda=(q+\kappa)/2$ on the radial representation
\(D^+_\lambda\). If

\[
g=\begin{pmatrix}a&b\\c&e\end{pmatrix},\quad
a,e>0,\quad b,c\geq0,\quad ae-bc=1,
\]

define

\begin{equation}\label{eq:10.1}
\mathcal U_\lambda(g)
=e^{-(c/a)\kappa}\,e^{2i(\log a)P}\,e^{-(b/a)q}.
\end{equation}

We suppress the subscript $\lambda$ in $\mathcal U_\lambda$ when
the weight is fixed. Every factor is a contraction or unitary, so
\(\|\mathcal U_\lambda(g)\|\leq1\). With $I_\nu$ the modified
Bessel function of the first kind, for \(c>0\) its kernel is

\begin{equation}\label{eq:10.2}
\mathcal U_\lambda(g)(p,v)=
\frac1c\exp[-(ap+ev)/c]
I_{2\lambda-1}(2\sqrt{pv}/c).
\end{equation}

To check this, the kernel of \(e^{-t\kappa}\) is
\(t^{-1}e^{-(p+v)/t}I_{2\lambda-1}(2\sqrt{pv}/t)\), obtained by
substitution in its heat equation and the Bessel approximate-identity
limit. It has the semigroup property by

\begin{equation}\label{eq:10.3}
\int_0^\infty e^{-\alpha v}I_\nu(\beta\sqrt v)I_\nu(\gamma\sqrt v)\,dv
=\alpha^{-1}e^{(\beta^2+\gamma^2)/(4\alpha)}
I_\nu(\beta\gamma/(2\alpha)),
\end{equation}

for positive parameters and $\nu>-1$
\cite[10.43.28]{DLMF}. Substitution in \eqref{eq:10.2}, using
$ae-bc=1$, gives $\mathcal U(g_1)\mathcal U(g_2)=\mathcal U(g_1g_2)$;
boundary cases follow by strong continuity. Transposing the kernel gives
\[
 g^\sharp=\begin{pmatrix}e&b\\c&a\end{pmatrix},
 \qquad \mathcal U(g)^*=\mathcal U(g^\sharp).
\]
For $\theta\in\mathbb R$ and $A\ge0$, put

\[
D_\theta=\operatorname{diag}(e^{\theta/2},e^{-\theta/2}),\qquad
M_A=\begin{pmatrix}\cosh(A/2)&\sinh(A/2)\\
\sinh(A/2)&\cosh(A/2)\end{pmatrix},
\]

we have \(\mathcal U(D_\theta)=e^{i\theta P}\) and
\(\mathcal U(M_A)=e^{-Ah}\). The latter follows either from the kernel
or its derivative at zero on the Laguerre basis.

For strictly positive \(b,c\) there is a unique factorization

\begin{equation}\label{eq:10.4}
g=D_uM_AD_v,\quad
A=2\operatorname{arsinh}\sqrt{bc},\quad
u=\tfrac12\log\frac{ab}{ce},\quad
v=\tfrac12\log\frac{ac}{be}.
\end{equation}

For $0<2\lambda-1<1$, the kernel behaves as $p^{\lambda-1/2}$
at zero, selecting the Friedrichs realization of
Section~\ref{sub:endpoint}. The Bessel integral converges for
$\nu>-1$, since its integrand is $O(v^\nu)$ at zero and decays
exponentially at infinity.

\subsection{Boundary distributions and cross contractions}\label{sub:cross}

Let \(g_R=g_\lambda\) and \(g_L=s_\lambda g_\lambda\), where
\(s_\lambda\) is the Hankel involution of order \(2\lambda-1\). The
latter is a boundary distribution, not an \(L^2(dp)\) function. Its
precise action on finite Laguerre sums, and on functions with the
corresponding smooth Frobenius expansion at zero, is $\langle g_L,f\rangle=\sqrt{\Gamma(2\lambda)} \lim_{p\downarrow0}p^{-(\lambda-1/2)}f(p)$. This follows from $L_n^{2\lambda-1}(0)$ in \eqref{eq:2.1}. Moreover,

\begin{equation}\label{eq:10.5}
\langle g_R,e^{-Ah}g_R\rangle=\sinh(A/2)^{-2\lambda},\qquad
\langle g_R,e^{-Ah}g_L\rangle=\cosh(A/2)^{-2\lambda}.
\end{equation}

Indeed the Laguerre coefficients have squared modulus
\(4^\lambda\Gamma(n+2\lambda)/(n!\Gamma(2\lambda))\); the second series
has the additional factor \((-1)^n\). Summing the binomial series proves
both identities for \(A>0\). The second has a finite limit at zero.

More generally the boundary functional just defined and the
small-argument expansion of \(I_{2\lambda-1}\) in \eqref{eq:10.2} give
\(\mathcal U(g)g_L=c^{-2\lambda}e^{-(a/c)p}g_R\) for \(c>0\). Pairing
this with \(g_R\) is an absolutely convergent Gamma integral equal to
\(a^{-2\lambda}\). Multiplying \(g\) on the left and right by the
positive triangular matrices for \(e^{-tq}\) and \(e^{-u\kappa}\) gives

\begin{equation}\label{eq:10.6}
\langle e^{-tq}g_R,\mathcal U(g)e^{-u\kappa}g_L\rangle
=(a+t c+u b+tu e)^{-2\lambda}.
\end{equation}

The right side extends to $g=1$ and is analytic near $t=u=0$.

For a species of weight $d=\lambda>1/2$, put
\(g_{R,k}=i^kq^kg_R\) and \(g_{L,k}=(-i)^k\kappa^kg_L\). Their
dimensions are \(d_k=d+k\). Differentiating \eqref{eq:10.6} shows that for
\(g=D_uM_AD_v\), \(A\geq0\),

\begin{equation}\label{eq:10.7}
|\langle g_{R,k},\mathcal U(g)g_{L,l}\rangle|
\leq C_{k,l}\exp[C_{k,l}(|u|+|v|)].
\end{equation}

The constant is independent of \(A\). Each derivative is a finite sum of
powers of \(b/a,c/a,e/a\) times \(a^{-2d}\); \eqref{eq:10.4} bounds the ratios by
exponentials of \(|u|+|v|\), using \(\tanh(A/2)\leq1\).

At $g=1$, define the cross-contraction kernel by $K_d(t,u):=\langle e^{-tq}g_R,e^{-u\kappa}g_L\rangle_{\rm Abel} =(1+tu)^{-2d}$. The subscript indicates the semigroup extension in \eqref{eq:10.6}.
More explicitly, for $\sigma>0$ let
\[
 C_{kl}^{(\sigma)}
 :=\langle e^{-\sigma h}g_{R,k},e^{-\sigma h}g_{L,l}\rangle,
 \qquad
 C_{kl}:=\lim_{\sigma\downarrow0}C_{kl}^{(\sigma)}
        =\langle g_{R,k},g_{L,l}\rangle_{\rm Abel}.
\]
The regularized vectors belong to $L^2(dp)$.
Their pairing is obtained by differentiating \eqref{eq:10.6} with
$g=M_{2\sigma}$. Its denominator remains nonzero near
$(g,t,u)=(1,0,0)$, so every fixed derivative converges as
$\sigma\downarrow0$.

The phases must be included before asserting reality.
Antilinearity in the first variable contributes $(-i)^k$, the
left jet contributes $(-i)^l$, and differentiation of the two
heat factors contributes $(-1)^{k+l}$. Their product is $i^{k+l}$.
Consequently $C_{kl} =i^{k+l}\left.\partial_t^k\partial_u^l K_d(t,u)\right|_{t=u=0}$. Put $(2d)_r=\Gamma(2d+r)/\Gamma(2d)$ for the rising factorial.
The Taylor expansion $(1+tu)^{-2d} =\sum_{r=0}^{\infty}\frac{(-1)^r(2d)_r}{r!}\,t^ru^r$ shows that the derivative vanishes unless $k=l$, and in that case
equals $(-1)^k k!(2d)_k$. The phase $i^{2k}=(-1)^k$ cancels this
sign. Thus
\[
 \boxed{C_{kl}=\delta_{kl}\,k!(2d)_k\in\mathbb R.}
\]
Different species have zero contractions, since the one-particle
representation and its heat operators preserve the species summands.

Conjugate symmetry gives the reversed Abel contraction
$\langle g_{L,l},g_{R,k}\rangle_{\rm Abel}=\overline{C_{kl}}=C_{kl}$.
These real contractions will identify the two Wick expansions in
Section~\ref{sub:locality}.

\subsection{Heat smoothing of the jets}\label{sub:energy}

For \(d_k=d+k>1/2\),

\begin{equation}\label{eq:10.8}
\|e^{-ah}g_{R,k}\|+\|e^{-ah}g_{L,k}\|
       \le C_k a^{-d_k},\qquad a>0.
\end{equation}

The estimate follows from one recurrence. For a fixed species $a$,
put $\delta=d_a+k$. The polynomials \(R_k\) defined by
\(p^kg_{a,R,0}=R_k(h_a)g_{a,R,0}\)
satisfy

\begin{equation}\label{eq:19.29}
R_0=1,\quad R_1(z)=2z,\quad
R_{k+1}(z)=2zR_k(z)+k(k+2d_a-1)R_{k-1}(z),
\end{equation}

by \eqref{eq:19.14} and \(2h=q+\kappa\). Thus \(R_k\) has degree \(k\). The
squared Laguerre coefficients of \(g_{a,R,0}\) are
\(4^{d_a}\Gamma(n+2d_a)/(n!\Gamma(2d_a))\). Multiplying them by
$R_k(n+d_a)^2e^{-2t(n+d_a)}$ bounds the squared heat-smoothed
norm by
$C\sum_{n\ge0}(n+1)^{2\delta-1}e^{-2t(n+d_a)}$.
For $0<t\le1$, comparison with the Gamma integral gives
$Ct^{-2\delta}$; for $t\ge1$, exponential decay gives the same
bound. Taking square roots proves \eqref{eq:10.8}. The Hankel
reflection commutes with $h_a$, so the left estimate is identical.

The same bound holds with \(g_{R,k}\) or \(g_{L,k}\) replaced by
\(\mathcal U(g)g_{R,k}\), \(\mathcal U(g)g_{L,k}\) or their
adjoint-semigroup versions, multiplied by $Ce^{C(|u|+|v|)}$.
Only positivity of the matrices is needed to control the dilation
parameters. For $a_0>0$, $A\ge0$, write
$g'=M_{a_0}D_uM_A=\left(\begin{smallmatrix}a'&b'\\c'&e'\end{smallmatrix}\right)$,
and let $\le_{\rm ent}$ mean entrywise inequality. Since every
entry of $M_{a_0}$ and $M_A$ is nonnegative,
\begin{equation}\label{eq:10.9}
 e^{-|u|/2}M_{a_0+A}
 \le_{\rm ent}g'\le_{\rm ent}e^{|u|/2}M_{a_0+A}.
\end{equation}
The two diagonal entries of $M_{a_0+A}$ agree, as do its two
off-diagonal entries. Comparing products of two entries gives
\begin{equation}\label{eq:10.10}
 e^{-2|u|}\le\frac{a'b'}{c'e'},\,
                 \frac{a'c'}{b'e'}\le e^{2|u|}.
\end{equation}
Thus both new dilation parameters have absolute value at most
$|u|$, by \eqref{eq:10.4}. The heat parameter does not decrease:
direct multiplication gives
\[
 b'c'=\sinh^2((a_0+A)/2)
       +\tfrac12\sinh(a_0)\sinh(A)(\cosh u-1),
\]
so the new middle parameter is at least $a_0+A$.

For $g=D_uM_AD_v$, the needed conclusion is $\|e^{-ah}\mathcal U(g)g_{R,k}\| \le C e^{C(|u|+|v|)}a^{-d_k}$. Indeed, factor $M_aD_uM_A$ as above. Its middle heat
parameter is at least $a+A$, and its right dilation parameter
has absolute value at most $|u|$. The final $D_v$ adds $v$;
on a jet of dimension $d_k$ this costs at most
$e^{d_k(|u|+|v|)}$. The left dilation is unitary, so
\eqref{eq:10.8} proves the estimate. The same argument applies
to left jets and to adjoints, which exchange and negate the two
dilation parameters. These one-particle estimates suffice for
the finite-particle calculation below.

\subsection{Continuity after positive heat}\label{sub:continuity}

For $g=\left(\begin{smallmatrix}a&b\\c&e\end{smallmatrix}\right)$
as in \eqref{eq:10.1}, the kernel and boundary functional give

\begin{equation}\label{eq:10.11}
\begin{aligned}
\mathcal U_\lambda(g)e^{-tq}g_R
 &=(e+ct)^{-2\lambda}
   \exp\!\left[-p\frac{b+at}{e+ct}\right]g_R,\\
\mathcal U_\lambda(g)e^{-t\kappa}g_L
 &=(c+et)^{-2\lambda}
   \exp\!\left[-p\frac{a+bt}{c+et}\right]g_R.
\end{aligned}
\end{equation}

For the first identity, insert the power series of \(I_{2\lambda-1}\) in
\eqref{eq:10.2} and integrate each term against \(v^{\lambda-1/2}e^{-tv}\). The
resulting exponential uses \(ae-bc=1\). The second identity follows by
multiplying \(g\) on the right by the matrix for \(e^{-t\kappa}\) and
applying the formula for \(\mathcal U(g)g_L\) preceding \eqref{eq:10.6}. The
formulas hold first for positive entries and \(t\geq0\), and by the
resulting limits where defined.

Compose on the left with any fixed positive heat \(M_\epsilon\),
\(\epsilon>0\). If \(g\) varies near the identity, the entries of
\(M_\epsilon g\) remain strictly positive. The denominators and the
exponential decay rates in \eqref{eq:10.11} are then bounded away from zero.
Differentiating a fixed number of times at \(t=0\) gives every right and
left jet, with the phases specified in Section~\ref{sub:cross}. The resulting
coefficients depend continuously on \(g\), and, for a fixed integer
$N\ge0$ and constants $C,c_*>0$, the functions have a common bound

\begin{equation}\label{eq:10.12}
C p^{\lambda-1/2}(1+p)^N e^{-c_*p},
\qquad c_*>0.
\end{equation}

Its square is integrable at zero for \(\lambda>0\) and at infinity by
exponential decay. Dominated convergence proves that
\(e^{-\epsilon h}\mathcal U_\lambda(g)g_{R,k}\) and its left analogue
vary continuously in \(L^2(dp)\). Fixed outside
dilations are handled by the same formulas and their continuous changes
of scale. Adjoint-semigroup versions follow from
\(\mathcal U(D_uM_AD_v)^*=\mathcal U(D_{-v}M_AD_{-u})\). This proves the
one-particle convergence required in the mixed-product limit.
The positive outside heat is part of the statement; the undamped
boundary vectors themselves remain distributions.

\subsection{The global-energy regulator}\label{sub:abel}

We use a global-energy cutoff to identify the mixed products with the bounded letters already constructed. Replace every elementary jet \(g_{R,k}\) by
\(e^{-\sigma h}g_{R,k}\), \(0<\sigma\leq1\), and define the left cutoff
by \(J\). These are \(L^2\) one-particle functions. Form each finite
Wick polynomial from these functions, retaining the original physical
length factor. The resulting heat sandwiches converge strongly, together
with their adjoints, to the bounded letters already constructed in
Sections~\ref{sub:schur} and \ref{sub:wick}. Their norms are bounded uniformly in \(\sigma\).

Put $c=\cosh(\sigma/2)$ and $s=\sinh(\sigma/2)$. The heat kernel gives

\begin{equation}\label{eq:9.15}
e^{-\sigma h}e^{-tq}g_d
=(c+st)^{-2d}\exp\left[-p\frac{s+ct}{c+st}\right]g_d.
\end{equation}

Differentiate finitely many times in \(t\) at zero. Every cutoff jet is
a finite sum of \(p^jg_de^{-p\tanh(\sigma/2)}\), \(j\leq k\), with
coefficients bounded for \(0<\sigma\leq1\). The top coefficient tends to
its uncut value and the lower ones tend to zero. In the right gauge,
dilation replaces the exponential cutoff by
\(e^{-p e^{-x/2}\tanh(\sigma/2)}\), of modulus at most one. For a fixed polynomial of total jet dimension $D$, a term of
total remaining momentum dimension \(D'\leq D\) has length factor
\(e^{-(D+D')x/2}\). Put $\rho=(D-D')/2$ and split the remaining jet dimension into
creation and annihilation parts $D'_C+D'_A=D'$. For $a,b>0$ and
fixed length $x$, the matter estimate between $e^{-a\mathsf W(x)}$ and
$e^{-b\mathsf W(x)}$ cancels the factor $e^{-D'x}$. The remaining
length factor is controlled by the scalar part of $\mathsf W$:
\[
 e^{-\rho x}e^{-2a e^{-x}}\leq C_\rho a^{-\rho},
 \qquad C_\rho=\sup_{v\geq0}v^\rho e^{-2v}<\infty,
\]
with $C_0=1$. Thus each cutoff term is bounded, uniformly in the
cutoff and $x$, by a constant times
$a^{-(D'_C+\rho)}b^{-D'_A}$. Choose an integer $r>D$ and integrate
this bound against the two Gamma densities in
\eqref{eq:resolvent-laplace}. Both integrals converge because
$D'_C+\rho\leq D$ and $D'_A\leq D$. The factorization
\eqref{eq:7.3} gives a uniform bound for the heat sandwiches.
At equal small heat times the power is
$t^{-D'-\rho}=t^{-(D+D')/2}\leq t^{-D}$ for $0<t\leq1$.

For fixed positive matter heat parameters and fixed length, each
cutoff jet converges in the weighted one-particle norms used in
\eqref{eq:6.1} and \eqref{eq:8.1}. The finite expansion therefore
converges in operator norm after matter heat regularization.
Dominated convergence in the Gamma integrals gives fiberwise norm
convergence of the resolvent-regularized fields. Their uniform bound
gives strong-$*$ convergence on $L^2(dx;\mathcal F)$ exactly as in
\eqref{eq:7.4a}--\eqref{eq:7.4b}; the fixed bounded factors $B_r$
then give strong-$*$ convergence of the heat sandwiches. The limits
agree with the original forms on compact tests, hence are the same
bounded letters. Reflection proves the left assertion.

\subsection{Product approximation of the joint heat semigroup}\label{sub:jointheat}

Use the scaled Hamiltonians \(A_R,A_L\) of Section~\ref{sec:commutation} and write

\begin{equation}\label{eq:11.1}
\mathscr S=A_R+A_L
=k^2+P^2/4+2e^{-x/2}H_m+4e^{-x}.
\end{equation}

The equality is an equality of positive closed forms; by strong
commutation it also specifies their self-adjoint sum. More explicitly,
the two kinetic terms add to \(\|k\psi\|^2+\|P\psi\|^2/4\), and the
matter potentials add using \(Q+K=2H_m\). The common form domain is $D(k)\cap D(P)\cap D(H_m^{1/2}e^{-x/4})\cap D(e^{-x/2})$. It is complete in the sum of these graph norms. Choose
$\chi\in C_c^\infty([0,\infty))$ with $0\le\chi\le1$ and
$\chi=1$ near zero. Compact length cutoffs followed by
$\chi(H_m/N)$, for integers $N\to\infty$, give a form core. To see the
latter assertion, \(P\) changes global matter energy by at most one and
its matrix coefficients are bounded by a constant times that energy.
Thus \([P,\chi(H_m/N)]\) is uniformly bounded and tends strongly to
zero; the other terms commute with the cutoff. For finitely many species, bounded global-energy subspaces are finite-dimensional. Smooth approximation in length then completes
the core argument.

For a finite species set $F\subset\mathcal I$, let $\pi_F$ be the
orthogonal projection of $\mathfrak h$ onto
$\bigoplus_{a\in F}\mathfrak h_a$. For countably many species, first
apply $E_F=\Gamma_s(\pi_F)$. These projections
commute with the positive form constituents and increase strongly to
one. The finite-species approximation converges in form norm; the
preceding energy cutoff is then applied inside each finite-species
subspace. This proves the core assertion without a global multiplicity
or partition-function assumption.

Put $A_0=k^2+P^2/4$ and $B_0=2e^{-x/2}H_m+4e^{-x}$.
For $t>0$ and an integer $n\geq1$, write $\delta=t/n$ and define
\begin{equation}\label{eq:joint-trotter}
 F_{t,n}=(e^{-\delta A_0}e^{-\delta B_0})^n,
 \qquad F_{0,n}=1.
\end{equation}
The closed-form product theorem \cite[equation (1.3)]{TK} gives
$F_{t,n}\to e^{-t\mathscr S}$ strongly. The form intersection is
dense by the core just established. The same theorem applied in
reverse order gives strong convergence of the adjoints. All
estimates below are uniform in $n$.

We write out the finite Gaussian integrals in this product.
For input length $y$ and output length $x$, put $x_0=y$, $x_n=x$,
and integrate over $x_1,\ldots,x_{n-1}\in\mathbb R$. Set
$\theta_0=0$ and integrate over $\theta_1,\ldots,\theta_n\in\mathbb R$.
With $p_t$ as in Section~\ref{sub:heat-estimates}, define
\begin{equation}\label{eq:joint-gaussian-densities}
\begin{aligned}
 d\mu_{t,n}^{x,y}
 &=\frac{\prod_{j=1}^n p_{2\delta}(x_j-x_{j-1})}{p_{2t}(x-y)}
          \,dx_1\cdots dx_{n-1},\\
 d\nu_{t,n}
 &=\prod_{j=1}^n p_{\delta/2}(\theta_j-\theta_{j-1})
          \,d\theta_1\cdots d\theta_n.
\end{aligned}
\end{equation}
Both measures have mass one by Gaussian convolution. For $n=1$,
the first integral has no variables and has value one.
The spectral theorem gives
$e^{-\delta P^2/4}=\int p_{\delta/2}(v)e^{ivP}\,dv$.
For $0\leq j<n$, put $a_j=2\delta e^{-x_j/2}$, and set
\begin{equation}\label{eq:11.3}
\begin{aligned}
 G_{t,n}
 &=D_{\theta_n-\theta_{n-1}}M_{a_{n-1}}\cdots
       D_{\theta_1-\theta_0}M_{a_0},\\
 E_{t,n}&=4\delta\sum_{j=0}^{n-1}e^{-x_j}.
\end{aligned}
\end{equation}
Here $D_\theta,M_a$ are the matrices of Section~\ref{sub:semigroup}.
For such a matrix $G$, write
$\mathcal U(G)=\bigoplus_{c\in\mathcal I}\mathcal U_{d_c}(G)$
on the one-particle space. Multiplying the $n$ heat kernels gives
the operator-valued kernel of $F_{t,n}$:
\begin{equation}\label{eq:11.2}
 \mathcal K_{t,n}(x,y)
 =p_{2t}(x-y)\int
      e^{-E_{t,n}}\Gamma_s(\mathcal U(G_{t,n}))
          \,d\mu_{t,n}^{x,y}\,d\nu_{t,n}.
\end{equation}
This is a finite-dimensional integral, taken strongly on each
vector. Multiplicativity in Section~\ref{sub:semigroup} proves the
displayed order of the matrix factors. Since the integrand is a
contraction, $\|\mathcal K_{t,n}(x,y)\|\leq p_{2t}(x-y)$.
Thus for every $c\geq0$, Gaussian convolution also gives
\begin{equation}\label{eq:joint-weighted-convolution}
 \|e^{c|x|}F_{t,n}e^{-c|x|}\|
 \leq \int_{\mathbb R}p_{2t}(v)e^{c|v|}\,dv
 \leq2e^{c^2t}.
\end{equation}
The same bounds hold for the adjoint kernels and pass to the heat
semigroup. In particular, on vectors with compact length support,
the product formula converges in every fixed exponential length
weight: use strong convergence on a bounded length interval and
the common Gaussian tail bound outside it.

\paragraph{Uniform Gaussian tails}
We need two estimates for these finite Gaussian integrals. Put
\[
 \beta_j=x_j-\bigl((1-j/n)y+(j/n)x\bigr),\qquad
 N_{t,n}=\max_{0\leq j\leq n}|\beta_j|,
 \qquad M_{t,n}=\max_{0\leq j\leq n}|\theta_j|.
\]
The centered coordinates $\beta_j$ have a density independent of
$x,y$. Completing the square identifies it with the distribution
of $\zeta_j-(j/n)\zeta_n$, where $\zeta_0=0$ and the increments
$\zeta_j-\zeta_{j-1}$ have product density
$\prod_{j=1}^n p_{2\delta}(\zeta_j-\zeta_{j-1})$.
This statement is just equality of finite Gaussian integrals;
their common covariance is $2\delta(\min(j,k)-jk/n)$.

Here is an elementary maximal estimate for Gaussian convolution.
Suppose $z_0=0$, and let $\omega$ be a product of centered Gaussian
measures on the increments $z_j-z_{j-1}$, with total variance $v>0$.
Split the set $\{\max_jz_j>R\}$ according to the first index at
which $z_j>R$. For $\lambda>0$, integrating the later increments
in $e^{\lambda z_n}$ contributes a factor at least one. The
Gaussian integral of $e^{\lambda z_n}$ is $e^{\lambda^2v/2}$.
Summing over those disjoint sets and minimizing in $\lambda$
therefore gives $\omega(\max_jz_j>R)\leq e^{-R^2/(2v)}$. Applying the same argument to $-z_j$ proves the two-sided bound
$2e^{-R^2/(2v)}$. For the dilation variables $v=t/2$; for the
$\zeta_j$ variables $v=2t$. Since
$\max_j|\zeta_j-(j/n)\zeta_n|\leq2\max_j|\zeta_j|$, we obtain
\begin{equation}\label{eq:finite-gaussian-tails}
 \nu_{t,n}(M_{t,n}>R)\leq2e^{-R^2/t},\qquad
 \mu_{t,n}^{x,y}(N_{t,n}>R)\leq2e^{-R^2/(16t)}.
\end{equation}
Consequently, for $\ell\geq0$, either maximum has exponential
integral at most
\begin{equation}\label{eq:finite-gaussian-moments}
 \begin{aligned}
 F_\ell(t)
 &:=1+2\ell\int_0^\infty e^{\ell R-R^2/(16t)}\,dR\\
 &\leq1+8\ell\sqrt{\pi t}\,e^{4\ell^2t}.
 \end{aligned}
\end{equation}
All these bounds are independent of $n$ and of the length
endpoints. They also give uniform integrability of the exponential
majorants: on a set where a maximum exceeds $R$, use the same
bound with a larger exponential coefficient and gain $e^{-R}$.

\paragraph{Bounds on the Cartan factors}
Finally, factor $G_{t,n}=D_uM_AD_v$ as in \eqref{eq:10.4}.
With $M=M_{t,n}$, one has
\begin{equation}\label{eq:11.4}
 |u|\leq2M,\qquad |v|\leq M,\qquad
 A\geq2e^{-M}\delta\sum_{j=0}^{n-1}e^{-x_j/2}.
\end{equation}
To prove this, write $G_{t,n}=D_{\theta_n}Z_n$. Direct multiplication
gives $Z_n=\exp(H_{n-1})\cdots\exp(H_0)$, where
\[
 H_j=\delta e^{-x_j/2}
       \begin{pmatrix}0&e^{-\theta_j}\\e^{\theta_j}&0\end{pmatrix}.
\]
Every entry is nonnegative. Comparing their exponential series
entrywise with the matrix having both off-diagonal entries
$\delta e^{-M}e^{-x_j/2}$ proves the lower bound on $A$.
For the dilation bounds, insert the factors one at a time. During
multiplication by $\exp(sH_j)$, $0\leq s\leq1$, the product of the
column ratios $r_1=Z_{11}/Z_{21}$ and $r_2=Z_{12}/Z_{22}$ satisfies
\[
 \frac d{ds}(r_1r_2)
 =\delta e^{-x_j/2}(r_1+r_2)
       (e^{-\theta_j}-e^{\theta_j}r_1r_2).
\]
The interval $[e^{-2M},e^{2M}]$ is invariant; the first positive
factor has its limiting ratio in this interval. This bounds the
first dilation parameter of $Z_n$ by $M$. Transposing the finite
product bounds the second one by $M$. The factor $D_{\theta_n}$
adds at most $M$ to the first parameter. This proves
\eqref{eq:11.4} by ordinary matrix estimates.

\subsection{A mixed-product estimate}\label{sub:mixed}

We use the one-particle direct sum $\mathcal U$ defined in
Section~\ref{sub:jointheat}. In this subsection $V_R,V_L$ denote
dressed forms in the original canonical gauge.

Let $X_R,X_L$ be finite Wick primaries at opposite boundaries,
of dimensions $D_R,D_L$, as constructed in
Section~\ref{sub:primaries}. Thus $[K,X_R]=[Q,X_L]=0$ as forms,
by \eqref{eq:19.15} and reflection. Set

\[
V_R(x)=2^{D_R}e^{-D_Rx/2}X_R,
\qquad V_L(x)=2^{D_L}e^{-D_Lx/2}X_L.
\]
For $u,v>0$, abbreviate the bounded heat sandwiches by
\[
 T_R(u,v)=e^{-uA_R}V_Re^{-vA_R},\qquad
 T_L(u,v)=e^{-uA_L}V_Le^{-vA_L}.
\]
Section~\ref{sec:bounds} proves their boundedness, and
Section~\ref{sub:ward} proves that $T_R(u,v)$ commutes with every
spectral function of $A_L$, and conversely for $T_L(u,v)$ and $A_R$.
Primarity is used here for the complete polynomials. Their
individual descendant jets need not have this commutation property.
These primary fields are exactly the ones used to generate the
boundary algebras; the verification below still treats each Wick
term separately.

\begin{lemma}\label{lem:mixed} For the Wick primaries just specified
and every \(a,b>0\), there is a finite
constant \(C_{a,b,X_R,X_L}\) such that

\begin{equation}\label{eq:11.5}
\sup_{0<\eta\leq1}
\|e^{-a\mathscr S}V_Re^{-\eta\mathscr S}V_Le^{-b\mathscr S}\|
\leq C_{a,b,X_R,X_L}.
\end{equation}

One may take $C_{a,b,X_R,X_L}=\|T_R(a,b)\|\,\|T_L(a,b)\|$.
The products for \(\eta>0\) mean products of already bounded
heat-sandwiched forms, splitting the middle heat in half. Their strong
limit as \(\eta\downarrow0\) is the bounded extension of
\(e^{-a\mathscr S}(V_RV_L)_{\mathrm{Wick}}e^{-b\mathscr S}\). Reversing
\(R,L\) gives the corresponding statement with the reversed Wick
product.

The subscript \emph{Wick} includes all cross contractions of
Section~\ref{sub:cross}.

\end{lemma}

\begin{proof}
Split the middle heat in half. By the bounded-letter convention,
the mixed product in \eqref{eq:11.5} equals
\[
 e^{-aA_L}T_R(a,\eta/2)e^{-\eta A_L/2}
 e^{-\eta A_R/2}T_L(\eta/2,b)e^{-bA_R}.
\]
Move only heat factors past letters on the opposite boundary.
Combining the same-boundary heat factors gives
\[
 T_R(a,b+\eta)T_L(a+\eta,b)
       =T_R(a,b)e^{-\eta\mathscr S}T_L(a,b).
\]
Thus its norm is at most $\|T_R(a,b)\|\,\|T_L(a,b)\|$, and
it converges strongly to $T_R(a,b)T_L(a,b)$. This proves the
uniform bound and existence of the limit on the full Hilbert space.

Claims~\ref{clm:mixed-wick}--\ref{clm:mixed-continuity} justify
removing the cutoff and middle heat on a dense finite-particle
space. Normal ordering then gives the full Wick expansion, with
the Abel contractions of Section~\ref{sub:cross}. The bound above
gives its unique bounded extension. Interchanging the boundaries
proves the reversed statement.
\end{proof}

\paragraph{Identification of the Wick limit}
We normal order the finite Gaussian integrand and find a common
integrable majorant on each fixed finite sum of particle sectors.

Let $\Pi_{\le N}$ be the orthogonal projection onto the sectors
with at most $N$ particles, acting as the identity on length space.
The joint heat and every second-quantized matter contraction preserve
these sectors. The fixed Wick polynomials change particle number
by a bounded amount.

For $0<\sigma\leq1$, apply the global-energy cutoff of
Section~\ref{sub:abel} to both polynomials, and denote the resulting
forms by $V_{R,\sigma},V_{L,\sigma}$. Every jet is now an $L^2$
one-particle function. For $n\geq1$ and $\eta\geq0$, define on
finite-particle vectors with compact length support
\begin{equation}\label{eq:mixed-approximants}
 C_{n,\sigma}(\eta)
 =F_{a,n}V_{R,\sigma}F_{\eta,n}V_{L,\sigma}F_{b,n}.
\end{equation}
At $\eta=0$, this expression contains the ordinary product of the
two cutoff fields. We bound $C_{n,\sigma}(\eta)\Pi_{\le N}$ uniformly
in $n,\sigma$ and $0\leq\eta\leq1$, for each fixed $N$, before
taking limits.

Insert \eqref{eq:11.2} for the three heat factors.
Label their data by $j=1,0,2$, with durations
$(t_1,t_0,t_2)=(a,\eta,b)$ and endpoints
$(z,x),(w,z),(y,w)$. Thus $z,w$ are the insertion lengths;
$G_j,E_j,M_j,N_j$ have the meanings of Section~\ref{sub:jointheat}.
For $\eta=0$, set $w=z$, $G_0=1$ and $E_0=M_0=N_0=0$,
and omit the middle integral.

\begin{proofclaim}[Wick expansion]\label{clm:mixed-wick}
Every cutoff matter integrand is a finite sum of normally ordered
terms. All scalar contractions join opposite boundaries, and the
middle second-quantized operator is a contraction.
\end{proofclaim}
\begin{proof}
Suppressing the cutoff subscript in the normal-ordering formulas,
the matter integrand is a finite sum of terms of the form
\begin{equation}\label{eq:11.6}
 \Gamma_s(\mathcal U(G_1))X_R
 \Gamma_s(\mathcal U(G_0))X_L
 \Gamma_s(\mathcal U(G_2)).
\end{equation}
The cutoff factors are additional positive matrices $M_\sigma$.
The estimates \eqref{eq:10.7}--\eqref{eq:10.8} are uniform as
these additional heat times decrease to zero.

For any contraction \(T\) on the one-particle space,
\(a(f)\Gamma_s(T)=\Gamma_s(T)a(T^*f)\) and
\(\Gamma_s(T)a^*(f)=a^*(Tf)\Gamma_s(T)\) on finite particle vectors. To
exhibit every contraction already in the elementary case, let \(f,g\) be
cutoff right and left vectors. The canonical commutation relations give

\begin{equation}\label{eq:11.6a}
\begin{aligned}
\Phi(f)\Gamma_s(T)\Phi(g)
={}&a^*(f)a^*(Tg)\Gamma_s(T)
   +a^*(f)\Gamma_s(T)a(g)\\
 &+a^*(Tg)\Gamma_s(T)a(T^*f)
   +\Gamma_s(T)a(T^*f)a(g)\\
 &+\langle f,Tg\rangle\Gamma_s(T),
\qquad \Phi(f)=a^*(f)+a(f).
\end{aligned}
\end{equation}

Iterating these identities normally orders each finite Wick term.
Every contraction joins an $R$ jet to a transformed $L$ jet,
so \eqref{eq:10.7} applies. The surviving creators are $R$ jets or
$\mathcal U(G_0)$-transformed $L$ jets; the annihilators are $L$
jets or adjoint-transformed $R$ jets. Between them lies the
contraction $\Gamma_s(\mathcal U(G_0))$.
\end{proof}

\begin{proofclaim}[Finite-particle matter bound]\label{clm:mixed-fock}
On inputs with at most $N$ particles, the normally ordered terms
admit \eqref{eq:11.7}, uniformly in the cutoff. The exponents
record the dimensions of the surviving jets.
\end{proofclaim}
\begin{proof}
To make the outside heat bounds explicit, factor
$G_j=D_{u_j}M_{A_j}D_{v_j}$ for $j=1,2$, as in \eqref{eq:10.4}.
Write \(d_u=\mathcal U(D_u)\), \(T=\mathcal U(G_0)\), and
\(\mathcal U(G_j)=d_{u_j}e^{-A_jh}d_{v_j},\qquad j=1,2.\)
Move \(d_{v_1}\) through the surviving creators and \(d_{u_2}\)
through the surviving annihilators. After removing outside unitaries,
each normally ordered term has the form
\[
 e^{-A_1H_m}\mathcal C'
 \Gamma_s(d_{v_1}T d_{u_2})\mathcal A'e^{-A_2H_m}.
\]
Here \(\mathcal C'\) is a product of creators with vectors
\(d_{v_1}f_R\) or \(d_{v_1}Tg_L\), and \(\mathcal A'\) is a
product of annihilators with vectors \(d_{-u_2}g_L\) or
\(d_{-u_2}T^*f_R\). The middle second quantization is a contraction.
Move the outside heats through the creators and annihilators, using
$\Gamma_s(e^{-Ah})a^*(f)=a^*(e^{-Ah}f)\Gamma_s(e^{-Ah})$
and its adjoint. Each surviving jet has acquired a positive heat.
On the $j$-particle sector the creation and annihilation norms are
at most $\sqrt{j+1}\|f\|$ and $\sqrt j\|f\|$, respectively.
Only finitely many values of $j$ occur for inputs in $\Pi_{\le N}$.
The one-particle bounds of Section~\ref{sub:energy} therefore bound
\eqref{eq:11.6}, restricted to those inputs, by a finite sum of

\begin{equation}\label{eq:11.7}
C_N e^{C(M_0+M_1+M_2)} A_1^{-r}A_2^{-s},
\qquad r+s\leq D_R+D_L.
\end{equation}

Here $r,s$ are the total dimensions of the surviving jets.
The missing dimension $m=D_R+D_L-r-s\geq0$ belongs to contracted
pairs.
\end{proof}

\begin{proofclaim}[Cancellation of the length factors]\label{clm:mixed-length}
The field dressing is absorbed by the outside matter heat and
the scalar potential. The resulting bounds are
\eqref{eq:11.8}--\eqref{eq:11.9}.
\end{proofclaim}
\begin{proof}
The length coordinates in the first finite product satisfy
$x_k^{(1)}\leq z+|x-z|+N_1$ for $0\leq k\leq n$. Hence
\eqref{eq:11.4} gives
\begin{equation}\label{eq:11.8}
 e^{-rz/2}A_1^{-r}
 \leq(2a)^{-r}
       \exp\bigl[rM_1+\tfrac r2(|x-z|+N_1)\bigr].
\end{equation}
The analogous bound applies to $A_2^{-s}$ at $w$. Moving a
mismatched dimension from $w$ to $z$ costs at most $e^{C|z-w|}$.
The remaining factor $e^{-mz/2}$ is absorbed by the scalar
potential in the first product, since
\[
 E_1=\frac{4a}{n}\sum_{k=0}^{n-1}e^{-x_k^{(1)}}
       \geq4a e^{-z-|x-z|-N_1}.
\]
Maximizing $v^{m/2}e^{-v}$ for $v\geq0$ gives
\begin{equation}\label{eq:11.9}
 e^{-mz/2}e^{-E_1}
 \leq\left(\frac{m}{8ea}\right)^{m/2}
          e^{\frac m2(|x-z|+N_1)},
\end{equation}
with the right prefactor interpreted as one when $m=0$.
\end{proof}

\begin{proofclaim}[A common convolution majorant]\label{clm:mixed-schur}
For every fixed $N$, the finite Gaussian products satisfy
\eqref{eq:mixed-approximant-bound}, including zero middle time.
\end{proofclaim}
\begin{proof}
Choose $\ell_*>0$ large enough for the finitely many Wick terms.
By the preceding two claims, the integrand norm on $\Pi_{\le N}$
is bounded by a constant $C_{N,a,b}$ times the three Gaussian length kernels
and an exponential of
\[
 \ell_*(|x-z|+|z-w|+|w-y|+M_0+M_1+M_2+N_1+N_2).
\]
Put $\gamma_t(v)=p_{2t}(v)e^{\ell_*|v|}$ for $t>0$.
Let $\mathcal C_{n,\sigma,\eta}(x,y)$ denote the length kernel of
$C_{n,\sigma}(\eta)$. Integrating the maxima by
\eqref{eq:finite-gaussian-moments}, then the two insertion lengths,
gives the simple convolution bound
\begin{equation}\label{eq:11.9a}
\begin{aligned}
 \|\mathcal C_{n,\sigma,\eta}(x,y)\Pi_{\le N}\|
 \le{}&C_{N,a,b}F_{\ell_*}(a)^2F_{\ell_*}(b)^2F_{\ell_*}(\eta)\\
 &\times(\gamma_a*\gamma_\eta*\gamma_b)(x-y).
\end{aligned}
\end{equation}
Here $*$ denotes convolution on $\mathbb R$, and $F_\ell$ is
defined in \eqref{eq:finite-gaussian-moments}. The fixed constant
$C_{N,a,b}$ absorbs the finite powers of the outside times in
\eqref{eq:11.8}--\eqref{eq:11.9}; it contains no negative power of
$\eta$. At $\eta=0$, omit $\gamma_\eta$ and set $F_{\ell_*}(0)=1$.
Since $\|\gamma_t\|_1\le2e^{\ell_*^2t}$, Young's convolution
inequality, applied to the pointwise Hilbert-space norm, proves
\begin{equation}\label{eq:mixed-approximant-bound}
 \sup_{\substack{n\geq1,\;0<\sigma\leq1\\0\leq\eta\leq1}}
       \|C_{n,\sigma}(\eta)\Pi_{\le N}\|<\infty
       \qquad\text{for each fixed }N.
\end{equation}
This is the only domination estimate needed for the Wick limit.
\end{proof}

\begin{proofclaim}[Passage to the heat semigroup]\label{clm:mixed-trotter}
For fixed $\sigma>0$ and $0\leq\eta\leq1$,
$C_{n,\sigma}(\eta)$ converges on every fixed finite sum of particle
sectors to the product of the cutoff fields and the three heat
operators. The same assertion holds with all factors adjointed
and their order reversed.
\end{proofclaim}
\begin{proof}
For fixed $\sigma>0$ and $0\leq\eta\leq1$, the product formula now
identifies the limit as $n\to\infty$. On each fixed finite sum of
particle sectors, the cutoff polynomials are bounded in the matter
variables and multiply the length variable by fixed exponentials.
The weighted convergence and bound
\eqref{eq:joint-weighted-convolution} therefore allow passage to
the limit through each insertion, first on compact length tests.
The result is
$e^{-a\mathscr S}V_{R,\sigma}e^{-\eta\mathscr S}
V_{L,\sigma}e^{-b\mathscr S}$, also for $\eta=0$.
Bound \eqref{eq:mixed-approximant-bound} extends the convergence
from compact length tests to every vector with at most $N$
particles. The number of particles can increase by at most the
fixed degrees of the Wick polynomials, so the same argument
applies to the reversed adjoint products.
\end{proof}

\begin{proofclaim}[Reduction to compact integration sets]\label{clm:mixed-tails}
For matrix elements between compact smooth length tests and
finite-particle Laguerre vectors, both regulator limits can be
studied on bounded integration sets for sufficiently small middle
time. The discarded tails are
uniformly small in the subdivision, cutoff and middle time.
\end{proofclaim}
\begin{proof}
We give the uniform continuity argument needed to interchange the
remaining limits. For fixed finite Gaussian coordinates, the
outside Cartan parameters $A_1,A_2$ are positive. The formulas of
Section~\ref{sub:continuity}, together with
\eqref{eq:10.6}, therefore define the limit of the normal-ordered
integrand as $\sigma\downarrow0$. Its integrable majorant is the
one just proved. Denote the resulting map on finite-particle
vectors by $C_{n,0}(\eta)$, including at $\eta=0$.
Its restriction to each $\Pi_{\le N}$ is bounded.

Take $\psi,\varphi$ to be finite sums of compact smooth length
functions times finite-particle Laguerre vectors, using finitely
many species. These vectors form a dense subspace. For $\eta>0$,
put $w=z+\sqrt\eta\,v$ in the middle Gaussian integral. Express
all internal length coordinates as their affine endpoint
interpolation plus their centered coordinates. In the middle
factor, divide the centered coordinates and the dilation variables
by $\sqrt\eta$. Their time-one Gaussian densities depend on $n$
but not on $\eta$ or on the endpoints.

Set $\eta_* =\min(1,b/4)$. The inequality
\begin{equation}\label{eq:11.9b}
 (z+\sqrt\eta\,v-y)^2
       \geq\tfrac12(z-y)^2-\eta v^2
\end{equation}
gives a common Gaussian majorant in $z,v$ for
$0<\eta\leq\eta_*$, after multiplication by all fixed linear
exponential factors. Indeed, the extra factor $e^{\eta v^2/(4b)}$
is integrable against the middle density $p_2(v)$, uniformly in
this range. The estimates \eqref{eq:finite-gaussian-tails} and
\eqref{eq:finite-gaussian-moments}, used with larger coefficients,
also control the tails of every maximum uniformly in $n$.
We may consequently restrict $z,v$, the outside maxima, and the
time-one middle maxima to bounded sets, with an arbitrarily small
error in these matrix elements, uniformly in $n,\sigma,\eta$.
\end{proof}

\begin{proofclaim}[Uniform removal of the regulators]\label{clm:mixed-continuity}
On the dense test space of Claim~\ref{clm:mixed-tails}, the
cutoff and middle-time limits obey \eqref{eq:mixed-uniform-limits}.
\end{proofclaim}
\begin{proof}
On each such restricted set, the outside matrices range over a
compact set of strictly positive matrices whose Cartan heat
parameters are bounded below by a positive constant. To see the
upper bound as well as the lower bound in \eqref{eq:11.4}, use,
for either outside duration $t\in\{a,b\}$,
\[
 \|Z_n\|\leq
       \exp\!\left(e^{M_{t,n}}\frac{t}{n}\sum_{j=0}^{n-1}e^{-x_j/2}\right).
\]
The middle matrices satisfy $G_0=1+O(\sqrt\eta)$ and
$E_0=O(\eta)$, uniformly in $n$. Replacing the endpoint $w$ by
$z$ in the last outside factor changes every length coordinate by
at most $\sqrt\eta|v|$. In the exponential product for $Z_n$,
each coefficient changes by $O((b/n)\sqrt\eta)$. Telescoping the
finite product, whose partial products have the displayed uniform
bound, proves that $G_2$ and $E_2$ change by $O(\sqrt\eta)$,
again uniformly in $n$.

Move the outside heat onto the surviving jets as in
Claim~\ref{clm:mixed-fock}. Formulas \eqref{eq:10.11}--\eqref{eq:10.12}
give uniform continuity of these jets in one-particle norm,
over the compact matrix sets and down to
$\sigma=0$. The cross-contractions are uniformly continuous by
\eqref{eq:10.6}. Second quantization is strongly continuous on
contractions, first on finite-particle sectors by tensor-product
continuity and then by density. On the compact families of vectors
just obtained this continuity is uniform. Thus every
normal-ordered matrix element is uniformly continuous in both
limits on the restricted sets. The common tail bounds remove the
restrictions. More precisely,
\begin{equation}\label{eq:mixed-uniform-limits}
\begin{aligned}
 &\lim_{\sigma\downarrow0}
   \sup_{\substack{n\geq1\\0\leq\eta\leq\eta_*}}
   |\langle\psi,(C_{n,\sigma}(\eta)-C_{n,0}(\eta))\varphi\rangle|=0,\\
 &\lim_{\eta\downarrow0}
   \sup_{\substack{n\geq1\\0\leq\sigma\leq1}}
   |\langle\psi,(C_{n,\sigma}(\eta)-C_{n,\sigma}(0))\varphi\rangle|=0.
\end{aligned}
\end{equation}
\end{proof}

\paragraph{Taking the limits in order}
For fixed positive cutoff, let $n\to\infty$ by
Claim~\ref{clm:mixed-trotter}. At fixed $\eta>0$, split the middle
heat in half and use the strong-$*$ cutoff convergence in
Section~\ref{sub:abel}. Each joint heat sandwich factors as
opposite heat times a single-boundary sandwich times opposite
heat, so its limit is the physical product in Lemma~\ref{lem:mixed}.
The second line of \eqref{eq:mixed-uniform-limits} allows
$\eta\downarrow0$ uniformly through the first two limits. At zero
middle time, the first line removes the cutoff in the ordinary
Wick expansion, including every cross-contraction. Thus the
sandwiched full Wick expression and $T_R(a,b)T_L(a,b)$ have the
same matrix elements on the dense test space. This is the
identification used in the short proof of the lemma. The
full-space bound comes from that proof, so the constants depending
on $N$ in the verification cause no loss.

\subsection{Bounded boundary locality}\label{sub:locality}

Take \(\mathsf R=e^{-aA_R}V_Re^{-bA_R}\) and \(\mathsf L=e^{-cA_L}V_Le^{-dA_L}\), with
all four times positive. Section~\ref{sub:ward} proves commutation of each bounded
letter with the opposite Hamiltonian. Consequently, splitting a positive
middle heat and commuting only these bounded factors gives

\begin{equation}\label{eq:11.10}
\mathsf R e^{-\eta\mathscr S}\mathsf L
=e^{-aA_R-cA_L}V_R e^{-\eta\mathscr S}V_L
e^{-bA_R-dA_L}.
\end{equation}

This is the same bounded-factor calculation as in the short proof
of Lemma~\ref{lem:mixed}, now with unequal outside times: the
right side reduces to $T_R(a,b+\eta)T_L(c+\eta,d)$.

This is an identity of bounded operators for \(\eta>0\). The right side
has a joint positive heat on both outside ends: factor out
\(e^{-\min(a,c)\mathscr S}\) on the left and
\(e^{-\min(b,d)\mathscr S}\) on the right. The lemma therefore
identifies its strong limit with the sandwiched Wick product. The left side
converges strongly to \(\mathsf R\mathsf L\). The corresponding identity for
\(\mathsf L e^{-\eta\mathscr S}\mathsf R\) has the same outside heats and reversed Wick
product.

\paragraph{The Wick identity}
For $X_R=:R_1\cdots R_r:$ and $X_L=:L_1\cdots L_s:$,
Wick's formula sums over partial matchings between the right and
left labels. Each term is the product of its cross-contractions
and the normally ordered unpaired fields.
There are no internal contractions because both inputs are normally
ordered. Reversing the two inputs leaves the matchings and the
normally ordered terms unchanged, and conjugates the contractions.
Their Abel limits are real by Section~\ref{sub:cross}. Thus $(V_RV_L)_{\rm Wick}=(V_LV_R)_{\rm Wick}$. Keeping labels distinct accounts for multiplicities when jets coincide.

At $G_0=1$ the middle propagation has disappeared, while the
outside heats remain positive. Lemma~\ref{lem:mixed} identifies
the limits with the full Wick products, whose equality gives
$\mathsf R\mathsf L=\mathsf L\mathsf R$. Hamiltonian powers commute by Section~\ref{sub:ward}.
Taking double commutants gives

\begin{equation}\label{eq:11.11}
M_L\subseteq M_R',\qquad M_R\subseteq M_L'.
\end{equation}

For countably many species each Wick polynomial still involves
only finitely many. The contractions control all spectators;
density of finite-species, finite-particle vectors gives the same
strong limits. No sum over primary-field constants is needed.

\section{The joint spectral decomposition}\label{sec:spectral}

The joint energy representation gives bounded vacuum-to-primary
columns. We use them first to prove factoriality and then to
construct the dense vacuum map for the trace.

\subsection{The Liouville transform}\label{sub:kl}

At fixed matter momentum the Hamiltonian has an exponential
potential. Its decaying eigenfunctions give the Liouville transform
with the following normalization.

For fixed total matter momentum \(q\geq0\), write

\[
L_q=-\frac12\partial_x^2+(q+2)e^{-x},\qquad
\zeta_s(q,x)=2K_{2is}\!\left(2\sqrt{2(q+2)}e^{-x/2}\right).
\]

Changing \(x\) by \(\log((q+2)/2)\) reduces this to the same Liouville
operator for every \(q\). The Kontorovich-Lebedev transform gives

\begin{equation}\label{eq:4.1}
L_q\zeta_s=\frac{s^2}{2}\zeta_s,\qquad
d\mu(s)=\rho(s)ds,\quad
\rho(s)=\frac{s\sinh(2\pi s)}{\pi^2}.
\end{equation}

In this subsection a hat denotes the length spectral transform,
distinct from the Fourier transform of a time test. The transform and
inverse are

\[
\widehat\psi(s)=\int_{\mathbb R}\zeta_s(q,x)\psi(x)dx,
\qquad
\psi(x)=\int_0^\infty\zeta_s(q,x)\widehat\psi(s)d\mu(s).
\]

The factors are checked by putting \(r=2\sqrt{2(q+2)}e^{-x/2}\) and
\(\nu=2s\) in the transform-inversion pair \cite[10.43.30-31]{DLMF}. The
Parseval identity, first for compact tests and then by completion, makes
this transform unitary. The transformed Friedrichs operator is
multiplication by \(s^2/2\).

Denote its direct integral over all momentum and particle sectors by
$\mathcal U:\mathcal H\to L^2(d\mu(s);\mathcal F)$, so that
$\mathcal U L\mathcal U^*=s^2/2$. For one species this gives

\begin{equation}\label{eq:4.2}
\mathcal H\cong
\bigoplus_{n=0}^\infty
L^2\!\left((0,\infty),d\mu(s);
L^2_{\mathrm{sym}}((0,\infty)^n,d\mathbf p)\right).
\end{equation}

Here \(H_R\) is multiplication by \(s^2/(4\phi_b)\) after also applying
\(U^*\). This is an energy representation with explicit multiplicity
spaces. For several species the same formula is applied on each finite occupation sector. The one-sided transform retains the parabolic momenta; we next diagonalize the other Hamiltonian.

\subsection{The insertion kernel}\label{sub:insertion}

For $d>0$, $t,s\in\mathbb R$ and total momenta $q,q'\ge0$, define

\begin{equation}\label{eq:4.3}
F_d(t,s;q',q)=2^d\int_{\mathbb R}e^{-dx}
\zeta_t(q',x)\zeta_s(q,x)dx.
\end{equation}

For \(q'\geq q\), put \(r=(q+2)/(q'+2)\in(0,1]\) and

\begin{equation}\label{eq:Gd}
G_d(t,s)=
\frac{\prod_{\sigma,\eta\in\{\pm1\}}
\Gamma(d+i\sigma t+i\eta s)}{\Gamma(2d)}
=\frac{|\Gamma(d+i(t+s))\Gamma(d+i(t-s))|^2}{\Gamma(2d)}.
\end{equation}

With ${}_2F_1(a,b;c;z)$ denoting Gauss's hypergeometric function,
one has

\begin{equation}\label{eq:4.4}
\boxed{
F_d(t,s;q',q)=(q'+2)^{-d}r^{is}G_d(t,s)
{}_2F_1\!\left(d+i(s+t),d+i(s-t);2d;1-r\right).}
\end{equation}

For \(q'<q\), use \(F_d(t,s;q',q)=F_d(s,t;q,q')\). Although \eqref{eq:4.4} is
written with complex factors, \eqref{eq:4.3} shows that it is real and has the
adjoint symmetry just stated.

\textbf{Derivation.} Set \(u=e^{-x/2}\) in \eqref{eq:4.3}. The result is

\[
2^{d+3}\int_0^\infty u^{2d-1}K_{2it}(au)K_{2is}(bu)du,
\quad a^2=8(q'+2),\quad b^2=8(q+2).
\]

Insert the integral representation \cite[10.32.10]{DLMF},

\[
K_\nu(z)=\tfrac12(z/2)^\nu
\int_0^\infty v^{-\nu-1}e^{-v-z^2/(4v)}dv.
\]

For imaginary orders and \(d>0\), absolute values are dominated by the
corresponding zero-order integral, so Fubini is valid. Perform the \(u\)
integral, then set \(v+w=R\), \(v/R=y\) in the remaining two integrals.
The \(R\) integral is a Gamma integral, and the \(y\) integral is
Euler's beta integral for \({}_2F_1\), obtained by multiplying
\cite[15.6.1]{DLMF} by \(\Gamma(c)\) to undo its regularized
hypergeometric normalization. More generally, for real
\(a\geq b>0\) and
\(\operatorname{Re}\lambda>|\operatorname{Re}\mu|+|\operatorname{Re}\nu|\),
the same absolutely convergent computation gives

\[
\begin{aligned}
\int_0^\infty u^{\lambda-1}K_\mu(au)K_\nu(bu)du
={}&\frac{2^{\lambda-3}a^{-\lambda}(b/a)^\nu}{\Gamma(\lambda)}
\prod_{\sigma,\eta=\pm1}
\Gamma\!\left(\frac{\lambda+\sigma\mu+\eta\nu}{2}\right)\\
&\times{}_2F_1\!\left(
\frac{\lambda+\mu+\nu}{2},
\frac{\lambda-\mu+\nu}{2};\lambda;1-b^2/a^2\right).
\end{aligned}
\]

Substitution of \(\lambda=2d\), \(\mu=2it\), \(\nu=2is\) gives \eqref{eq:4.4},
including its prefactor. The kernel and measure are completely fixed in
this representation.

\subsection{Convergence of the spectral integrals}\label{sub:majorants}

The vertical-line asymptotic \cite[5.11.9]{DLMF}, combined with
continuity on a bounded interval, gives for fixed \(d>0\)

\begin{equation}\label{eq:5.1}
|\Gamma(d+iu)|^2\leq C_d(1+|u|)^{2d-1}e^{-\pi|u|},
\qquad u\in\mathbb R.
\end{equation}

Using \(|s+t|+|s-t|=2\max(s,t)\) yields

\begin{equation}\label{eq:5.2}
G_d(s,t)\leq C'_d
(1+s+t)^{2d-1}(1+|s-t|)^{2d-1}
e^{-2\pi\max(s,t)}.
\end{equation}

This single estimate covers \(s\simeq t\), either off-diagonal region,
and the origin. Negative exponents cause no singularity because every
power has base at least one. Moreover,

\[
\rho(s)\sim\frac2\pi s^2\quad(s\downarrow0),
\qquad \rho(s)\leq C(1+s)e^{2\pi s}.
\]

Every fixed polynomial times \(e^{cs-as^2}\) is integrable for \(a>0\),
by completing the square. For the vacuum-to-primary direct-integral block one obtains the
exact squared norm

\begin{equation}\label{eq:5.3}
\operatorname*{ess\,sup}_{r>0}
e^{-2ar^2}\int_0^\infty e^{-2as^2}G_d(r,s)d\mu(s)<\infty.
\end{equation}

This proves that block is bounded, including for \(d\leq1/2\). The full Fock-space boundedness was proved separately in Section~\ref{sec:bounds}.

There is also a global bound for the exact kernel \eqref{eq:4.3}:
the representation \cite[10.32.9]{DLMF} gives
\(|K_{iu}(z)|\leq K_0(z)\) for \(z>0\), and hence

\begin{equation}\label{eq:5.4}
|F_d(t,s;q',q)|\leq F_d(0,0;q',q).
\end{equation}

For $q'\geq q$ this is \eqref{eq:4.4} with $s=t=0$.
The particle-number factors require the estimates of
Section~\ref{sec:bounds}.

\subsection{The radial hypergeometric transform}\label{sub:jacobi}

For a primary of weight \(\lambda>1/2\) and parameters $r,s>0$,
define \(v=p/2\) and

\begin{equation}\label{eq:12.1}
\begin{aligned}
\chi_{r,s}^{\lambda}(p)
={}&\frac{v^{\lambda-1/2}\sqrt{G_\lambda(r,s)}}{\sqrt{2\Gamma(2\lambda)}}
(1+v)^{is}\\
&\times{}_2F_1(\lambda+is+ir,\lambda+is-ir;2\lambda;-v).
\end{aligned}
\end{equation}

\begin{proposition}[Normalized radial transform]\label{prop:jacobi}
For fixed $\lambda>1/2$ and $s>0$, the real kernel
\eqref{eq:12.1} defines a unitary operator
\begin{equation}\label{eq:jacobi}
 \begin{split}
 (\mathcal J_{\lambda,s}h)(r)
     &=\int_0^\infty\chi_{r,s}^{\lambda}(p)h(p)\,dp,\\
 \mathcal J_{\lambda,s}:L^2(dp)&\longrightarrow L^2(d\mu(r)).
 \end{split}
\end{equation}
Its kernel satisfies the insertion identity \eqref{eq:12.2} below.
The parameter $s$ is fixed and $r$ is the transform variable.
\end{proposition}

Fix $\lambda>1/2$ and $s>0$.

\begin{proofclaim}[Regular solution and insertion identity]\label{clm:jacobi-insertion}
The kernel $\chi_{r,s}^{\lambda}$ is real, solves
$B_{\lambda,s}\chi_{r,s}^{\lambda}=r^2\chi_{r,s}^{\lambda}/2$,
has the Friedrichs behavior at zero,
and satisfies \eqref{eq:12.2}.
\end{proofclaim}
\begin{proof}
Substitution into the hypergeometric equation gives
\(B_{\lambda,s}\chi_{r,s}^{\lambda}=r^2\chi_{r,s}^{\lambda}/2\). Its behavior at zero is the Friedrichs behavior $p^{\lambda-1/2}$, including in the limit-circle range.
It is real: the integral expression \eqref{eq:4.3} gives

\begin{equation}\label{eq:12.2}
g_\lambda(p)F_\lambda(s,r;p,0)
=\sqrt{G_\lambda(r,s)}\,\chi_{r,s}^{\lambda}(p).
\end{equation}

This also derives \eqref{eq:12.1} from the previously normalized insertion
kernel, using the hypergeometric fractional-linear transformation.

\end{proof}

\begin{proofclaim}[Spectral density]\label{clm:jacobi-density}
The connection coefficients of the regular solution give the
density $d\mu(r)$ in Stone's formula, as in \eqref{eq:12.4b}.
\end{proofclaim}
\begin{proof}
The hypergeometric connection formula at infinity gives two terms
\(p^{-1/2\pm ir}\). The coefficient of \(v^{-\lambda-ir}\) before the
prefactor in \eqref{eq:12.1} is

\begin{equation}\label{eq:12.3}
\frac{\Gamma(2\lambda)\Gamma(-2ir)}
{\Gamma(\lambda+is-ir)\Gamma(\lambda-is-ir)}.
\end{equation}

After multiplying by that prefactor and changing from \(v\) to \(p\),
the squared modulus of either coefficient is \(|\Gamma(2ir)|^2\). Thus
the incoming/outgoing normalization is

\begin{equation}\label{eq:12.4}
2\pi|\Gamma(2ir)|^2
=\frac{\pi^2}{r\sinh(2\pi r)}=\rho(r)^{-1}.
\end{equation}

Here is the resolvent calculation fixing this spectral normalization.
Write \(\chi(p)=\chi_{r,s}^{\lambda}(p)\) and let \(f_+(r,p)\) be the
outgoing Jost solution normalized by \(p^{-1/2+ir}\) at infinity,
including its differentiated asymptotic. For fixed real \(r>0\), the
real regular solution is

\[
\chi=A f_++\overline A\,\overline{f_+},
\qquad |A|^2=|\Gamma(2ir)|^2.
\]

Put \(a_0(p)=p(p+2)/2\) and \(W_{a_0}(u,v)=a_0(uv'-u'v)\). The
asymptotics give

\begin{equation}\label{eq:12.4a}
W_{a_0}(\chi,f_+)=ir\,\overline A.
\end{equation}

For \(\mathcal E=r^2/2\), denote by \(R_\pm(\mathcal E;p,q)\) the
boundary values of the kernel of
\((B_{\lambda,s}-\mathcal E\mp i\varepsilon)^{-1}\) as
\(\varepsilon\downarrow0\). Put $p_<=\min(p,q)$ and
$p_>=\max(p,q)$. The separated-endpoint Green formula gives

\[
R_+(\mathcal E;p,q)
=-\frac{\chi(p_<)f_+(r,p_>)}{ir\,\overline A},
\qquad R_-=\overline{R_+},
\]

Therefore

\begin{equation}\label{eq:12.4b}
R_+-R_-=\frac{i\,\chi(p)\chi(q)}{r|A|^2},
\qquad
\frac{R_+-R_-}{2\pi i}\,d\mathcal E
=\chi(p)\chi(q)\rho(r)\,dr.
\end{equation}

In the last identity \(d\mathcal E=r\,dr\) and
\(|\Gamma(2ir)|^2=\pi/[2r\sinh(2\pi r)]\). This is Stone's formula with
its measure and wave-number factor explicitly included.
\end{proof}

\begin{proofclaim}[Completeness of the spectral resolution]\label{clm:jacobi-complete}
The density in Claim~\ref{clm:jacobi-density} exhausts the
spectral measure of $B_{\lambda,s}$. 
\end{proofclaim}
\begin{proof}
To justify completeness, use the Friedrichs condition of Section~\ref{sub:endpoint} at zero. The potential in \eqref{eq:9.11} is exponentially integrable at infinity. Positivity follows from the closed form $B_{\lambda,s}=\tfrac12P^2+\kappa_\lambda+s^2/(p+2)$; The
Jost solution is obtained there by the convergent Volterra equation and
has its differentiated wave asymptotics. The zero-end regular solution
is \eqref{eq:12.1}, whose connection coefficient is nonzero. Consequently the
Green kernels converge to the displayed boundary values locally
uniformly on compact positive-energy intervals and compact interior
coordinate sets. Stone's formula, paired with compactly supported
interior vectors, then gives an absolutely continuous spectral measure
on each such interval. These test vectors are dense, so no
positive-energy singular spectral subspace is omitted. Nonnegativity
excludes negative spectrum; the two nondecaying asymptotic waves exclude
positive eigenvectors. A zero eigenvector is excluded by the positive
form, since \(\kappa_\lambda=s_\lambda q s_\lambda\) has trivial kernel.
A residual spectral measure supported only at zero would be an atom,
hence that same excluded eigenprojection. This verifies the endpoint and
resolvent hypotheses of the spectral-transform framework in \cite[Chapter 9, Sections 9.2, 9.6]{T}.
\end{proof}

\begin{proofclaim}[Surjectivity]\label{clm:jacobi-onto}
The isometry defined by this spectral resolution is onto
$L^2(d\mu(r))$. The corresponding direct-integral transform is
also unitary.
\end{proofclaim}
\begin{proof}
Fix $\lambda,s$ and put $\mathcal T=\mathcal J_{\lambda,s}$.
The spectral resolution gives
\[
 \mathcal T^*\mathcal T=1,\qquad
 \mathcal T^*M_{\mathbf1_I(r^2/2)}\mathcal T
 =\mathbf1_I(B_{\lambda,s})
\]
for every Borel set \(I\subset[0,\infty)\), where \(M\) denotes
multiplication on \(L^2(d\mu(r))\). Since the right side is a
projection, compression by the isometry \(\mathcal T\) shows that $M_{\mathbf1_I(r^2/2)}\mathcal T =\mathcal T\mathbf1_I(B_{\lambda,s})$. Thus the closed range of \(\mathcal T\) reduces every scalar
multiplication projection and is \(L^2(S,d\mu)\) for a measurable set
\(S\). If \(S^c\) had positive measure, the transforms of a countable
locally dense collection of compact interior tests would all vanish
almost everywhere there. Continuity in \(p\) would then imply
\(\chi_{r,s}^{\lambda}(p)=0\) for every \(p>0\) at such \(r\),
contradicting its nonzero principal coefficient at zero. Hence
\(S^c\) is null, and \(\mathcal T\) is unitary onto
\(L^2(d\mu(r))\). The kernel depends measurably on its parameters,
so the direct-integral transform is unitary.
\end{proof}

\begin{proof}[Proof of Proposition~\ref{prop:jacobi}]
Claim~\ref{clm:jacobi-insertion} gives the stated kernel and
insertion identity. Claims~\ref{clm:jacobi-density} and
\ref{clm:jacobi-complete} make the transform an isometry, and
Claim~\ref{clm:jacobi-onto} proves that it is onto. These are all
the assertions of the proposition.
\end{proof}

Combining this transform with the primary decomposition, the full
physical Hilbert space is now rigorously identified as

\begin{equation}\label{eq:12.5}
\mathcal H\simeq L^2(\mu)\ \oplus\
\bigoplus_{i\in\mathcal P} L^2(\mu\otimes\mu),
\end{equation}

with \((H_L,H_R)=(E(r),E(r))\) on the vacuum summand and \((E(r),E(s))\)
on primary summand \(i\), where \(E(s)=s^2/(4\phi_b)\).
Here $L^2(\mu)$ abbreviates $L^2((0,\infty),d\mu)$, and a vector
has components $\psi=(\psi_0,(\psi_i)_{i\in\mathcal P})$ in the
vacuum and primary summands.

\subsection{Vacuum wave packets and bounded columns}\label{sub:columns}

With $\mathcal P$ as in Section~\ref{sub:decomposition}, write
\begin{equation}\label{eq:packet-space}
 \mathcal H=\mathcal H_{\mathrm{vac}}\oplus\mathcal H_{\mathrm{nv}},
 \quad \mathcal H_{\mathrm{vac}}=L^2(\mu),\quad
 \mathcal H_{\mathrm{nv}}=L^2(\mu\otimes\mu;\ell^2(\mathcal P)).
\end{equation}
Let $(e_i)$ be the standard basis of $\ell^2(\mathcal P)$,
$\iota_0:L^2(\mu)\to\mathcal H$ the vacuum inclusion, and
$P_0=\iota_0\iota_0^*$.
The generalized ket $|0;s\rangle$ has energy $E(s)$ on both
boundaries and normalization $\delta(s-s')/\rho(s)$.
The ket $|i;r,s\rangle$ has energies $E(r),E(s)$.
All column formulas below act on incoming wave packets.

For $a,b>0$ use the bounded operators already constructed in
Section~\ref{sec:bounds}, and their reflected operators, to define
\begin{equation}\label{eq:column-definition}
 \begin{aligned}
 T_{i,R}^{a,b}&:=T_{i;a,b}=e^{-aH_R}O_{i,R}e^{-bH_R},&
 T_{i,L}^{a,b}&=JT_{i,R}^{a,b}J,\\
 C_{i,R}^{a,b}&=T_{i,R}^{a,b}\iota_0,&
 C_{i,L}^{a,b}&=T_{i,L}^{a,b}\iota_0.
 \end{aligned}
\end{equation}
The $T$'s act on $\mathcal H$. The $C$'s are bounded maps from
$\mathcal H_{\mathrm{vac}}$ to $\mathcal H$; they are the columns.
For brevity put $G_i=G_{\lambda_i}$. The Gamma product gives
\begin{equation}\label{eq:column-positive-kernel}
 G_i(r,s)=
 \frac{|\Gamma(\lambda_i+i(r+s))|^2
       |\Gamma(\lambda_i+i(r-s))|^2}{\Gamma(2\lambda_i)}>0,
 \qquad r,s>0.
\end{equation}

\begin{proposition}\label{prop:bounded-columns}
For every $f\in L^2(\mu)$, the right column has only primary component
$i$, and in \eqref{eq:packet-space} it is
\begin{equation}\label{eq:12.6}
 (C_{i,R}^{a,b}f)(r,s)
 =e^{-aE(s)-bE(r)}\sqrt{G_i(r,s)}\,f(r)e_i.
\end{equation}
Its squared norm as an operator is
\begin{equation}\label{eq:column-norm}
 \|C_{i,R}^{a,b}\|^2
 =\operatorname*{ess\,sup}_{r>0}e^{-2bE(r)}
   \int_0^\infty e^{-2aE(s)}G_i(r,s)\,d\mu(s)<\infty.
\end{equation}
There are measurable functions $u_i$ of modulus one such that
\begin{equation}\label{eq:left-column-phase}
 (C_{i,L}^{a,b}f)(r,s)
 =u_i(r,s)e^{-aE(r)-bE(s)}\sqrt{G_i(r,s)}\,f(s)e_i.
\end{equation}
These are identities of bounded maps and $L^2$ wave packets.
\end{proposition}
\begin{proof}
First consider the expression on the right of \eqref{eq:12.6}.
Tonelli's theorem gives
\begin{equation}\label{eq:column-packet-norm}
 \|C_{i,R}^{a,b}f\|^2
 =\int_0^\infty |f(r)|^2e^{-2bE(r)}
   \left[\int_0^\infty e^{-2aE(s)}G_i(r,s)\,d\mu(s)\right]d\mu(r)
\end{equation}
for this candidate map, allowing infinity before estimating it.
By \eqref{eq:5.2}, with $M_i=4\lambda_i-2>0$,
\begin{equation}\label{eq:column-majorant}
 G_i(r,s)\rho(s)
 \leq C_i(1+r)^{M_i}(1+s)^{M_i+1}.
\end{equation}
Indeed $e^{-2\pi\max(r,s)}e^{2\pi s}\leq1$, and both polynomial
factors in \eqref{eq:5.2} are bounded by powers of $1+r+s$.
The Gaussian $e^{-2aE(s)}$ makes the integral in $s$ finite, and
$e^{-2bE(r)}(1+r)^{M_i}$ is bounded. Thus the candidate map is bounded.
Equation \eqref{eq:column-packet-norm} gives \eqref{eq:column-norm};
for the reverse norm inequality, choose characteristic wave packets
on finite-measure subsets of level sets of the bracketed multiplier.

We next identify this map with the column of the specified physical
letter. The vector $X_{i,R}\Omega$ in the formal creation formula is a
distribution, not a matter Hilbert-space vector. Section~\ref{sub:primaries}
identifies its radial creation distribution as $g_{\lambda_i}$ in
exactly the chosen primary copy. The other Wick terms annihilate the
vacuum. In the one-sided right energy representation of
Section~\ref{sub:kl}, write $\widehat\psi=\mathcal U U^*\psi$ for
the transform of a vector in the original canonical gauge. The cutoff
field calculation gives the weak kernel formula
\begin{equation}\label{eq:column-before-jacobi}
 \bigl(\widehat{T_{i,R}^{a,b}\iota_0f}\bigr)_i(s,p)
 =e^{-aE(s)}\int_0^\infty
   g_{\lambda_i}(p)F_{\lambda_i}(s,r;p,0)
   e^{-bE(r)}f(r)\,d\mu(r).
\end{equation}
This transform precedes radial diagonalization. The formula is first
paired with tests of compact right
energy and interior matter momentum support. The one-particle cutoffs
can then be removed by the bounded-letter convergence proved in
Section~\ref{sec:bounds}; their creation distributions converge on
these tests. For compact incoming spectral support the remaining
integrals on such tests are absolutely convergent, so Fubini applies.

Insert \eqref{eq:12.2} into \eqref{eq:column-before-jacobi} only after
integrating against $e^{-bE(r)}f(r)$. For each fixed $s$, the resulting
integral is the inverse radial transform of the function $r\longmapsto e^{-aE(s)-bE(r)}\sqrt{G_i(r,s)}\,f(r)$. It is an $L^2(\mu)$ function, and its squared norm integrated over $s$
is finite by \eqref{eq:column-packet-norm}. The unitary radial transform
therefore identifies the physical column with \eqref{eq:12.6} on a
dense space of wave packets. Boundedness of both maps extends the
identity to all of $L^2(\mu)$. The absence of other
primary components follows first in the distributional creation formula
and then in this bounded extension.

The vacuum transform has a real kernel, so $J\iota_0=\iota_0J_0$,
where $J_0f=\overline f$. By Section~\ref{sub:primaries}, $J$ preserves
each primary copy, including its multiplicity label. It exchanges
$H_L,H_R$, and the joint spectral multiplicity in each copy is one.
Consequently, composing $J$ with conjugation followed by the flip of
$r,s$ gives a unitary commuting with both scalar spectral multipliers.
That unitary is multiplication by a measurable function of modulus one.
Thus
\begin{equation}\label{eq:reflection-unfixed}
 (J\psi)_i(r,s)=u_i(r,s)\overline{\psi_i(s,r)},
 \qquad |u_i(r,s)|=1\quad\text{almost everywhere}.
\end{equation}
Now $C_{i,L}^{a,b}=JC_{i,R}^{a,b}J_0$ proves
\eqref{eq:left-column-phase} and its boundedness, with
$\|C_{i,L}^{a,b}\|=\|C_{i,R}^{a,b}\|$.
\end{proof}

\subsection{Factoriality from bounded columns}\label{sub:column-factor}

\begin{proposition}\label{prop:column-factor}
The algebras $M_R$ and $M_L$ are factors.
\end{proposition}
\begin{proof}
Let $z\in Z(M_R)$. Bounded locality implies that it commutes
with both Hamiltonians. In \eqref{eq:12.5}, the vacuum projection is
\begin{equation}\label{eq:vacuum-joint-projection}
 P_0=\mathbf1_{\{(u,v):u=v\}}(H_L,H_R):
\end{equation}
the diagonal has zero product measure in each nonvacuum summand.
Thus $z\iota_0=\iota_0M_c$ for some $c\in L^\infty(\mu)$, since the
vacuum spectrum is simple. On the nonvacuum part, joint spectral
commutation makes $z$ a bounded measurable operator field
$Z(r,s)$ on $\ell^2(\mathcal P)$.

Commutation with the full bounded letters gives
\begin{equation}\label{eq:right-column-intertwining}
 zC_{i,R}^{a,b}=C_{i,R}^{a,b}M_c,\qquad
 zC_{i,L}^{a,b}=C_{i,L}^{a,b}M_c.
\end{equation}
Apply these to $f_\beta(r)=e^{-\beta E(r)}$.
The right coefficient is strictly positive, so
$Z(r,s)e_i=c(r)e_i$ almost everywhere.
The left coefficient has nonzero modulus, and similarly gives
$Z(r,s)e_i=c(s)e_i$. Countability permits one common null set.
Hence $Z(r,s)=c(r)1=c(s)1$ almost everywhere.
Fubini, averaging $s$ over any finite positive-measure interval,
shows that $c$ is constant. This proves factoriality of $M_R$,
and reflection proves it for $M_L$.

\end{proof}

This is Kolchmeyer's spectral factoriality argument
\cite[equations (5.12)--(5.19)]{K}, written with bounded
intertwiners on wave packets.

\subsection{The reflection phase from bounded mixed products}\label{sub:reflection-phase}

We determine the phase in \eqref{eq:reflection-unfixed}.

\begin{proposition}\label{prop:reflection-phase}
In the spectral representation \eqref{eq:packet-space},
\begin{equation}\label{eq:12.8}
 (J\psi)_0(r)=\overline{\psi_0(r)},\qquad
 (J\psi)_i(r,s)=\overline{\psi_i(s,r)}.
\end{equation}
In particular $u_i=1$ almost everywhere in
\eqref{eq:left-column-phase}.
\end{proposition}

\begin{proofclaim}[Normalized matter contraction]\label{clm:phase-abel}
For $\epsilon>0$,
\begin{equation}\label{eq:regulated-matter-pairing}
 \left\langle e^{-\epsilon H_m/2}X_{i,L}\Omega,
                 e^{-\epsilon H_m/2}X_{i,R}\Omega\right\rangle
 =\cosh(\epsilon/2)^{-2\lambda_i}\longrightarrow1.
\end{equation}
\end{proofclaim}
\begin{proof}
Multiply the $n$th Laguerre coefficient by
$e^{-\epsilon(\lambda_i+n)/2}$.
The two resulting vectors have squared norm
$\sinh(\epsilon/2)^{-2\lambda_i}$; their pairing is the
alternating series in \eqref{eq:10.5}.
Different primary copies are orthogonal.
\end{proof}

\begin{proofclaim}[Pairing of the bounded columns]\label{clm:phase-mixed}
For every $t>0$, the columns $C_{i,L}^{t,t}$ and $C_{i,R}^{t,t}$
satisfy \eqref{eq:12.7} on compactly supported energy wave packets.
\end{proofclaim}
\begin{proof}
Fix $t>0$, put $h=t/(2\phi_b)$, and abbreviate
$T_R=T_{i,R}^{t,t}$, $T_L=T_{i,L}^{t,t}$ and
$C_R=C_{i,R}^{t,t}$, $C_L=C_{i,L}^{t,t}$. Both $T_R,T_L$ are bounded
self-adjoint operators. Write $\mathscr S=A_R+A_L$ as in
Section~\ref{sub:jointheat}. The bounded-factor identity
\eqref{eq:11.10}, with left and right interchanged, gives for $\eta>0$
\begin{equation}\label{eq:phase-mixed-identity}
 T_L e^{-\eta\mathscr S}T_R
 =e^{-h\mathscr S}O_{i,L}e^{-\eta\mathscr S}
                         O_{i,R}e^{-h\mathscr S}.
\end{equation}
The right side means a product of bounded heat-sandwiched forms, as
specified in that identity. The left side tends strongly to $T_LT_R$.
Lemma~\ref{lem:mixed} identifies the WOT limit of the right side with
the bounded realization of the full mixed Wick product between the
two outside joint heats. This is the step relating the Abel contraction
\eqref{eq:regulated-matter-pairing} to the physical bounded operators.

Take $f,g\in L^2(\mu)$ with compact spectral support in $(0,\infty)$.
Since $T_L=T_L^*$, the limit just described evaluates
$\langle C_Lf,C_Rg\rangle=\langle\iota_0f,T_LT_R\iota_0g\rangle$.
On the vacuum summand $e^{-h\mathscr S}$ is multiplication by
$e^{-2tE}$. The vacuum contraction of the mixed Wick product has
length factor $2^{2\lambda_i}e^{-\lambda_i x}$, and its scalar
coefficient is one by \eqref{eq:regulated-matter-pairing}. Therefore
\begin{equation}\label{eq:12.7}
 \langle C_Lf,C_Rg\rangle
 =\iint e^{-2t(E(r)+E(s))}G_i(r,s)
          \overline{f(s)}g(r)\,d\mu(r)d\mu(s).
\end{equation}
Here the length integral is
\[
 2^{2\lambda_i}\int_{\mathbb R}e^{-\lambda_i x}
       \zeta_s(0,x)\zeta_r(0,x)\,dx=G_i(r,s),
\]
by \eqref{eq:4.3}--\eqref{eq:4.4}. On compact spectral sets Fubini is
justified by $|K_{2ir}(z)|\leq K_0(z)$: the resulting length integral
is finite at both ends since $\lambda_i>0$. The integral in
\eqref{eq:12.7} is absolutely convergent; \eqref{eq:column-majorant}
also controls it after the outside heats. Approximation by the test
vectors in Lemma~\ref{lem:mixed} extends its WOT identity to these
vacuum wave packets.
\end{proof}

\begin{proof}[Proof of Proposition~\ref{prop:reflection-phase}]
Fix $t>0$ and abbreviate $C_L=C_{i,L}^{t,t}$ and
$C_R=C_{i,R}^{t,t}$. Claim~\ref{clm:phase-mixed} computes their
pairing without an unknown phase. On the other hand, the already proved column formulas give
\begin{equation}\label{eq:phase-packet-pairing}
 \langle C_Lf,C_Rg\rangle
 =\iint\overline{u_i(r,s)}e^{-2t(E(r)+E(s))}G_i(r,s)
          \overline{f(s)}g(r)\,d\mu(r)d\mu(s).
\end{equation}
Subtract \eqref{eq:12.7} from \eqref{eq:phase-packet-pairing}. On each
compact rectangle the difference kernel is integrable. Testing with
characteristic functions of measurable subsets in the two coordinates
shows that its integral over every measurable rectangle is zero.
Uniqueness of the corresponding finite complex measure makes the
kernel zero almost everywhere on that compact rectangle. A countable
exhaustion and the strict positivity of $G_i$ give $u_i=1$ almost
everywhere. Countability of $\mathcal P$ permits a common null set for
all primary copies, proving \eqref{eq:12.8} on the whole Hilbert space.
\end{proof}

\subsection{The vacuum map and density}\label{sub:vacuum}

Let $f_\beta(s)=e^{-\beta E(s)}$ and $\xi_\beta=\iota_0f_\beta$.
Then $J\xi_\beta=\xi_\beta$, and monotone convergence gives
\begin{equation}\label{eq:thermal-packet-norm}
 \|\xi_\beta\|^2=\int_0^\infty e^{-2\beta E(s)}\,d\mu(s)<\infty,
 \qquad \|\xi_\beta\|^2\longrightarrow\infty
 \quad(\beta\downarrow0).
\end{equation}
On the nonunital heat-word algebra $\mathfrak A_0$, define

\begin{equation}\label{eq:12.9}
\Lambda(a)=\lim_{\beta\downarrow0}a\xi_\beta.
\end{equation}

Write a heat word as $a=Be^{-\delta H_R}$ with $B$ bounded
and $\delta>0$, absorbing any Hamiltonian powers into half
their adjoining heat interval. Dominated convergence gives
$f_{\delta+\beta}\to f_\delta$ in $L^2(\mu)$, hence
$\Lambda(a)=B\iota_0f_\delta$.
By linearity $\Lambda$ exists on $\mathfrak A_0$.
In particular, \eqref{eq:12.6} gives

\begin{equation}\label{eq:12.10}
\Lambda(e^{-aH_R}O_{i,R}e^{-bH_R})
=e^{-aE(s)-bE(r)}\sqrt{G_{\lambda_i}(r,s)}e_i.
\end{equation}

To distinguish the domains of the two vacuum maps, put
\(\mathfrak A_{0,L}=J\mathfrak A_0J\) and define

\begin{equation}\label{eq:12.10a}
\Lambda_R(a)=\Lambda(a)\quad(a\in\mathfrak A_0),
\qquad
\Lambda_L(b)=\lim_{\beta\downarrow0}b\xi_\beta
\quad(b\in\mathfrak A_{0,L}).
\end{equation}

The left limit exists in norm by the reflected heat-tail argument, or
directly by \(J\xi_\beta=\xi_\beta\). Both maps are linear on their
respective represented algebras. We continue to write \(\Lambda\) for
\(\Lambda_R\) when only right words occur.

Together with the vacuum heat vectors, the vectors in \eqref{eq:12.10} are total.
Fix a primary summand $i$ and put $G=G_{\lambda_i}$. If a vector
in this summand is orthogonal to all of them, multiply its conjugate
by \(\sqrt G\) and fixed positive heat factors.
Cauchy-Schwarz and \eqref{eq:5.2} produce a finite complex measure whose
two-variable Laplace transform vanishes. Push it forward under
\((r,s)\mapsto(e^{-E(r)},e^{-E(s)})\) and use polynomial density on
\([0,1]^2\). The measure is zero, and \(G>0\) gives that the original
vector was zero. The same one-variable argument treats the vacuum
summand.

Writing \(a^{\mathrm{op}}=Ja^*J\), \eqref{eq:12.8}-\eqref{eq:12.10} show for every
generating letter, including positive heat operators,

\begin{equation}\label{eq:12.11}
\lim_{\beta\downarrow0}a\xi_\beta
=\lim_{\beta\downarrow0}a^{\mathrm{op}}\xi_\beta
\quad\text{in norm}.
\end{equation}

Equivalently, $\Lambda_R(a)=\Lambda_L(a^{\mathrm{op}})$
for generating letters.

\section{The trace and the boundary factors}\label{sec:trace}

The dense vacuum map and reflection identity make the heat-word
algebra a Hilbert algebra. Its trace theorem gives the normal
semifinite trace and mutual commutants.

\subsection{Reflection on the word algebra}\label{sub:reflection}

By Section~\ref{sub:locality}, \(\mathfrak A_0\) commutes with \(J\mathfrak A_0J\). We
can therefore propagate \eqref{eq:12.11} to products, using only bounded
operators and norm limits:

\begin{equation}\label{eq:13.2}
\begin{aligned}
\Lambda(ab)
&=a\Lambda(b)=\lim_\beta ab^{\mathrm{op}}\xi_\beta
=b^{\mathrm{op}}\Lambda(a)\\
&=\lim_\beta b^{\mathrm{op}}a^{\mathrm{op}}\xi_\beta
=\lim_\beta(ab)^{\mathrm{op}}\xi_\beta.
\end{aligned}
\end{equation}

Linearity gives this for every element. Since \(J\xi_\beta=\xi_\beta\),
it follows that

\begin{equation}\label{eq:13.3}
J\Lambda(a)=\Lambda(a^*),\qquad
\Lambda(ab)=a\Lambda(b)=b^{\mathrm{op}}\Lambda(a).
\end{equation}

The map $\Lambda$ is injective. If \(\Lambda(a)=0\), the last equality
implies that \(a\) kills the dense space \(\Lambda(\mathfrak A_0)\),
hence \(a=0\).

\subsection{Algebraic cyclicity}\label{sub:cyclicity}

\begin{proposition}\label{prop:cyclicity}
For $a\in\mathfrak A_0$, the limit
$\tau_0(a)=\lim_{\beta\downarrow0}\langle\xi_\beta,a\xi_\beta\rangle$
exists and is finite. It defines a linear functional which is faithful
on squares and satisfies $\tau_0(ab)=\tau_0(ba)$ for all
$a,b\in\mathfrak A_0$.
\end{proposition}
\begin{proof}
Each $a\in\mathfrak A_0$ is a sum of terms
$e^{-\epsilon H_R}Be^{-\delta H_R}$ with positive heat times and
bounded $B$. The heat tails applied to $\xi_\beta$ converge in
norm, so $\tau_0(a)$ exists. Moreover
$\tau_0(a^*b)=\langle\Lambda(a),\Lambda(b)\rangle$.
Injectivity of $\Lambda$ gives faithfulness on squares, and
\eqref{eq:13.3} gives $\tau_0(a^*a)=\tau_0(aa^*)$.
Applying this to $a+zb$ yields
\begin{equation}\label{eq:13.3a}
 z[\tau_0(a^*b)-\tau_0(ba^*)]
 +\overline z[\tau_0(b^*a)-\tau_0(ab^*)]=0.
\end{equation}
Taking $z=1,i$ proves cyclicity.
\end{proof}

\subsection{The normal semifinite trace}\label{sub:normaltrace}

\begin{theorem}\label{thm:trace}
The space $\Lambda(\mathfrak A_0)$ is a Hilbert algebra whose left
von Neumann algebra is $M_R$. Its canonical trace is faithful, normal
and semifinite, and $M_R'=JM_RJ=M_L$.
\end{theorem}
\begin{proof}
Transport multiplication and involution to the dense space
\(\Lambda(\mathfrak A_0)\subset\mathcal H\). Left multiplication by
\(\Lambda(a)\) is the original bounded operator \(a\). Moreover

\begin{equation}\label{eq:13.4}
\langle\Lambda(ab),\Lambda(c)\rangle
=\langle\Lambda(b),\Lambda(a^*c)\rangle,
\qquad \|\Lambda(a^*)\|=\|\Lambda(a)\|.
\end{equation}

The involution extends to $J$, and products are dense since
$\Lambda(e^{-\delta H_R}a)=e^{-\delta H_R}\Lambda(a)\to\Lambda(a)$.
Together with \eqref{eq:13.4}, these are the Hilbert-algebra
axioms \cite[Definition 2.1]{G}. Left multiplication is the
physical action; the bounded fulfillment gives the trace theorem.

The Hilbert-algebra trace theorem \cite[Theorem 2.3]{G} now gives a
faithful normal semifinite trace \(\tau\) on the canonical
algebra \(M_R\), satisfying

\begin{equation}\label{eq:13.5}
\tau(a^*b)=\langle\Lambda(a),\Lambda(b)\rangle,
\qquad a,b\in\mathfrak A_0.
\end{equation}

Its commutant theorem gives

\begin{equation}\label{eq:13.6}
M_R'=JM_RJ=M_L,\qquad M_L'=M_R.
\end{equation}

The Tomita involution closes to $J$, so its modular operator is one.
\end{proof}

\subsection{The spectral measure and the thermal limit}\label{sub:thermal}

For \(a>0\), \eqref{eq:13.5} applied to \(e^{-aH_R/2}\) gives

\begin{equation}\label{eq:13.7}
\tau(e^{-aH_R})=\int_0^\infty e^{-aE(s)}d\mu(s).
\end{equation}

Normality supplies a spectral measure \(I\mapsto\tau(\mathbf1_I(H_R))\),
finite on bounded intervals by \eqref{eq:13.7}. Uniqueness of its Laplace
transform gives

\begin{equation}\label{eq:13.8}
\tau(f(H_R))=\int_0^\infty f(E(s))d\mu(s)
\quad\text{for every nonnegative Borel }f.
\end{equation}

For an explicit proof of uniqueness, multiply both measures by
\(e^{-aE}\) to make them finite and push them forward by
\(E\mapsto e^{-E}\). Equality of all integer moments, including the
zeroth, and polynomial density on \([0,1]\) give equality of the
measures.

Fix a countable partition \((I_m)\) of the energy half-line into bounded
intervals and put \(p_m=\mathbf1_{I_m}(H_R)\). In the physical
Hilbert-algebra completion, the \(L^2\) vector of \(p_m\) is the vacuum
vector $v_m:=\iota_0[s\mapsto\mathbf1_{I_m}(E(s))]$. To verify this
identification, approximate the indicator in \(L^2(\mu)\) by bounded
finite combinations of positive exponentials; \eqref{eq:13.7} gives the same
\(L^2(\tau)\) convergence. Then for \(x\in(M_R)_+\),

\begin{equation}\label{eq:13.9}
\begin{aligned}
\tau(x)
&=\sum_m\tau(p_mxp_m)
=\sum_m\langle v_m,xv_m\rangle.
\end{aligned}
\end{equation}

Here normality applies to the increasing finite sums of
$x^{1/2}p_mx^{1/2}$, and traciality gives the first equality.

The original thermal-vector prescription also follows. The vacuum
compression of \(x\) commutes with \(H_L\) and is a multiplier
\(c_x(s)\geq0\), because the vacuum spectrum is simple. Equations \eqref{eq:13.9}
and monotone convergence give

\begin{equation}\label{eq:13.10}
\tau(x)=\int c_x(s)d\mu(s)
=\lim_{\beta\downarrow0}\langle\xi_\beta,x\xi_\beta\rangle.
\end{equation}

\subsection{Factoriality and type}\label{sub:factor}

\begin{theorem}\label{thm:factor}
The algebras $M_R$ and $M_L$ are factors of type
$\mathrm{II}_\infty$. If $0\ne p=p^*=p^2\in M_R$ and $\tau(p)<\infty$,
then $pM_Rp$ is a factor of type $\mathrm{II}_1$.
\end{theorem}
\begin{proof}
Proposition~\ref{prop:column-factor} already proves that both algebras
are factors. Theorem~\ref{thm:trace} supplies their faithful normal
semifinite traces. For a bounded energy interval $I$ of positive
measure, set $p_I=\ind_I(H_R)$. Equation~\eqref{eq:13.8} gives

\begin{equation}\label{eq:14.1}
0<\tau(p_I)=\int_{E^{-1}(I)}\frac{s\sinh(2\pi s)}{\pi^2}ds<\infty,
\qquad \tau(1)=\infty.
\end{equation}

There are nonzero projections with arbitrarily small trace, since

\begin{equation}\label{eq:14.2}
\tau(\mathbf1_{(0,\delta^2/(4\phi_b))}(H_R))
\sim\frac{2\delta^3}{3\pi}\quad(\delta\downarrow0).
\end{equation}

A semifinite type I factor has a positive minimum trace among nonzero
projections, because its faithful normal semifinite trace is a positive
scalar multiple of the usual operator trace. Equation \eqref{eq:14.2} excludes
it. Semifiniteness excludes type III. A type \(\mathrm{II}_1\) factor
has finite identity trace for every faithful normal semifinite trace,
whereas \eqref{eq:14.1} gives an infinite value. Consequently

\begin{equation}\label{eq:14.3}
M_R\text{ and }M_L\text{ are type }\mathrm{II}_\infty\text{ factors}.
\end{equation}

For $0\ne p$ of finite trace, $pM_Rp$ is finite and has no minimal
projections. It is a factor: a central projection $0<q<p$ would give
$qM_R(p-q)=0$, contradicting the fact that both projections have
central support one. Its normalized trace is $\tau(\cdot)/\tau(p)$.
\end{proof}

\section{Finite corners, normal states and entropy}\label{sec:corners}

Finite energy projections give finite corners. We describe their
representations and normal states, using the trace constructed above.

\subsection{Simultaneous energy cutoffs}\label{sub:corners}

For $\Lambda>0$, put
$e=\mathbf1_{[0,\Lambda)}(H_R)$,
$f=JeJ=\mathbf1_{[0,\Lambda)}(H_L)$, and $P_\Lambda=ef$.
The projections commute, and $e,f$ have finite nonzero traces
in their respective boundary algebras.

Under the unitary \(\mathcal H=L^2(M_R,\tau)\) proved in Section~\ref{sec:trace},
\(e\) acts by left multiplication and \(f\) by right multiplication by
\(e\). Thus, first on bounded $L^2$ elements $x$ and then by
continuity,

\begin{equation}\label{eq:15.1}
P_\Lambda x=exe,\qquad
P_\Lambda\mathcal H=L^2(eM_Re,\tau|_{eM_Re}).
\end{equation}

Indeed, $x\mapsto exe$ is an $L^2$ contraction, and the bounded
elements of $eM_Re$ are dense in its $L^2$ space.
Compression of $a\in M_R$ acts there by left multiplication by
$eae$. Hence

\begin{equation}\label{eq:15.2}
\begin{aligned}
P_\Lambda M_RP_\Lambda&\simeq eM_Re,\\
(P_\Lambda M_RP_\Lambda)'_{B(P_\Lambda\mathcal H)}
&=P_\Lambda M_LP_\Lambda.
\end{aligned}
\end{equation}

These are type $\mathrm{II}_1$ factors in standard form, with
identity $P_\Lambda$. Rescaling the trace only rescales the
$L^2$ inner product. Thus the simultaneous cutoff gives the
standard representation of a finite corner
\cite[Section 7]{K}.

\subsection{Normal representations}\label{sub:representations}
Put $C=eM_Re$ and normalize its trace by $\tau_C=\tau/\tau(e)$.
A representation here is a normal unital $*$-homomorphism.
Every separable such representation is a compression
\begin{equation}\label{eq:15.3}
 p\bigl(L^2(C,\tau_C)\otimes\ell^2(\mathbb N)\bigr),\qquad
 p\in C^{\op}\bar\otimes B(\ell^2(\mathbb N))
\end{equation}
by a projection in the commutant \cite[Theorem 10.1.1]{J}.
Indeed, split the representation into cyclic summands.
A cyclic normal state has a density $h\in L^1(C,\tau_C)_+$,
and $\pi(a)\xi\mapsto ah^{1/2}$ embeds its cyclic space
isometrically in $L^2(C)$. The range reduces $C$.
Taking sums gives \eqref{eq:15.3}; equivalence of representations
is equivalence of the corresponding commutant projections.

There is no nonzero isometric intertwiner of both boundary actions
from $L^2(C)$ into a direct sum of tensor products of normal
representations of $C$ and $C^{\op}$.
Such an intertwiner would extend left and right multiplication to
a faithful normal representation of the finite factor
$C\bar\otimes C^{\op}$. Its image would be $B(L^2(C))$, since
the joint commutant of the two actions is $Z(C)=\mathbb C$.
This contradicts finiteness. On the uncut space, the same commutant calculation gives
$(M_R\cup M_L)''=B(\mathcal H)$.

\subsection{Normal states and trace ideals}\label{sub:states}

For \(M=M_R\), define

\begin{equation}\label{eq:15.4}
\mathfrak n_\tau=\{a\in M:\tau(a^*a)<\infty\},\qquad
\mathfrak m_\tau=\operatorname{span}\{a^*b:a,b\in\mathfrak n_\tau\}.
\end{equation}

The trace extends linearly to $\mathfrak m_\tau$.
The completions of $\mathfrak n_\tau$ and $\mathfrak m_\tau$ in
$\|a\|_2=\tau(a^*a)^{1/2}$ and $\|a\|_1=\tau(|a|)$ are
$L^2(M,\tau)$ and $L^1(M,\tau)$, respectively.

A positive self-adjoint $h$ is affiliated with $M$ when its
spectral projections belong to $M$; it belongs to $L^1(M,\tau)$
when $\tau(h)<\infty$.

Every normal state has a unique density

\begin{equation}\label{eq:15.5}
h\in L^1(M,\tau)_+,\quad \tau(h)=1,\qquad
\varphi(a)=\tau(ha),\quad a\in M.
\end{equation}

This is the semifinite predual theorem \cite[Corollary 5.14]{H}.
Truncating and normalizing $h$ gives bounded densities whose
states converge in norm to $\varphi$.

Every heat-separated word belongs to \(\mathfrak m_\tau\). Namely, write
it as \(e^{-\epsilon H_R}B e^{-\delta H_R}\); the outside factors are in
\(L^2\), and the \(L^2\cdot M\cdot L^2\subset L^1\) inequality gives
finiteness. Thus the finite algebraic trace of Section~\ref{sec:trace} is exactly the
restriction of \(\tau\).

\subsection{Uniqueness of the trace and thermal entropy}\label{sub:entropy}

To compare faithful normal semifinite traces $\sigma,\tau$, choose
a nonzero projection $q$ finite for both. Uniqueness of the normalized
trace on $qM_Rq$ gives $\sigma=c\tau$ there.
Choose partial isometries $v_j$ with orthogonal ranges summing
to $1$ and $v_j^*v_j\le q$. Such a countable family exists since
$q$ has central support $1$. Normality and traciality give
\begin{equation}\label{eq:15.6}
 \sigma(a)=\sum_j\sigma(v_j^*av_j)
          =c\sum_j\tau(v_j^*av_j)=c\tau(a),\qquad a\ge0.
\end{equation}

For a density $h$, define $S_\tau(h)=-\tau(h\log h)$ by spectral
calculus, with $0\log0=0$, whenever at least one sign part is
integrable. If $\tau_c=c\tau$, the same state has density $h/c$, and

\begin{equation}\label{eq:15.7}
-\tau_c\bigl((h/c)\log(h/c)\bigr)
=-\tau(h\log h)+\log c.
\end{equation}

This proves the state-independent additive normalization ambiguity. 

For $\beta>0$, define the partition function
$Z(\beta):=\tau(e^{-\beta H_R})$. The thermal density
\(h_\beta=Z(\beta)^{-1}e^{-\beta H_R}\) is well-defined. Differentiating the elementary
Gaussian integral for \(\cosh(bs)\) gives

\begin{equation}\label{eq:15.8}
Z(\beta)=\frac{4\phi_b^{3/2}}{\sqrt\pi\,\beta^{3/2}}
\exp\left(\frac{4\pi^2\phi_b}{\beta}\right).
\end{equation}

Gaussian domination justifies the derivative and
\(\tau(h_\beta H_R)<\infty\). Hence its entropy is the finite number
\(\log Z(\beta)-\beta\partial_\beta\log Z(\beta)\).

Under the standard-form identification, the vacuum vector
$\xi_{\beta/2}$ is $e^{-\beta H_R/2}\in L^2(M_R,\tau)$. Consequently
\begin{equation}\label{eq:thermaldensity}
 \langle\xi_{\beta/2},x\xi_{\beta/2}\rangle
 =\tau(e^{-\beta H_R/2}xe^{-\beta H_R/2})
 =\tau(e^{-\beta H_R}x),\qquad x\in M_R.
\end{equation}
After division by $Z(\beta)$, the vector state therefore has density
$h_\beta$. This establishes the canonical thermal-state formula as
well as the high-temperature trace limit.

This establishes assertions (1)--(5) of Theorem~\ref{thm:main}.
Assertion (6) is proved in Theorem~\ref{thm:nonhyper} below.

\section{The sharp weight threshold and positive-energy scalars}\label{sec:smallweight}

The restriction $d>1/2$ concerns JT point fields, not unitarity
of the matter theory. We describe the positive-energy scalars
below this threshold and prove that their heat sandwiches are
unbounded.

\subsection{The scalar mass, boundary condition and positive energy}
\label{sub:smallweight-ads}

For AdS radius $R$, a scalar of lowest weight $d$ has lowest
global energy $d/R$ and mass
\begin{equation}\label{eq:smallweight-mass}
 m^2R^2=d(d-1),\qquad
 \Delta_\pm=\frac12\pm\sqrt{\frac14+m^2R^2}.
\end{equation}
The range $0<d<1/2$ corresponds to $d=\Delta_-$ and
$-1/4<m^2R^2<0$, above the Breitenlohner--Freedman bound
\cite{BF}. We specify the positive self-adjoint spatial
operator and its boundary condition, as in \cite{IW}.

For the rest of this subsection fix $0<d<1/2$. We work on the
universal covering spacetime, with metric
\[
 ds^2=R^2\sec^2\rho\,(-dt^2+d\rho^2),\qquad
 t\in\mathbb R,\quad -\pi/2<\rho<\pi/2.
\]
The spatial Klein--Gordon expression is
\begin{equation}\label{eq:smallweight-spatial}
 \mathcal A_d=-\frac{d^2}{d\rho^2}
             +\frac{d(d-1)}{\cos^2\rho}.
\end{equation}
At each endpoint its two characteristic behaviors are $z^d$ and
$z^{1-d}$, where $z$ is the distance to that endpoint. Both are
square-integrable. The alternative, or generalized Neumann,
condition selects the $z^d$ branch at both endpoints. Equivalently,
put $r_d(\rho)=(\cos\rho)^d$ and impose
\begin{equation}\label{eq:smallweight-bc}
 \lim_{\rho\to\pm\pi/2}W(f,r_d)(\rho)=0,
 \qquad W(f,r_d)=f r_d'-f'r_d,
\end{equation}
on the maximal domain of \eqref{eq:smallweight-spatial}: $f,f'$ are
locally absolutely continuous and $f,\mathcal A_df\in L^2(d\rho)$.
For an eigenfunction $f=a_\pm z^d+b_\pm z^{1-d}+\cdots$, this
condition is $b_\pm=0$. It is different from the Friedrichs
condition, which selects $z^{1-d}$. The two limit-circle endpoints
and the separated real conditions \eqref{eq:smallweight-bc} define
a self-adjoint extension by the singular Sturm--Liouville extension
theorem; see \cite[Theorem 3.11]{GLN}. Denote it by
$\mathcal A_d^{\mathrm{alt}}$.

\begin{proposition}\label{prop:smallweight-positive}
For every $0<d<1/2$, $\mathcal A_d^{\mathrm{alt}}$ is strictly positive,
with eigenvalues $(n+d)^2$, $n\geq0$. Its positive-frequency solution
space carries a unitary lowest-weight representation of the universal
covering group $\widetilde{\mathrm{SL}}(2,\mathbb R)$ with lowest weight $d$.
Bosonic Fock quantization has an invariant vacuum and nonnegative
global Hamiltonian.
\end{proposition}
\begin{proof}
Let $C_n^{(d)}$ be the Gegenbauer polynomial of degree $n$, choose
$c_n>0$, and set
\begin{equation}\label{eq:smallweight-modes}
 \begin{split}
 u_n(\rho)&=c_n(\cos\rho)^d C_n^{(d)}(\sin\rho),\\
 c_n^2&=\frac{2^{2d-1}n!(n+d)\Gamma(d)^2}
                  {\pi\Gamma(n+2d)}.
 \end{split}
\end{equation}
The substitution $y=\sin\rho$ reduces their orthogonality to the
weight $(1-y^2)^{d-1/2}$ on $(-1,1)$; the normalization and
differential equation are those in \cite[Tables 18.3.1 and 18.8.1]{DLMF}.
Polynomials are dense for this finite positive measure, so the $u_n$
form a complete orthonormal basis of $L^2(d\rho)$. Substitution in
\eqref{eq:smallweight-spatial} gives
$\mathcal A_du_n=(n+d)^2u_n$, and the endpoint expansion of the
polynomial shows \eqref{eq:smallweight-bc}. Consequently
\begin{equation}\label{eq:smallweight-domain}
 \begin{split}
 D(\mathcal A_d^{\mathrm{alt}})
 &=\left\{\sum_{n\geq0}a_nu_n:
                  \sum_{n\geq0}(n+d)^4|a_n|^2<\infty\right\},\\
 \mathcal A_d^{\mathrm{alt}}\sum_na_nu_n
 &=\sum_n(n+d)^2a_nu_n.
 \end{split}
\end{equation}
In particular, $\mathcal A_d^{\mathrm{alt}}\geq d^2\mathbf1$.

The positive-frequency solutions $v_n(t,\rho)=\frac{e^{-i(n+d)t}}{\sqrt{2(n+d)}}u_n(\rho)$ are orthonormal for the Klein--Gordon inner product
$i\int(\overline v\,\partial_tw-\partial_t\overline v\,w)\,d\rho$.
Let $\mathcal K_d$ be their Hilbert completion. Its global generator
$h_dv_n=(n+d)v_n$ is positive and has its maximal diagonal domain.
After a choice of phases, the AdS raising and lowering generators on
the finite span of the $v_n$ are
\begin{equation}\label{eq:smallweight-ladder}
 \begin{split}
 L_+v_n&=\sqrt{(n+1)(n+2d)}\,v_{n+1},\\
 L_-v_n&=\sqrt{n(n+2d-1)}\,v_{n-1},\qquad L_-v_0=0.
 \end{split}
\end{equation}
All squared coefficients are positive when they occur. Directly,
$\langle L_+v,w\rangle=\langle v,L_-w\rangle$ on this common core,
$[h_d,L_\pm]=\pm L_\pm$ and $[L_-,L_+]=2h_d$.
With $X=(L_++L_-)/2$ and $Y=(L_+-L_-)/(2i)$, the sum of squares
$h_d^2+X^2+Y^2=2h_d^2-d(d-1)\mathbf1$ is essentially self-adjoint
on the finite span. Nelson's integrability theorem \cite{N} therefore
gives the unitary representation of the simply connected group.
The coefficients also show directly that each finite-span vector
is analytic for the Lie generators. Second quantization gives an
invariant Fock vacuum and
$R^{-1}d\Gamma(h_d)\geq(d/R)N\geq0$.
\end{proof}

This is the generalized Neumann theory of
\cite[Sections IV.C and IV.E, equations (78)--(79)]{HSB},
with $\lambda=1-d$ and representation $D_{1-\lambda}^+=D_d^+$.
The universal covering group allows every $d>0$.

\subsection{The endpoint in the momentum realization}
\label{sub:smallweight-momentum}

For $d<1/2$ the same momentum differential expression requires
a different endpoint condition.

The Laguerre functions \eqref{eq:2.1} remain an orthonormal basis for
every $d>0$, since $2d-1>-1$. The unitary map $v_n\mapsto f_n^d$
therefore gives the momentum realization on $L^2(\mathbb R_+,dp)$.
For $0<d<1/2$ one must change the extension in
\eqref{eq:19.3}. More precisely, define the unitary involution
$s_df_n^d=(-1)^nf_n^d$, let $q$ be maximal multiplication by $p$,
and put
\begin{equation}\label{eq:smallweight-kappa}
 \kappa_d=s_dqs_d,\qquad D(\kappa_d)=s_dD(q).
\end{equation}
It is a nonnegative self-adjoint operator. The Laguerre recurrence
identifies its expression on the finite Laguerre span as
$-\partial_p p\partial_p+(d-1/2)^2/p$, with endpoint behavior
$p^{d-1/2}$. It also gives $(q+\kappa_d)/2=h_d$ on that span.

For $0<d<1/2$, $f_0^d\propto p^{d-1/2}e^{-p}$ has infinite
energy for the form \eqref{eq:19.3}. The Friedrichs realization
instead selects $p^{1/2-d}$, of lowest weight $1-d$.

\subsection{Unbounded fields and unbounded JT heat sandwiches}
\label{sub:smallweight-unbounded}

For real $f\in C_c^\infty(\mathbb R)$ set
$w_f=g_d\widehat f$, where $\widehat f(p)=\int e^{ipt}f(t)\,dt$.
Since $\widehat f$ decays rapidly,
\begin{equation}\label{eq:smallweight-smearing}
 \|w_f\|^2=\frac1{\Gamma(2d)}
       \int_0^\infty p^{2d-1}|\widehat f(p)|^2\,dp<\infty
 \qquad(d>0).
\end{equation}
The field $\Phi_d(f)=a^*(w_f)+a(w_f)$ is essentially self-adjoint
on finite-particle vectors by \cite{N}. If $w_f\ne0$, then
$\eta_n=(a^*(w_f/\|w_f\|))^n\Omega/\sqrt{n!}$ satisfies
\begin{equation}\label{eq:smallweight-field-unbounded}
 \|\Phi_d(f)\eta_n\|^2=(2n+1)\|w_f\|^2,
\end{equation}
so $\Phi_d(f)$ is unbounded.

For one species write
$\mathcal H_n=L^2(\mathbb R_x)\otimes
L^2(\mathbb R_+,dp)^{\otimes_s n}$ and
$\mathcal H_{\mathrm{fin}}=\bigoplus_n^{\mathrm{alg}}\mathcal H_n$.
The positive form \eqref{eq:3.3} defines $L$ for every $d>0$.

\begin{proposition}\label{prop:smallweight-heat-operator}
For $d>0$ and $t>0$, the cutoff limit of
\[
 T_t:=T_{d,t}=e^{-tL}\bigl(2^de^{-dx}\Phi_d(0)\bigr)e^{-tL}
\]
defines a symmetric operator on $\mathcal H_{\mathrm{fin}}$ which is
essentially self-adjoint there. If $0<d\leq1/2$, its self-adjoint
closure is unbounded. The same assertions hold after conjugation by
$U$, for the symmetric right JT heat sandwich.
\end{proposition}
\begin{proof}
For all $a>0$ and $d>0$,
\(\|e^{-aq}g_d\|^2=(2a)^{-2d}.\)
Pull-through and the fixed-particle-number creation and annihilation
bounds therefore give, on the matter $n$-particle sector, the
following estimates, where $\|\cdot\|_{n\to m}$ is the operator
norm of the block from $n$ to $m$ particles:
\begin{equation}\label{eq:smallweight-block-matter}
 \begin{split}
 \|e^{-AQ}a^*(g_d)e^{-BQ}\|_{n\to n+1}
       &\leq\sqrt{n+1}\,(2A)^{-d},\\
 \|e^{-AQ}a(g_d)e^{-BQ}\|_{n\to n-1}
       &\leq\sqrt n\,(2B)^{-d}.
 \end{split}
\end{equation}
These inequalities first hold with momentum cutoffs; for fixed
$A,B>0$ the blocks converge in operator norm as the cutoffs are
removed, because $e^{-Aq}g_d,e^{-Bq}g_d\in L^2(dp)$.

Choose an integer $r>d$ and apply the resolvent factorization of
Section~\ref{sub:schur} to these fixed-particle-number blocks.
For the Gamma integration variables $s,v>0$ and a fixed length
$x$, the factor $2^de^{-dx}$ cancels
$(2st e^{-x})^{-d}$ in the creation bound and
$(2vt e^{-x})^{-d}$ in the annihilation bound. Integration against
the Gamma densities and the estimate for $B_r(t)$ therefore give
strong cutoff limits $T_{m,n}:\mathcal H_n\to\mathcal H_m$,
$m=n\pm1$, satisfying
\begin{equation}\label{eq:smallweight-block-JT}
 \begin{split}
 \|T_{n+1,n}\|&\leq C_{t,d}\sqrt{n+1},\qquad
 \|T_{n-1,n}\|\leq C_{t,d}\sqrt n,\\
 C_{t,d}&=\gamma_r(d)b_r(t)^2t^{-d}<\infty.
 \end{split}
\end{equation}
The fiberwise norm convergence of the matter blocks and the
integrable Gamma majorants give strong convergence of each block
and its adjoint, by \eqref{eq:7.4a}--\eqref{eq:7.4b}. This proof
uses $r>d>0$ and does not use the full-Fock annihilation estimate,
which would require $d>1/2$. The blocks thus define an operator on
$\mathcal H_{\mathrm{fin}}$, agree with the original form on
finite-particle tests, and satisfy $T_{n+1,n}^*=T_{n,n+1}$.

The raising and lowering parts separately give
\begin{equation}\label{eq:smallweight-N-bound}
 \|T_t\psi\|\leq2C_{t,d}\|(N+1)^{1/2}\psi\|,
 \qquad \psi\in\mathcal H_{\mathrm{fin}}.
\end{equation}
If $\psi$ has at most $n_0$ particles, each application increases
that number by at most one, so
\[
 \|T_t^m\psi\|\leq(2C_{t,d})^m
       \sqrt{\frac{(n_0+m)!}{n_0!}}\,\|\psi\|,
 \qquad m=0,1,2,\ldots.
\]
Consequently $\sum_{m\geq0}s^m\|T_t^m\psi\|/m!<\infty$ for every $s>0$.
The dense domain consists of analytic vectors, and Nelson's theorem
\cite{N} proves essential self-adjointness. Proposition~\ref{prop:threshold}
below provides normalized finite-particle test vectors with arbitrarily
large diagonal matrix elements when $0<d\leq1/2$; hence this closure
is unbounded. Finally, \eqref{eq:3.1}--\eqref{eq:3.2} give
\[
 e^{-\varepsilon H_R}O_Re^{-\varepsilon H_R}=UT_tU^*,
 \qquad t=\varepsilon/(2\phi_b),
\]
as forms and then as the operators just constructed. Their
self-adjoint closures are conjugate by $U$.
\end{proof}

The threshold comes from
\[
 \|p^{-1/2}g_d\|_{L^2(0,1)}^2
 =\frac1{\Gamma(2d)}\int_0^1p^{2d-2}\,dp=\infty
 \qquad(0<d\le1/2).
\]
Coherent states with bounded total momentum turn this divergence
into unbounded matrix elements after heat evolution.

\subsection{Sharpness of the weight condition}\label{sub:threshold}

\begin{proposition}\label{prop:threshold}
For one species of weight $0<d\le1/2$ and every $t>0$, the form
$e^{-tL}Ve^{-tL}$, where $V=2^de^{-dx}\Phi_d(0)$,
has no bounded extension.
\end{proposition}

The amplitude is chosen to make the field large at fixed total
momentum. Equality in weighted Cauchy--Schwarz suggests an amplitude
proportional to $g_d(p)/p$; a lower momentum cutoff makes it a vector,
and the normalization below fixes its momentum expectation. For
$0<\delta<1$, set

\begin{equation}\label{eq:18.1}
I_\delta=\int_\delta^1\frac{g_d(p)^2}{p}dp,
\qquad
h_\delta(p)=\frac{g_d(p)}{p\sqrt{I_\delta}}
\mathbf1_{[\delta,1]}(p).
\end{equation}

For each such $\delta$, $h_\delta\in\mathfrak h$. Define its
normalized coherent vector by $\eta_\delta=e^{-\|h_\delta\|^2/2} \bigoplus_{n=0}^\infty\frac{h_\delta^{\otimes n}}{\sqrt{n!}}$. It satisfies

\begin{equation}\label{eq:18.2}
\langle\eta_\delta,Q\eta_\delta\rangle=1,
\qquad
\int g_d(p)h_\delta(p)dp=\sqrt{I_\delta}.
\end{equation}

The first equality follows from \(\int p|h_\delta|^2=1\). But

\[
I_\delta=\frac1{\Gamma(2d)}
\begin{cases}
\displaystyle\frac{1-\delta^{2d-1}}{2d-1},&0<d<1/2,\\[2mm]
\log(1/\delta),&d=1/2,
\end{cases}
\]

which diverges as \(\delta\downarrow0\).

\begin{proofclaim}[Lower bound after matter heat]\label{clm:threshold-matter}
For every $A,B>0$, the coherent matrix element has the lower
bound \eqref{eq:18.4}, with its divergent factor $\sqrt{I_\delta}$.
\end{proofclaim}
\begin{proof}
For \(A,B>0\), the coherent-vector formula and the pull-through
identities give

\begin{equation}\label{eq:18.3}
\begin{aligned}
\langle\eta_\delta,e^{-AQ}\Phi_d(0)e^{-BQ}\eta_\delta\rangle
={}&\exp\!\left[\int|h_\delta(p)|^2
(e^{-(A+B)p}-1)dp\right]\\
&\times\int g_d(p)h_\delta(p)
(e^{-Ap}+e^{-Bp})dp.
\end{aligned}
\end{equation}

This is initially computed with a momentum cutoff containing
\([\delta,1]\); its value is independent of enlarging that cutoff. The
inequality \(e^{-v}-1\geq-v\) and \eqref{eq:18.2} bound the exponential from
below by \(e^{-(A+B)}\). Since the support lies below one,

\begin{equation}\label{eq:18.4}
\langle\eta_\delta,e^{-AQ}\Phi_d(0)e^{-BQ}\eta_\delta\rangle
\geq\sqrt{I_\delta}e^{-(A+B)}(e^{-A}+e^{-B}).
\end{equation}
\end{proof}

\begin{proofclaim}[Lower bound after JT heat]\label{clm:threshold-jt}
With a fixed nonnegative compact length test, the positive-kernel
matrix elements have the lower bound \eqref{eq:18.5}. Its positive
constant is independent of $\delta$.
\end{proofclaim}
\begin{proof}
Take a nonnegative normalized $\chi\in C_c^\infty((0,1))$,
positive on a compact interval $J\subset(0,1)$, and put
$\Psi_\delta=\chi\otimes\eta_\delta$. Let $I=(-1,2)$, and let
$T_I=-\tfrac12\partial_x^2$ with Dirichlet boundary conditions on
$I$. Denote the positive kernel of $e^{-tT_I}$ by $p_t^I(x,z)$.
Since $(q+2)e^{-x}\leq e(q+2)$ on $I$, positivity and the product
formula give the scalar kernel comparison
\[
 s_t(q;x,z)\geq e^{-te(q+2)}p_t^I(x,z),\qquad x,z\in I.
\]
Here one first restricts to the Dirichlet semigroup on $I$, which
is dominated by the full-line semigroup, and then bounds each
potential factor below by $e^{-te(q+2)/n}$. All comparisons are
in the cone of nonnegative functions, not in the order of
self-adjoint operators.

The momentum kernels of the cutoff field and the components of
$\eta_\delta$ are nonnegative. Apply this lower bound to both heat
factors, restrict the insertion length to $J$, and use
\eqref{eq:18.4} with $A=B=te$. It follows that
\begin{equation}\label{eq:18.5}
\begin{aligned}
\langle\Psi_\delta,e^{-tL}Ve^{-tL}\Psi_\delta\rangle
&\geq c_{t,d,\chi}\sqrt{I_\delta}\longrightarrow\infty,\\
c_{t,d,\chi}&=2^{d+1}e^{-7te}
 \int_J e^{-dz}|(e^{-tT_I}\chi)(z)|^2\,dz>0.
\end{aligned}
\end{equation}
The constant is positive by strict positivity of the Dirichlet
heat kernel and is independent of $\delta$.
\end{proof}

\begin{proof}[Proof of Proposition~\ref{prop:threshold}]
For fixed $\delta$, truncate the coherent vector in particle
number and approximate $h_\delta$ by nonnegative compact smooth
functions in $(0,\infty)$. The nonnegative integrands in
\eqref{eq:18.3} converge, so normalized finite-particle tests
approximate \eqref{eq:18.5}. Their matrix elements are unbounded.
\end{proof}
The same argument applies with other species in the vacuum.

\subsection{Bounded functions of the low-weight fields}\label{sub:smallweight-outlook}
For $0<d<1/2$, Proposition~\ref{prop:smallweight-heat-operator}
still gives canonical self-adjoint operators
\begin{equation}\label{eq:smallweight-closed-sandwich}
 A^{(d)}_{\varepsilon,R}
 =\overline{e^{-\varepsilon H_R}O_Re^{-\varepsilon H_R}}.
\end{equation}
Their spectral functions define candidate boundary algebras
\begin{equation}\label{eq:smallweight-spectral-algebras}
 \mathcal M_R^{(d)}
 =\{e^{itH_R},e^{isA^{(d)}_{\varepsilon,R}}:
             s,t\in\mathbb R,\ \varepsilon>0\}'',
 \qquad \mathcal M_L^{(d)}=J\mathcal M_R^{(d)}J.
\end{equation}
This uses bounded spectral functions as in \cite[Section 3.2]{PW}.

\section{A conditional-expectation obstruction to injectivity}\label{sec:tensor}

The following criterion is the algebraic part of the
noninjectivity proof. Sections~\ref{sec:energy}--\ref{sec:nonhyper}
construct the required operators; the quantitative version will
also be used in the scaling limit.

Injectivity of $N\subset B(\mathcal K)$ gives a conditional
expectation $P:B(\mathcal K)\to N$, which is unital, completely
positive and $N$-bimodular \cite{Tom,C}.

\begin{lemma}\label{lem:rowcriterion}
Let $N\subset B(\mathcal K)$ be a concrete von Neumann algebra
acting unitally,
with a faithful normal tracial state $\sigma$. Suppose
$0\ne b=b^*\in N$, $u\in N$ is unitary, and
$c\in B(\mathcal K)$ satisfies
\begin{equation}\label{eq:abstractrow}
 b=c+c^*,\qquad
 \sum_{j=1}^n u^jcc^*u^{-j}\le R^2\,1
       \quad\hbox{for every }n\ge1,
\end{equation}
where $R<\infty$. Then $N$ is noninjective and nonhyperfinite.

\end{lemma}
\begin{proof}
Suppose that $N$ is injective and choose a conditional
expectation $P:B(\mathcal K)\to N$. Put $a=P(c)$ and
$c_j=u^jcu^{-j}$. Bimodularity gives
$P(c_j)=u^jau^{-j}$. Schwarz's inequality and positivity give
\begin{align}\label{eq:expectationrow}
 \sum_{j=1}^n u^jaa^*u^{-j}
 &=\sum_{j=1}^n P(c_j)P(c_j)^*\notag\\
 &\le P\!\left(\sum_{j=1}^n c_jc_j^*\right)
 \le R^2\,1.
\end{align}
Taking the normalized trace yields
$n\sigma(aa^*)\le R^2$ for every $n$. Hence
$\sigma(aa^*)=0$, and faithfulness gives $a=0$. But $P$ fixes
$b\in N$, so
\(b=P(b)=P(c)+P(c)^*=a+a^*=0,\)
a contradiction. Thus $N$ is noninjective. Every hyperfinite
von Neumann algebra is injective, so $N$ is nonhyperfinite.
\end{proof}

Necessarily $c\notin N$: otherwise the trace applied directly to
\eqref{eq:abstractrow} would give $c=0$.

\subsection{A quantitative tensor version}\label{sub:tensor-version}

\begin{lemma}\label{lem:tensortrace}
Let $(N,\sigma)$ be a finite hyperfinite von Neumann algebra with
separable predual and faithful normal tracial state. If
$x_1,\ldots,x_n\in N$ are self-adjoint, then
\begin{equation}\label{eq:hypernecessary}
 \sum_{j=1}^n\sigma(x_j^2)
 \le\left\|\sum_{j=1}^n x_j\otimes x_j^{\op}
        \right\|_{N\otimes_{\min}N^{\op}}.
\end{equation}
\end{lemma}
\begin{proof}
Choose increasing finite dimensional unital subalgebras
$D_m\subset N$ with strongly dense union, and let $E_m$ be the
trace preserving conditional expectations. They are the
orthogonal projections in $L^2(N,\sigma)$, so $E_m(x)\to x$ in
$L^2$. On $D_m\otimes D_m^{\op}$ the functional
$a\otimes b^{\op}\mapsto\sigma(ab)$ is the unit-vector state
for left and right multiplication on $L^2(D_m,\sigma)$.
Finite dimensionality makes it a state for the minimal tensor
norm. Composing with the unital completely positive map
$E_m\otimes E_m^{\op}$ gives a state $\omega_m$ on
$N\otimes_{\min}N^{\op}$. Its value at
$x\otimes y^{\op}$ is $\sigma(E_m(x)E_m(y))$. Apply its norm
bound to the sum in \eqref{eq:hypernecessary} and pass to the
$L^2$ limit.
\end{proof}

\begin{proposition}\label{prop:tensorcriterion}
Under the hypotheses of Lemma~\ref{lem:rowcriterion}, put
$b_j=u^jbu^{-j}$ and $v=\sigma(b^2)>0$. Then
\[
 \sum_{j=1}^n\sigma(b_j^2)=nv,\qquad
 \left\|\sum_{j=1}^n b_j\otimes b_j^{\op}\right\|_{\min}
       \le2R^2(\sqrt n+1).
\]
For separable predual this contradicts
\eqref{eq:hypernecessary} once $nv>2R^2(\sqrt n+1)$.
\end{proposition}
\begin{proof}
Put $c_j=u^jcu^{-j}$ and $b_j=c_j+c_j^*$. The row bound gives
\[
 \Big\|\sum c_jc_j^*\Big\|\le R^2,
 \qquad \Big\|\sum c_j^*c_j\Big\|\le nR^2.
\]
For finite families of bounded operators $a_j$ and $d_j$ on the
two tensor factors, the row-column inequality is
\begin{equation}\label{eq:rowcolumn}
 \Big\|\sum a_j\otimes d_j\Big\|
 \le\Big\|\sum a_ja_j^*\Big\|^{1/2}
       \Big\|\sum d_j^*d_j\Big\|^{1/2}.
\end{equation}
Represent $N^{\op}$ on the conjugate Hilbert space
$\overline{\mathcal K}$ by $x^{\op}\mapsto\overline{x^*}$, where
$\overline y$ is the operator induced by $y$ on the conjugate Hilbert
space.
For self-adjoint $b_j$, its image is $\overline{b_j}$.
Expanding the tensor sum, \eqref{eq:rowcolumn} and adjunction give
\begin{align*}
 \Big\|\sum c_j\otimes\overline{c_j}\Big\|&\le\sqrt nR^2,&
 \Big\|\sum c_j^*\otimes\overline{c_j^*}\Big\|&\le\sqrt nR^2,\\
 \Big\|\sum c_j\otimes\overline{c_j^*}\Big\|&\le R^2,&
 \Big\|\sum c_j^*\otimes\overline{c_j}\Big\|&\le R^2.
\end{align*}
Both representations are faithful, so their spatial tensor
product computes the minimal norm. Thus
\begin{equation}\label{eq:tensorupper}
 \Big\|\sum_{j=1}^n b_j\otimes b_j^{\op}\Big\|_{\min}
       \le2R^2(\sqrt n+1).
\end{equation}
Traciality and faithfulness give
$\sigma(b_j^2)=\sigma(b^2)=v>0$. If $N$ were hyperfinite,
Lemma~\ref{lem:tensortrace} would imply
$nv\le2R^2(\sqrt n+1)$ for every $n$, which is impossible.
\end{proof}

\section{Uniform bounds in the energy representation}\label{sec:energy}

We need two kernel estimates: a pointwise bound at small momentum
and an $L^2$ bound at large momentum.

\subsection{The energy gauge and the canonical input}\label{sub:energy-gauge}

We apply the preceding criterion to the right boundary algebra in
the gauge already used for the heat estimates. Throughout
Sections~\ref{sec:energy}--\ref{sec:loss}, put
\begin{equation}\label{eq:energy-gauge}
 M=U^*M_RU,\qquad L=2\phi_bU^*H_RU
      =-\tfrac12\partial_x^2+(Q+2)e^{-x},\qquad
 \mathscr J=U^*JU.
\end{equation}
Here $U$ is the decomposable unitary of
Section~\ref{sub:hamiltonians}, and $J$ is the antiunitary
defined on all Fock sectors by
\eqref{eq:reflection-fock}--\eqref{eq:reflection-occupation}.
The transported trace is again denoted by $\tau$:
$\tau(x)=\tau_{M_R}(UxU^*)$ for $x\in M_+$. Theorems
\ref{thm:trace} and~\ref{thm:factor} give
$M'=\mathscr J M\mathscr J$ and the type
$\mathrm{II}_\infty$ property.

For a chosen elementary species of weight $d>1/2$, write
\begin{equation}\label{eq:pointform}
 g_d(p)=\frac{p^{d-1/2}}{\sqrt{\Gamma(2d)}},\qquad
 V_d(x)=2^de^{-dx}\bigl(a^*(g_d)+a(g_d)\bigr),\qquad
 A_{d,t}=e^{-tL}V_de^{-tL}\in M.
\end{equation}
The last expression is the canonical bounded self-adjoint
extension proved in Section~\ref{sec:bounds}, for every $t>0$.
The algebra also contains every bounded Borel function of $L$.
The common test space $\D$ is the finite-species, finite-particle
space of jointly smooth compactly supported length and momentum
functions used in Section~\ref{sub:hamiltonians}; momentum
supports stay away from zero. Our Fock convention is
\begin{equation}\label{eq:annihilation}
 (a(h)\psi)_n(\mathbf p)=\sqrt{n+1}
   \int_0^\infty\overline{h(u)}\psi_{n+1}(u,\mathbf p)\,du.
\end{equation}
Thus creation is linear and annihilation is antilinear in its
one-particle argument. The operator $Q$ always includes the
positive momenta of all spectator species.

The trace formula of Section~\ref{sub:thermal} becomes
\begin{equation}\label{eq:traceinput}
 \tau(H(L))=\int_0^\infty H(s^2/2)\,d\mu(s)
 \qquad(H\ge0\text{ Borel}).
\end{equation}
In the canonical $L^2(M,\tau)$ representation, the vector of
$\ind_J(L)$ is $\ind_J(s^2/2)\Omega$ for every bounded Borel
energy set $J$. This follows from the vacuum map and the spectral
trace formula in Sections~\ref{sub:vacuum} and~\ref{sub:thermal}.
Finally, Section~\ref{sub:distributions} proves that for $d=1$
and every $r>1$ the point form has its compatible extension
\begin{equation}\label{eq:graphinput}
 |q_{V_1}(u,v)|\le C_r\|(1+L)^r u\|\|(1+L)^r v\|.
\end{equation}

\begin{remark}\label{rem:projection}
For $x=e^{-tL}$, the space $xMx$ is not strongly closed:
$e_n=\ind_{[0,n]}(L)=x(e^{2tL}e_n)x\to1$ strongly, whereas
$xBx=1$ would imply $1\le\|B\|e^{-2tn}$ on
$\ind_{[n,n+1]}(L)\mathcal H$. Finite corners use projections.
\end{remark}

\subsection{Scalar and transform estimates}\label{sub:kernel-estimates}

We use the one-sided Kontorovich--Lebedev transform $\mathcal U$
of Section~\ref{sub:kl}, retaining the parabolic momentum
variables. It identifies $\mathcal H$ with
$L^2((0,\infty),d\mu(s);\mathcal F)$ and $L$ with multiplication
by $s^2/2$. In this representation the real insertion kernel is
$F_d(s,r;q',q)$ from \eqref{eq:4.3}, with the adjoint symmetry
$F_d(s,r;q',q)=F_d(r,s;q,q')$. For the Gamma factor
\eqref{eq:Gd}, set
\begin{equation}\label{eq:Gzero}
 G_0=\frac{\Gamma(d)^4}{\Gamma(2d)},\qquad
                  0<G_d(r,s)\le G_0.
\end{equation}
The upper bound follows from Euler's Gamma integral.

\begin{lemma}\label{lem:pointkernel}
For $q,p\ge0$ and real $r,s$,
\begin{equation}\label{eq:pointbound}
 |F_d(s,r;q+p,q)|\le G_0(q+2)^{-d}.
\end{equation}
\end{lemma}
\begin{proof}
The representation
$K_{i\alpha}(z)=\int_0^\infty e^{-z\cosh t}\cos(\alpha t)\dd t$
gives $|K_{i\alpha}(z)|\le K_0(z)$, and $K_0$ is decreasing.
Thus the absolute integral is at most $F_d(0,0;q,q)$.
Changing variables in \eqref{eq:4.3} and using
\begin{equation}\label{eq:besselproduct}
 \int_0^\infty z^{2d-1}K_0(z)^2\dd z
       =2^{2d-3}\frac{\Gamma(d)^4}{\Gamma(2d)}
\end{equation}
gives $F_d(0,0;q,q)=G_0(q+2)^{-d}$. For completeness,
\eqref{eq:besselproduct} follows by inserting the two Schwinger
integrals for $K_0$, integrating first in $z$, and changing the
two remaining variables to their sum and ratio. Tonelli applies
because the integrands are nonnegative.
\end{proof}

For the $L^2$ bound, the Liouville transform replaces the outgoing
energy variable by length. A positive Bessel majorant then becomes
an integrable convolution kernel in logarithmic momentum.

\begin{corollary}\label{cor:kernelmap}
For every $q\ge0$ and fixed $s>0$, the operator defined initially
for $h\in C_c^\infty(0,\infty)$ by
\begin{equation}\label{eq:kernelmap}
 (A_{q,s}h)(r)=\int_0^\infty
   g_d(p)F_d(s,r;q+p,q)h(p)\dd p
\end{equation}
extends with norm at most $\sqrt{G_0}$ from $L^2(dp)$ to $L^2(d\mu(r))$,
uniformly in $q$ and $s$.
\end{corollary}
\begin{proof}
Write $\mathcal U_q$ for the unitary fixed-$q$ Liouville transform
of Section~\ref{sub:kl}, from $L^2(dx)$ to $L^2(d\mu(r))$.
For a test function $h$, its length-space input is
\[
 (B_{q,s}h)(x)=2^de^{-dx}\int_0^\infty
             g_d(p)\zeta_s(q+p,x)h(p)\dd p.
\]
The integral definition \eqref{eq:4.3} gives
$A_{q,s}h=\mathcal U_qB_{q,s}h$. Fubini is justified by the
positive Bessel-product bound in Lemma~\ref{lem:pointkernel};
the estimate below also proves that $B_{q,s}h\in L^2(dx)$.

Put $h_{\log}(y)=e^{y/2}h(e^y)$, so
$\|h_{\log}\|_{L^2(dy)}=\|h\|_{L^2(dp)}$.
As in Lemma~\ref{lem:pointkernel}, $|K_{2is}(z)|\le K_0(z)$
and $K_0$ is decreasing. Since $q+p+2\ge p$, this gives
\[
\begin{aligned}
 |B_{q,s}h(x)|
 &\le\frac{2^{d+1}e^{-dx}}{\sqrt{\Gamma(2d)}}
       \int_0^\infty p^{d-1/2}
       K_0(2\sqrt{2p}\,e^{-x/2})|h(p)|\dd p\\
 &=(k_d*|h_{\log}|)(x),
\end{aligned}
\]
where $*$ is convolution on $\mathbb R$ and
\[
 k_d(v)=\frac{2^{d+1}}{\sqrt{\Gamma(2d)}}e^{-dv}
              K_0(2\sqrt2\,e^{-v/2}).
\]
The required integral follows directly from the positive
representation $K_0(z)=\tfrac12\int_0^\infty
t^{-1}e^{-t-z^2/(4t)}\dd t$ and Tonelli:
\[
\begin{aligned}
 \int_0^\infty z^{2d-1}K_0(z)\dd z
 &=\frac12\int_0^\infty\frac{e^{-t}}t
       \int_0^\infty z^{2d-1}e^{-z^2/(4t)}\dd z\dd t\\
 &=2^{2d-2}\Gamma(d)^2.
\end{aligned}
\]
Indeed the inner integral is $2^{2d-1}t^d\Gamma(d)$.
Substituting $z=2\sqrt2\,e^{-v/2}$ therefore yields
\[
 \|k_d\|_1
 =\frac{2^{d+2}(2\sqrt2)^{-2d}}{\sqrt{\Gamma(2d)}}
      2^{2d-2}\Gamma(d)^2
 =\frac{\Gamma(d)^2}{\sqrt{\Gamma(2d)}}=\sqrt{G_0}.
\]
Young's convolution inequality and Liouville unitarity now give
$\|A_{q,s}h\|_2=\|B_{q,s}h\|_2
\le\sqrt{G_0}\|h\|_2$. Density gives the extension.
Restrictions of the input momentum or output energy are
orthogonal projections and preserve the same bound.
\end{proof}

The convolution proof supplies the norm bound. We also record the
exact Jacobi identification, which will be used for the norm
identity in Section~\ref{sub:exact-norm}. The normalized kernel
$\chi^d_{r,s}$ is \eqref{eq:12.1} with $\lambda=d$, and
Proposition~\ref{prop:jacobi} gives the unitary $\mathcal J_{d,s}$.
Here $s$ is fixed and $r$ is the transform variable.
Put $a_q=(q+2)/2$. Translation of the length variable gives $F_d(s,r;q+p,q)=a_q^{-d}F_d(s,r;p/a_q,0)$. Since $g_d(p)=a_q^{d-1/2}g_d(p/a_q)$,
\begin{equation}\label{eq:dilatedkernel}
 g_d(p)F_d(s,r;q+p,q)
   =a_q^{-1/2}\sqrt{G_d(r,s)}\chi^d_{r,s}(p/a_q).
\end{equation}
The dilation $(\delta_{a_q}h)(w)=a_q^{1/2}h(a_qw)$ is unitary.
Hence $A_{q,s}=M_{\sqrt{G_d(\cdot,s)}}\mathcal J_{d,s}\delta_{a_q}$.

\section{A bound on the full Fock space}\label{sec:fockbound}

An energy cutoff leaves every particle sector present. We must
therefore control the factor $\sqrt{n+1}$ in
\eqref{eq:annihilation} uniformly in $n$.

Split the removed momentum at the sum of the spectator momenta
plus $2$. Below this threshold the momentum fractions sum to at
most one; above it there is at most one particle, controlled by
Corollary~\ref{cor:kernelmap}.

\subsection{The column operator and the sector norms}\label{sub:fock-columns}

Fix $S>0$ and write
\begin{equation}\label{eq:corner}
 h=S^2/2,\quad e=\ind_{[0,h]}(L),\quad
 m_S=\int_0^S\dd\mu(s),\quad
 \mathcal K=e\Hh=L^2((0,S),d\mu(r);\F).
\end{equation}
We first write the proof for one elementary species of weight $d>1/2$.
For $n\ge1$, let
\[
 \F_n=L^2_{\mathrm{sym}}((0,\infty)^n,d\mathbf p),\qquad
 \F_0=\C\Omega,\qquad
 \mathcal K_n=L^2((0,S),d\mu(r);\F_n).
\]
Thus $\F=\bigoplus_{n\ge0}\F_n$ and
$\mathcal K=\bigoplus_{n\ge0}\mathcal K_n$. We use the usual
Lebesgue norm on each symmetric sector; the normalization of bosonic
annihilation is the factor $\sqrt{n+1}$, not an additional factorial
in this norm.

For a fixed incoming energy parameter $s\in(0,S)$, define on finite
sums of smooth compactly supported momentum tests
\begin{equation}\label{eq:Ts}
 (T_s\psi)_n(r;\mathbf p)=\sqrt{n+1}
   \int_0^\infty g_d(u)F_d(s,r;q+u,q)
                 \psi_{n+1}(u,\mathbf p)\dd u,
 \qquad q=\sum_{j=1}^np_j.
\end{equation}
Here $T_s\Omega=0$, the test momenta stay away from zero, and $u$
is the removed momentum. Hence the incoming and outgoing total
momenta are $q+u$ and $q$. Lemma~\ref{lem:decomposition} derives
this kernel from the field.

The column $T_s:\mathcal F_{n+1}\to\mathcal K_n$ fixes the
incoming energy and integrates the outgoing energy.
We integrate these columns in \eqref{eq:cBochner}.

\subsection{The estimate independent of particle number}\label{sub:fock-estimate}

\begin{lemma}\label{lem:fullfock}
Each $T_s$ extends to a bounded operator from $\F$ to
$\mathcal K$, and
\begin{equation}\label{eq:Bconstant}
 \sup_{0<s<S}\|T_s\|\le B_{d,S}:=\sqrt{\alpha_dm_S}+\sqrt{G_0},
 \qquad
 \alpha_d=\frac{G_0^2}{\Gamma(2d)}
              \left(\frac1{2d-1}+\frac1{2d}\right).
\end{equation}
\end{lemma}

\begin{proof}
Put $R=q+2$ and split \eqref{eq:Ts} at $u=R$.
For the lower part, Lemma~\ref{lem:pointkernel} gives
\begin{equation}\label{eq:softintegral}
 \int_0^R\frac{R+u}{u}
       |g_d(u)F_d(s,r;q+u,q)|^2\,du
 \leq\frac{G_0^2R^{-2d}}{\Gamma(2d)}
       \int_0^R(R+u)u^{2d-2}\,du=\alpha_d.
\end{equation}
Weighted Cauchy--Schwarz, followed by integration in $r$ and the
spectator momenta, therefore bounds the squared sector norm by
\[
 (n+1)\alpha_dm_S
 \int\frac{u}{q+u+2}\mathbf1_{\{u<q+2\}}
                  |\psi_{n+1}(u,\mathbf p)|^2\,du\,d\mathbf p.
\]
Symmetry replaces $(n+1)$ times this selected-particle integral by
the sum over all possible removed particles. Its multiplier is
\begin{equation}\label{eq:softbose}
 \sum_{j=1}^{n+1}
 \frac{p_j}{\sum_\ell p_\ell+2}
 \mathbf1_{\{p_j<\sum_{\ell\ne j}p_\ell+2\}}\leq1.
\end{equation}
Thus $\|T_s^<\psi_{n+1}\|^2\leq\alpha_dm_S\|\psi_{n+1}\|^2$.

For $u\geq R$, apply Corollary~\ref{cor:kernelmap} with the input
restricted to $[R,\infty)$ and the output to $(0,S)$:
\[
 \int_0^S\left|\int_R^\infty
       g_d(u)F_d(s,r;q+u,q)f(u)\,du\right|^2d\mu(r)
       \leq G_0\int_R^\infty|f(u)|^2\,du.
\]
After integrating spectators and using symmetry, the Bose factor
becomes
\begin{equation}\label{eq:hardbose}
 \sum_{j=1}^{n+1}
 \mathbf1_{\{p_j\geq\sum_{\ell\ne j}p_\ell+2\}}\leq1.
\end{equation}
Indeed two such inequalities would give
$0\geq2\sum_{\ell\ne i,j}p_\ell+4$.
Hence $\|T_s^\geq\psi_{n+1}\|^2\leq G_0\|\psi_{n+1}\|^2$.
Distinct input sectors have orthogonal outputs for each part.
Summing sectors and then using the triangle inequality gives $\|T_s\psi\|\leq(\sqrt{\alpha_dm_S}+\sqrt{G_0})\|\psi\|$. Density gives the asserted bounded extension. All constants are
independent of $s\in(0,S)$ and of particle number.
\end{proof}

\subsection{The meaning and use of the bound}\label{sub:fock-use}
For $d=1$,
\begin{equation}\label{eq:d1constant}
 G_0=1,\qquad \alpha_1=\tfrac32,\qquad B_{1,S}=1+\sqrt{3m_S/2}.
\end{equation}
The estimate controls one incoming-energy column on the full Fock
space. Integrating the columns gives a bounded annihilation part;
Parseval will control all its translates together. The constant
depends on the energy window. Spectator species are included in
Section~\ref{sub:spectator-species}.

\section{Nonhyperfiniteness of the finite corners}\label{sec:nonhyper}

We now apply Lemma~\ref{lem:rowcriterion}. The column bound gives
a bounded compressed field, and Fourier analysis supplies the
unitary translates.

\subsection{The compressed field and its creation part}\label{sub:compressed-field}

For $\Psi\in\mathcal K$, define
\begin{equation}\label{eq:cBochner}
 c^*\Psi=\int_0^S T_s\Psi(s)\dd\mu(s),\qquad c=(c^*)^*.
\end{equation}
This is a Bochner integral. To verify measurability, first use
compact momentum tests: the kernel is continuous in $s>0$ and
\eqref{eq:pointbound} gives an integrable majorant on their
supports. A countable dense family and the uniform bound
\eqref{eq:Bconstant} extend strong measurability to all vectors.
Simple-section approximation then treats $\Psi(s)$. Moreover,
\begin{equation}\label{eq:cnorm}
 \int_0^S\|T_s\Psi(s)\|\dd\mu(s)
      \le B_{d,S}\sqrt{m_S}\|\Psi\|,
 \qquad \|c\|\le B_{d,S}\sqrt{m_S}.
\end{equation}

Put $C=eMe$ and $\tau_C=\tau/m_S$. For $t>0$,
$D_t=e^{tL}e\in M$ is bounded by $e^{th}$ and supported on $e$.
Thus the canonical heat field defines
\begin{equation}\label{eq:bmember}
 b=D_tA_{d,t}D_t\in C.
\end{equation}

\begin{lemma}\label{lem:decomposition}
On $\mathcal K$, $b=c+c^*$, independently of $t$. Furthermore
$b\ne0$, and
\begin{equation}\label{eq:variance}
 v:=\tau_C(b^2)
      =\frac1{m_S}\int_0^S\int_0^S G_d(r,s)\dd\mu(r)\dd\mu(s)>0.
\end{equation}
\end{lemma}

\begin{proof}
Take input and output wave packets of compact energy and interior
momentum support in adjacent particle sectors. In their matrix
elements the absolute length integrand is bounded by a constant times
\begin{equation}\label{eq:majorant}
 e^{-dx}K_0(4e^{-x/2})^2.
\end{equation}
This is integrable at both ends, uniformly in the momentum cutoff.
The inverse Liouville transform and Fubini therefore identify the
annihilation block of the cutoff heat field with
\[
 \sqrt{n+1}\iint e^{-t(r^2+s^2)/2}
    g_{d,N}(u)F_d(s,r;q+u,q)
    \Psi_{n+1}(s;u,\mathbf p)\,du\,d\mu(s).
\]
The arguments reflect the incoming momentum $q+u$ and outgoing
momentum $q$. Dominated convergence removes the cutoff in these
pairings; the strong-$*$ convergence of Section~\ref{sub:schur}
identifies the result with the block of the canonical $A_{d,t}$.
Self-adjointness gives its creation block.

On $0<r,s<S$, the two bounded factors $e^{tL}e$ cancel
$e^{-t(r^2+s^2)/2}$. Thus the annihilation part of $b$ is
\eqref{eq:cBochner}, and the creation part is its adjoint.
Both are bounded by \eqref{eq:cnorm}; equality on the dense wave
packets gives $b=c+c^*$ everywhere, independently of $t$.

For a vacuum packet $\varphi\otimes\Omega$, only creation acts.
Equation~\eqref{eq:12.2} identifies its output at energy $s$ as $\mathcal J_{d,s}^* \bigl(\mathbf1_{(0,S)}\sqrt{G_d(\cdot,s)}\,\varphi\bigr)$. Unitarity gives
\begin{equation}\label{eq:vaccolumn}
 \|b(\varphi\otimes\Omega)\|^2
 =\int_0^S\int_0^S
     G_d(r,s)|\varphi(r)|^2\,d\mu(r)\,d\mu(s).
\end{equation}
This is positive for $\varphi\ne0$. The $L^2(M,\tau)$ vector of
$e$ is $\mathbf1_{(0,S)}\Omega$; substituting this packet gives
$\tau(b^2)$, and division by $\tau(e)=m_S$ proves \eqref{eq:variance}.
\end{proof}

Only $b$ is asserted to belong to $M$. Its separated creation
piece $c$ is an auxiliary bounded operator on $e\Hh$.

\subsection{Time translates and the Fourier estimate}\label{sub:fourier-obstruction}

Choose the time translations as Fourier phases on the energy
interval. Parseval will bound all translates at once.

\begin{theorem}\label{thm:nonhyper}
For the canonical free bosonic model with an elementary species
of weight $d>1/2$, the finite von Neumann algebra $eMe$ is
noninjective and nonhyperfinite for every $S>0$.
Consequently the canonical type $\mathrm{II}_\infty$ boundary
factor, and its opposite boundary factor, are nonhyperfinite and
noninjective. Every nonzero finite corner of either factor is a
nonhyperfinite type $\mathrm{II}_1$ factor.
\end{theorem}

\begin{lemma}[Hilbert-space-valued Parseval]\label{lem:vector-parseval}
For a separable Hilbert space $\mathcal V$ and
$F\in L^2((0,h),dE;\mathcal V)$, put
$\widehat F_h(j)=h^{-1/2}\int_0^h e^{-2\pi ijE/h}F(E)\,dE$.
Then
\[
 \sum_{j\in\mathbb Z}\|\widehat F_h(j)\|^2
 =\int_0^h\|F(E)\|^2\,dE.
\]
Thus the squared norms of the unnormalized coefficients sum to
$h\int_0^h\|F(E)\|^2\,dE$.
\end{lemma}
\begin{proof}
The integrals exist by Cauchy--Schwarz. Expand $F$ in an orthonormal
basis of $\mathcal V$, apply scalar Parseval to each coordinate,
and sum by Tonelli's theorem.
\end{proof}

Set $E=s^2/2$ and $h=S^2/2$.
On $(0,h)$,
\begin{equation}\label{eq:nudensity}
 d\mu(s)=\nu(E)\dd E,\qquad
 \nu(E)=\frac{\sinh(2\pi\sqrt{2E})}{\pi^2},\qquad
 \nu_{\max}=\frac{\sinh(2\pi S)}{\pi^2}.
\end{equation}
For $\Psi\in\mathcal K$, set $\psi(E)=\Psi(\sqrt{2E})$ and
$\mathcal T_E=T_{\sqrt{2E}}:\mathcal F\to\mathcal K$. Then
\[
 \|\Psi\|^2=\int_0^h\nu(E)\|\psi(E)\|^2\,dE,\qquad
 c^*\Psi=\int_0^hF_\Psi(E)\,dE,\quad
 F_\Psi(E)=\nu(E)\mathcal T_E\psi(E).
\]
The function $F_\Psi$ is strongly measurable by the argument
following \eqref{eq:cBochner}. Its values lie in the whole output
space $\mathcal K$, and the spectral density remains inside it.

\begin{lemma}[Fourier bound for the translated columns]\label{lem:fourier-row}
Let $u=e^{2\pi iL/h}e\in C$ and $c_j=u^jcu^{-j}$ for
$j\in\mathbb Z$. Then the identity and bound
\eqref{eq:parseval} hold. In particular, for every $n\ge1$,
\[
 \sum_{j=1}^n c_jc_j^*\le h\nu_{\max}B_{d,S}^2e.
\]
Here $e$ is the identity on $\mathcal K$, and $u$ is unitary
on that space.
\end{lemma}
\begin{proof}
Fix $\Psi\in\mathcal K$. The column bound
$\|\mathcal T_E\|\le B_{d,S}$ and
$0\le\nu(E)\le\nu_{\max}$ give
\begin{align}\label{eq:integrandL2}
 \int_0^h\|F_\Psi(E)\|_{\mathcal K}^2\,dE
 &=\int_0^h\nu(E)^2\|\mathcal T_E\psi(E)\|_{\mathcal K}^2\,dE\notag\\
 &\le B_{d,S}^2\int_0^h\nu(E)^2\|\psi(E)\|_{\mathcal F}^2\,dE\notag\\
 &\le\nu_{\max}B_{d,S}^2\|\Psi\|_{\mathcal K}^2.
\end{align}
Thus $F_\Psi\in L^2((0,h),dE;\mathcal K)$, as required
by Lemma~\ref{lem:vector-parseval}. Since $L$ acts as multiplication
by $E$ in the incoming energy variable, $(u^{-j}\Psi)(\sqrt{2E})=e^{-2\pi ijE/h}\psi(E)$. Linearity of $\mathcal T_E$ and
$c_j^*=u^jc^*u^{-j}$ therefore yield
\begin{equation}\label{eq:fouriercolumns}
 c_j^*\Psi
   =u^j\int_0^h e^{-2\pi ijE/h}F_\Psi(E)\,dE.
\end{equation}
In the notation of Lemma~\ref{lem:vector-parseval}, this says
$c^*u^{-j}\Psi=\sqrt h\,\widehat{F_\Psi}_h(j)$.
Taking norms removes the output unitary $u^j$; applying that
lemma with $\mathcal V=\mathcal K$ now gives every equality in
\begin{align}\label{eq:parseval}
 \sum_{j\in\mathbb Z}\|c_j^*\Psi\|^2
 &=\sum_{j\in\mathbb Z}\|c^*u^{-j}\Psi\|^2\notag\\
 &=h\sum_{j\in\mathbb Z}\|\widehat{F_\Psi}_h(j)\|_{\mathcal K}^2\notag\\
 &=h\int_0^h\|F_\Psi(E)\|_{\mathcal K}^2\dd E
 \le R_{d,S}^2\|\Psi\|^2,
\end{align}
where
\begin{equation}\label{eq:Rfinite}
 R_{d,S}^2=h\nu_{\max}B_{d,S}^2=h\nu_{\max}
                  \bigl(\sqrt{G_0}+\sqrt{\alpha_dm_S}\bigr)^2.
\end{equation}
Restricting the sum to $1\leq j\leq n$ proves the operator
inequality.
\end{proof}

\begin{proof}[Proof of Theorem~\ref{thm:nonhyper}]
The representation of $eMe$ on $e\Hh$ is faithful.
Lemma~\ref{lem:decomposition} gives $0\ne b=c+c^*\in eMe$,
and Lemma~\ref{lem:fourier-row} gives the uniform row bound.
The conditional-expectation argument of
Lemma~\ref{lem:rowcriterion} therefore proves the first assertion.

A corner of an injective algebra is injective. Hence $M$ is
noninjective and nonhyperfinite. Reflection
transports this conclusion to the opposite boundary. Finally,
in a semifinite factor any nonzero finite projection is full;
amplifying its corner by $B(\ell^2)$ recovers the type
$\mathrm{II}_\infty$ factor. Injectivity is preserved by this
amplification. Thus no nonzero finite corner can be injective or
hyperfinite. Its type is $\mathrm{II}_1$.
\end{proof}

\subsection{Several species}\label{sub:spectator-species}

Choose a species $a$ and regard all others as spectators in
$\mathcal F=\Gamma_s(\mathfrak h_a)\otimes
\Gamma_s(\bigoplus_{b\ne a}\mathfrak h_b)$.
Let $q_0\ge0$ be the total momentum of the other species. In
\eqref{eq:Ts} replace $q$ by $q_0+\sum_{j=1}^np_j$ and let
$n+1$ count only particles of species $a$. The Jacobi bound is
uniform in this enlarged $q$. The sum in
\eqref{eq:softbose} is at most
$\sum_jp_j/(q_0+\sum_jp_j+2)\le1$, and at most one momentum
can satisfy $p_j\ge q_0+\sum_{\ell\ne j}p_\ell+2$.
Thus the full Fock bound holds with the same constant
$B_{d_a,S}$. Integrating over spectators and summing the
orthogonal sectors proves it also for countably many species.
The total vacuum still gives \eqref{eq:vaccolumn}. Therefore
Theorem~\ref{thm:nonhyper} holds for all nonempty finite or
countable free families of weights $d_a>1/2$, under the stated
canonical input. 

\subsection{Sixteen translates at sufficiently small energy}\label{sub:sixteen}

There is also a finite witness with a fixed number of terms at
small energy. As $S\downarrow0$,
\[
 m_S=\frac{2S^3}{3\pi}+O(S^5),\qquad
 h\nu_{\max}=\frac{S^3}{\pi}+O(S^5),\qquad
 \frac{v}{m_SG_0}\longrightarrow1.
\]
Consequently $R_{d,S}^2/v\to3/2$. For $n=16$, the necessary
hyperfinite inequality would be
$16v\le10R_{d,S}^2$, whereas $10R_{d,S}^2/v\to15<16$.
It therefore fails for all sufficiently small $S>0$. 

\subsection{Physical meaning of nonhyperfiniteness}
\label{sub:nonhyper-meaning}

The obstruction also bounds simultaneous approximation by
finite-dimensional subalgebras. Assume $d_a>1/2$ throughout.

Keep $C=eMe$, $\tau_C$, $b$, $u=e^{2\pi iL/h}e$ and $h=S^2/2$
from Section~\ref{sub:fourier-obstruction}. Set
$b_j=u^jbu^{-j}$ and $v=\tau_C(b^2)>0$.
Proposition~\ref{prop:tensorcriterion} gives
\begin{equation}\label{eq:physical-tensor-bound}
 \left\|\sum_{j=1}^n b_j\otimes b_j^{\op}\right\|_{\min}
 \le2R_{d,S}^2(\sqrt n+1),\qquad n\ge1.
\end{equation}
For a unital finite-dimensional subalgebra $D\subset C$, let
$E_D:C\to D$ be the trace-preserving expectation, equivalently
the orthogonal projection in $L^2(C,\tau_C)$.

\begin{proposition}\label{prop:finite-algebra-error}
For every such $D$ and every $n\ge1$,
\begin{equation}\label{eq:finite-algebra-error}
 \frac1n\sum_{j=1}^n\|b_j-E_D(b_j)\|_{2,C}^2
 \ge v-2R_{d,S}^2(n^{-1/2}+n^{-1}),
 \qquad \|x\|_{2,C}^2=\tau_C(x^*x).
\end{equation}
\end{proposition}
\begin{proof}
The tracial pairing is a state on $D\otimes_{\min}D^{\op}$,
and $E_D\otimes E_D^{\op}$ is a unital completely positive
contraction. Hence \eqref{eq:physical-tensor-bound} gives
$\sum_j\tau_C(E_D(b_j)^2)\le2R_{d,S}^2(\sqrt n+1)$.
Subtract this from $\sum_j\tau_C(b_j^2)=nv$ and use
orthogonality of $E_D$.
\end{proof}

For $n=16$, the lower bound divided by $v$ is
$1-5R_{d,S}^2/(8v)\to1/16$ as $S\downarrow0$.
Thus sixteen translates have a positive simultaneous
approximation error in every sufficiently small window.

\section{The high energy scaling}\label{sec:scale}

From now on there is one massless species, $d=1$.
We rescale around high energy to obtain a first-order limiting
Hamiltonian and specify the observables and states to be compared.

In Sections~\ref{sec:scale}--\ref{sec:loss} there is one species
and $d=1$. Thus
$g(p)=\sqrt p$ and $V=2e^{-x}\Phi(g)$, where
$\Phi(h)=a^*(h)+a(h)$. Let $S_k$ and $K_k$ be as in
\eqref{eq:introscale}, for the scalar parameter $k\ge1$, and set
$M_k=S_k^*MS_k$. The length derivative in this scaling is denoted
by $D=-i\partial_y$.
The self-adjoint realization of $K_k$ is defined by this unitary
conjugation; its spectrum is $[-k/2,\infty)$. On $\D$,
\begin{equation}\label{eq:rescaledfield}
 k^{-1}S_k^*VS_k=2e^{-y}\Phi(g)=:V^{(0)}.
\end{equation}
The remaining first-order potential is removed by a pointwise
unitary phase. Conjugating the translation generator by that phase
specifies a self-adjoint limiting operator, including its domain:
\begin{equation}\label{eq:Kdomain}
 K_\infty=GDG^*,\qquad G(y)=e^{i(Q+2)e^{-y}},\qquad
 D(K_\infty)=G H^1(\R;\F).
\end{equation}
Here $H^1(\mathbb R;\mathcal F)$ is the vector-valued Sobolev space
of square-integrable functions with one square-integrable weak
derivative. The space $\D$ is a core: $G$ and $G^*$ preserve the jointly
smooth compact tests in each finite sector, and these are a core
for $D$. For $\psi\in\D$,
\begin{equation}\label{eq:coreK}
 (K_k-K_\infty)\psi=(2k)^{-1}D^2\psi\longrightarrow0.
\end{equation}
The identity
\[
 (K_k-i)^{-1}(K_\infty-i)\psi-\psi
       =(K_k-i)^{-1}(K_\infty-K_k)\psi
\]
on the dense space $(K_\infty-i)\D$, followed by the resolvent bound,
proves
\begin{equation}\label{eq:Kconverge}
 K_k\sres K_\infty,\qquad e^{itK_k}\longrightarrow e^{itK_\infty}
       \quad\text{strongly for every }t\in\R.
\end{equation}

\subsection{Bounded observables}\label{sub:scaling-observables}

Fix real $\chi\in C_c^\infty(\R)$ with $0\le\chi\le1$,
equal to one on $[-1,1]$ and supported in $[-2,2]$. Put
\begin{equation}\label{eq:Ck}
 r_k=k^{1/4},\qquad C_k=\chi(K_k/r_k).
\end{equation}
For $f\in C_c^\infty(\R,\R)$, define
\begin{equation}\label{eq:Bk}
 B_k(f)=\int_\R f(t)e^{itK_k}
            [C_k V^{(0)}C_k]e^{-itK_k}\dd t.
\end{equation}
The middle term has a unique bounded interpretation from the
canonical heat letter. In fact, with
$c_k(L)=\chi((L-k^2/2)/(kr_k))$, for any $a>0$,
\begin{equation}\label{eq:boundedcut}
 c_k(L)Vc_k(L)
  =[c_k(L)e^{aL}](e^{-aL}Ve^{-aL})[e^{aL}c_k(L)].
\end{equation}
The outside spectral functions are bounded. Dividing by $k$
and conjugating by $S_k$ gives the bracket in
\eqref{eq:Bk}. Thus $B_k(f)$ is bounded and self-adjoint in
$M_k$. The integral is a strong operator integral. Our uniformly
bounded observable families are
\begin{equation}\label{eq:boundedfamilies}
 U_k(t)=e^{itK_k},\qquad W_k(f,u)=e^{iuB_k(f)},
       \qquad t,u\in\R.
\end{equation}
In the original variables this uses the field $V/k$, time
$t/k$ for $L$, and the window
$L=k^2/2+O(k^{5/4})$. Its relative width tends to zero.
Since $L=2\phi_bU^*H_RU$, the corresponding canonical boundary
time is $2\phi_b t/k$.

\subsection{Normal states}\label{sub:scaling-states}

Choose $0\ne\eta\in C_c^\infty(\R)$ with $\|\eta\|_2=1$,
and define
\begin{equation}\label{eq:states}
 \Psi=G(\eta\otimes\Omega),\quad \Psi_k=S_k\Psi,
 \qquad \omega_k(x)=\langle\Psi_k,x\Psi_k\rangle\quad(x\in M).
\end{equation}
These are normal states. Since $\Psi\in\D$,
\begin{equation}\label{eq:stateenergy}
 \langle\Psi_k,L\Psi_k\rangle
       =k^2/2+k\langle\Psi,K_\infty\Psi\rangle+O(1),\qquad
 \|(L-k^2/2)\Psi_k\|=O(k).
\end{equation}
They therefore have relative energy fluctuations of order
$k^{-1}$. No path-integral identification of these vectors is
needed.

\section{Convergence of the smeared fields}\label{sec:convergence}

We first prove a weighted propagation estimate, then convergence
on a common field core and of the unitary exponentials.

\begin{lemma}\label{lem:propagation}
For $\psi\in\D$ and $T<\infty$,
\begin{equation}\label{eq:weightedprop}
 \sup_{|t|\le T}\left(\|(e^{-itK_k}-e^{-itK_\infty})\psi\|
       +\|e^{-y}(e^{-itK_k}-e^{-itK_\infty})\psi\|\right)
       \longrightarrow0.
\end{equation}
\end{lemma}
\begin{proof}
Work on a fixed particle sector and compact momentum set, where
$A=Q+2$ is bounded and $A\ge2$. Put $V_A=Ae^{-y}$,
$P_k=D+k$ and $H_k=K_k+k/2=P_k^2/(2k)+V_A$. Integration by
parts on compact tests gives
\begin{equation}\label{eq:ibp}
 2\operatorname{Re}\left\langle\frac{P_k^2}{2k}v,V_Av\right\rangle
  =\frac1k\|V_A^{1/2}P_kv\|^2
        -\frac1{2k}\langle v,V_Av\rangle.
\end{equation}
Expanding $\|H_kv\|^2$, then solving the resulting quadratic
inequality for $\|V_Av\|$, yields
\begin{equation}\label{eq:potentialgraph}
 \|V_Av\|\le\|H_kv\|+\frac1{2k}\|v\|
       \le\|K_kv\|+\left(\frac k2+\frac1{2k}\right)\|v\|.
\end{equation}
The scalar exponential Schr\"odinger operators have a compact
test operator core, and direct integration over the momentum
set gives the same core here. Closure of multiplication by
$V_A$ extends \eqref{eq:potentialgraph} to $D(K_k)$ on this
reducing subspace.

Write $E=D^2/2$, $v_0(t)=e^{-itK_\infty}\psi$ and $v_1(t)=-i\int_0^t e^{-i(t-s)K_\infty}Ev_0(s)\dd s$. Since $K_\infty=GDG^*$, these functions have common compact length and
momentum supports for $|t|\le T$, with uniformly bounded
derivatives of every fixed order. The residual of
$v_0+k^{-1}v_1$ in the $K_k$ evolution is
$-k^{-2}Ev_1$. Both this residual and its $K_k$ image are
$O(k^{-2})$. Duhamel's formula shows that the error and its
$K_k$ image are $O(k^{-2})$. Applying
\eqref{eq:potentialgraph} makes its $e^{-y}$-weighted norm
$O(k^{-1})$. The correction $k^{-1}v_1$ has the same weighted
bound. This proves \eqref{eq:weightedprop}.
\end{proof}

\subsection{An exact norm identity}\label{sub:exact-norm}

Temporarily distinguishing the created particle from its
spectators lets us compute its norm by the Jacobi isometry.
We restore bosonic symmetrization after controlling the energy tails.

Fix a real test function $f\in C_c^\infty(\mathbb R)$ throughout
the field estimates, and recall the Fourier convention
\begin{equation}\label{eq:fourierf}
 \widehat f(v)=\int_\R f(t)e^{itv}\dd t.
\end{equation}
For $d=1$, \eqref{eq:Gd} is
\begin{equation}\label{eq:G1}
 G_1(r,s)=\frac{\pi(r+s)}{\sinh\pi(r+s)}
          \frac{\pi(r-s)}{\sinh\pi(r-s)},
\end{equation}
with continuous values at zero. At fixed output energy $r$ and
spectator momentum $q$, Proposition~\ref{prop:jacobi} gives
\begin{equation}\label{eq:exactnorm}
 \int_0^\infty\left|\int_0^\infty
       \sqrt p F_1(r,s;q+p,q)a(s)\dd\mu(s)\right|^2\dd p
       =\int_0^\infty G_1(r,s)|a(s)|^2\dd\mu(s).
\end{equation}
Indeed the map inside the left norm is
$\delta_{a_q}^*\mathcal J_{1,r}^*M_{\sqrt{G_1(r,\cdot)}}$.
The fixed Jacobi parameter is now $r$, and its transform
variable is $s$. This use of the adjoint transform avoids an
incorrect exchange of the two energy parameters.

Define the uncut, rescaled, unsymmetrized creation operator
$\widetilde{\mathsf T}_k^+(f)$ on the unscaled length space by requiring
its sectorwise kernel in the energy representation $\mathcal U$ to be
\begin{equation}\label{eq:Tkpluskernel}
 \frac1k\sqrt p F_1(r,s;q+p,q)
                  \widehat f\!\left(\frac{r^2-s^2}{2k}\right).
\end{equation}
The target of $\widetilde{\mathsf T}_k^+(f)$ has one distinguished
momentum and $n$ symmetric spectator momenta. The actual creator
$\mathsf T_k^+(f)$ is $\sqrt{n+1}$ times its orthogonal symmetrization.
The sector operators are initially defined on spectral tests and then
closed. Their adjoints define $\mathsf T_k^-(f)$ for real $f$.
Transport these operators to the scaling coordinates and henceforth put
$T_k^\pm(f)=S_k^*\mathsf T_k^\pm(f)S_k$; use
$\widetilde T_k^+(f)$ for the same transport of the unsymmetrized
creator, with $S_k$ acting on the length variable also in its target.
In these coordinates, \eqref{eq:exactnorm} gives
\begin{equation}\label{eq:unsymnorm}
 \|\widetilde T_k^+(f)\psi_n\|^2
       =\langle\psi_n,w_k(K_k)\psi_n\rangle,
\end{equation}
where, for $\xi\ge-k/2$ and $s=\sqrt{k^2+2k\xi}$,
\begin{equation}\label{eq:wk}
 w_k(\xi)=\frac1{k^2}\int_0^\infty
       \rho(r)G_1(r,s)
       \left|\widehat f\!\left(\frac{r^2-s^2}{2k}\right)\right|^2\dd r.
\end{equation}
To compute its limit use
\[
 \frac{\sinh(2\pi r)}{\sinh\pi(r+s)\sinh\pi(r-s)}
       =\coth\pi(r-s)+\coth\pi(r+s).
\]
With $v=(r^2-s^2)/(2k)$, so $r\,dr=k\,dv$, this yields
\begin{equation}\label{eq:wkchange}
 w_k(\xi)=2\int_{-k/2-\xi}^\infty
       v\,[\coth\pi(r-s)+\coth\pi(r+s)]
                  |\widehat f(v)|^2\dd v,
 \quad r=\sqrt{s^2+2kv}.
\end{equation}
All singularities are removable. We write $z_+=\max\{z,0\}$ for
the positive part of a real number. Locally uniformly in $\xi$,
$r-s\to v$ and $r+s\to\infty$ for fixed $v$, and hence
\begin{equation}\label{eq:cf}
 w_k(\xi)\longrightarrow
 c_f:=4\int_\R\frac{v}{1-e^{-2\pi v}}|\widehat f(v)|^2\dd v.
\end{equation}
The fraction at zero is $1/(2\pi)$. The integrands obey the
uniform estimate
\begin{equation}\label{eq:wboundintegrand}
 0\le 2v[\coth\pi(r-s)+\coth\pi(r+s)]
   \le C\bigl(|v|+(r+s)/k\bigr)
   \le C(1+|v|+\sqrt{\xi_+}).
\end{equation}
For the first inequality use
$|z\coth(\pi z)|\le|z|+1/\pi$ and
$2kv=(r-s)(r+s)$; for the last use
$r+s\le2s+\sqrt{2kv_+}$. Since $\widehat f$ is Schwartz,
\begin{equation}\label{eq:wbound}
 0\le w_k(\xi)\le C_f(1+|\xi|).
\end{equation}
Compact tests satisfy $\sup_k\|K_k^j\psi\|<\infty$ for each
fixed integer $j$. Strong resolvent convergence, local uniform
convergence in \eqref{eq:cf}, and the tail bound obtained from
the second spectral moment give
\begin{equation}\label{eq:unsymnormlimit}
 \|\widetilde T_k^+(f)\psi_n\|^2
       \longrightarrow c_f\|\psi_n\|^2\qquad(\psi_n\in\D).
\end{equation}

\subsection{Identification and convergence of the limiting field}\label{sub:limit-field}

The phase in $G$ acts on a creation wavefunction through the
one-particle momentum. Thus, as a form,
\[
 G(y)^*V^{(0)}(y)G(y)
       =2e^{-y}\Phi\bigl(e^{-ip e^{-y}}g(p)\bigr).
\]
The free limiting flow translates $y$ to $y+t$. Integrating the
translated expression against $f(t)$ gives the wavefunction below;
its dependence on $y$ is precisely a one-particle dilation.
For the fixed real test $f$ and a real dilation parameter $a$, define
\begin{equation}\label{eq:hf}
 h_f(p)=2\sqrt p\int_\R f(t)e^{-t}e^{-ipe^{-t}}\dd t,
 \qquad (d_a h)(p)=e^{a/2}h(e^ap),\quad \mathsf D_a=\Gamma_s(d_a).
\end{equation}
Thus $d_a$ is the one-particle dilation and $\mathsf D_a$ its Fock
second quantization; the scalar $a$ here is not a species index.
The function $h_f$ is in $L^2(dp)$: it is $O(\sqrt p)$ at
zero, and integration by parts gives rapid decay after that
factor at infinity. Put
\begin{equation}\label{eq:limitfield}
 F(f)=G\left[\int_\R^\oplus\Phi(d_{-y}h_f)\dd y\right]G^*.
\end{equation}
Define its creation and annihilation parts on finite-particle vectors by
\begin{equation}\label{eq:limit-creation-definition}
 F^+(f)=G\left[\int_{\mathbb R}^{\oplus}a^*(d_{-y}h_f)\,dy\right]G^*,
 \qquad F^-(f)=F^+(f)^*,
\end{equation}
so that $F(f)=F^+(f)+F^-(f)$ there.
The decomposable field has its self-adjoint Fock realization.
On the $n$-particle sector it has norm at most
$2\|h_f\|\sqrt{n+1}$. The finite-particle subspace is an
invariant space of analytic vectors. Approximation within each
sector shows that $\D$ is also an operator core.

Write $\widetilde h(\xi)=(2\pi)^{-1/2}\int_0^\infty
p^{-1/2-i\xi}h(p)\,dp$ for the unitary Mellin transform from
$L^2(dp)$ to $L^2(d\xi)$, initially on compact tests and then by
$L^2$ extension. It gives
\begin{equation}\label{eq:mellin}
 \widetilde h_f(\xi)=\frac2{\sqrt{2\pi}}e^{-i\pi/2}
       e^{-\pi\xi/2}\Gamma(1-i\xi)\widehat f(-\xi),
 \qquad
 |\widetilde h_f(\xi)|^2
       =\frac{4\xi}{e^{2\pi\xi}-1}|\widehat f(-\xi)|^2.
\end{equation}
The integral can first be computed with an Abel factor
$e^{-\varepsilon p}$. Its removal in $L^2$ is justified by
\eqref{eq:hf}; equivalently, on the compact $t$ support the
Mellin integral gives the same boundary value. The Gamma
identity $|\Gamma(1+i\xi)|^2=\pi\xi/\sinh(\pi\xi)$ and
Plancherel imply
\begin{equation}\label{eq:hfnorm}
 \|h_f\|^2=c_f.
\end{equation}

\begin{proposition}\label{prop:fieldcore}
With $T_k(f)=T_k^+(f)+T_k^-(f)$ on the indicated sector
domains,
\begin{equation}\label{eq:Tcoreconv}
 T_k(f)\psi\longrightarrow F(f)\psi\qquad(\psi\in\D).
\end{equation}
Moreover, on each $n$-particle sector,
\begin{equation}\label{eq:sectorgraph}
 \|T_k(f)\psi_n\|
       \le C_f\sqrt{n+1}\|(1+K_k^2)^{1/2}\psi_n\|,
 \qquad \psi_n\in D(K_k),
\end{equation}
with $C_f$ independent of $k\ge1$ and $n$.
\end{proposition}
\begin{proof}
Conjugation by $G$ reduces the limiting evolution to translation,
whose time-smearing is \eqref{eq:limitfield}. Between compact tests only a compact
set of inserted momenta contributes, so
Lemma~\ref{lem:propagation} justifies convergence of these
matrix elements. The same argument applies with the created
momentum distinguished, before symmetrization.

The unsymmetrized creators therefore converge weakly on $\D$
to the corresponding part of \eqref{eq:limitfield}. Their
norms converge by \eqref{eq:unsymnormlimit}, and the limiting
norm is $\|h_f\|\|\psi_n\|$ by \eqref{eq:hfnorm}. Weak
convergence and convergence of norms imply strong convergence.
Apply bounded symmetrization to obtain convergence of the
actual creators. For annihilation on a compact test, the
annihilated momentum remains in a fixed compact set under
$K_k$ and $K_\infty$. On that set its one-particle coefficient is
square integrable. Use \eqref{eq:weightedprop} for the input,
ordinary strong convergence for the output propagator, and
integrate over $\supp f$. This proves convergence of the
annihilation part in norm.

The creator bound in \eqref{eq:sectorgraph} follows from
\eqref{eq:unsymnorm} and \eqref{eq:wbound}. For the adjoint
bound, insert an output spectral multiplier $b(K_k)$ in
\eqref{eq:unsymnorm}. The integrand in
\eqref{eq:wkchange} acquires $|b(\xi+v)|^2$. Choose
$b(z)=(1+z^2)^{-1/2}$. The estimate
$\sqrt{\xi_+}\le\sqrt{|\xi+v|}+\sqrt{|v|}$ and the Schwartz
decay of $\widehat f$ show that this integral is uniformly
bounded in $k,\xi$. Output spectral multipliers commute with
symmetrization, since the output Hamiltonian depends on the
momenta through their sum. Thus
$(1+K_k^2)^{-1/2}T_k^+(f)$ is bounded between the appropriate
sectors by $C_f\sqrt{n+1}$. Taking adjoints proves that
$D(K_k)$ on the output sector belongs to the annihilator
domain, with the required bound. Its dense adjoint domain
also proves closability of the initial sectorwise creator.

Finally, the energy-kernel definition agrees with the
time-integrated form. At fixed $k$, insert bounded spectral
cutoffs into \eqref{eq:graphinput}; the spectral theorem gives
\eqref{eq:Tkpluskernel}. Remove the cutoffs in that graph
topology. Compact tests and their time translates belong to
every power domain of $L$ (after transport by $S_k$), and the
cutoffs commute with the evolution. The sector estimates just
proved identify the limiting form with $T_k^+(f)+T_k^-(f)$.
\end{proof}

\begin{theorem}\label{thm:boundedconv}
For every real $f\in C_c^\infty(\R)$,
\begin{equation}\label{eq:boundedconv}
 B_k(f)\sres F(f),\qquad
 W_k(f,u)\longrightarrow e^{iuF(f)}
       \quad\text{strongly with adjoints, for every }u\in\R.
\end{equation}
\end{theorem}
\begin{proof}
The spectral bound and $r_k\to\infty$ give $C_k\to1$
strongly, first on $\D$ and then by contraction on all vectors.
Since $\sup_k\|K_k^2\psi\|<\infty$ for $\psi\in\D$,
\eqref{eq:sectorgraph} implies $\|T_k(f)(1-C_k)\psi\|\le C_{f,\psi}/r_k\longrightarrow0$. The time and spectral functional calculi commute, so on these
vectors $B_k(f)=C_kT_k(f)C_k$. Proposition~\ref{prop:fieldcore}
then gives $B_k(f)\psi\to F(f)\psi$. Since $\D$ is a core
for the self-adjoint $F(f)$, the resolvent argument following
\eqref{eq:coreK} applies. Functional calculus for strong
resolvent convergence gives the exponential limits, including
their adjoints.
\end{proof}

\section{The continuous core and convergence of states}\label{sec:core}

The limiting fields generate a half-line current algebra.
Their covariance under the limiting Hamiltonian identifies
the full limiting algebra as its continuous core.

Let $\A$ be the negative half-line current algebra on $\F$:
\begin{equation}\label{eq:currentalgebra}
 \A=\{e^{i\Phi(\sqrt p\widehat F(p))}:
               F\in C_c^\infty(( -\infty,0),\R)\}'',
 \qquad \widehat F(p)=\int_\R F(x)e^{ipx}\dd x.
\end{equation}
In \eqref{eq:hf}, the substitution $x=-e^{-t}$ gives
$F(x)=2f(-\log(-x))$. It is a bijection between the two real
test-function spaces. Thus the $h_f$ generate exactly $\A$.
The vacuum $\Omega$ is cyclic and separating and $\A$ is the
hyperfinite type $\mathrm{III}_1$ factor. Let
$\varphi(a)=\langle\Omega,a\Omega\rangle$ be its vacuum state.
Its modular automorphism group is
\begin{equation}\label{eq:modulardilation}
 \sigma_s^\varphi(a)=\mathsf D_{2\pi s}a\mathsf D_{-2\pi s}.
\end{equation}
These are the current-net and geometric modular results; see
\cite[Sections 2--3]{LW} and \cite[Section 3.3]{CKLW}.
The positive sign in \eqref{eq:modulardilation} is the
negative-half-line convention. Hyperfiniteness also follows
from the split property: the current vacuum character
$\prod_{n\ge1}(1-e^{-\beta n})^{-1}$ is finite for every
$\beta>0$, giving split inclusions. An exhaustion by
intermediate type I factors gives injectivity of a local
algebra. The type $\mathrm{III}_1$ and modular assertions
are the standard conformal-net conclusions in the cited
results.

After conjugation by $G^*$, the Hamiltonian generates translation
in $y$ and the matter algebra at $y$ is
$\mathsf D_{-y}\A\mathsf D_y$. This is the regular crossed-product
representation, as the covariance calculation below shows.

\begin{theorem}\label{thm:core}
The von Neumann algebra generated by the limits in
\eqref{eq:Kconverge} and \eqref{eq:boundedconv} is
\begin{equation}\label{eq:Mlimit}
 M_\infty=\{e^{itK_\infty},e^{iuF(f)}:t,u\in\R,
                         f\in C_c^\infty(\R,\R)\}''
      \cong\A\rtimes_{\sigma^\varphi}\R.
\end{equation}
It is a hyperfinite factor of type $\mathrm{II}_\infty$.
\end{theorem}
\begin{proof}
On $L^2(\R_y;\F)$ put
\begin{equation}\label{eq:regular}
 (\pi(a)\psi)(y)=\mathsf D_{-y}a\mathsf D_y\psi(y),\qquad
 (\lambda(t)\psi)(y)=\psi(y+t).
\end{equation}
For $\alpha_t=\Ad \mathsf D_{-t}$, direct calculation gives
\(\lambda(t)\pi(a)\lambda(t)^*=\pi(\alpha_t(a)).\)
This is the faithful regular covariant representation of the
crossed product, with the indicated sign for translations.
By \eqref{eq:Kdomain} and \eqref{eq:limitfield}, $M_\infty=G[\pi(\A)\cup\lambda(\R)]''G^*$. Equation \eqref{eq:modulardilation} says
$\alpha_t=\sigma_{-t/(2\pi)}^\varphi$, so a reparametrization
of $\R$ gives \eqref{eq:Mlimit}. The continuous core of a
type $\mathrm{III}_1$ factor is a type $\mathrm{II}_\infty$
factor \cite{Takesaki}. Since $\R$ is amenable and $\A$ is
injective, its crossed product is injective
\cite[Theorem 4.6]{McKee}. Its predual is separable, so
\cite{C} gives hyperfiniteness.
\end{proof}

To specify state convergence on one algebra, let
$\mathfrak S\subset\prod_{k\in\mathbb N}B(\Hh)$ be the unital
$C^*$-algebra, in the supremum norm, generated by the sequences
in \eqref{eq:boundedfamilies}. We use integer $k$ here only to
simplify notation. Uniformly bounded families with strong-$*$
limits form a norm-closed $*$-algebra. Indeed products converge
strongly on bounded sets, and a uniform-norm approximation
allows the strong limit to pass to the closure. Hence
\begin{equation}\label{eq:eval}
 \operatorname{ev}_\infty:\mathfrak S\longrightarrow B(\Hh),
 \qquad (a_k)\longmapsto\mathrm{s}^*\text{-}\lim_{k\to\infty}a_k
\end{equation}
is a unital $*$-homomorphism, with image generating $M_\infty$.

\begin{proposition}\label{prop:statelimit}
The states $\widehat\omega_k((a_j))=\langle\Psi,a_k\Psi\rangle =\omega_k(S_ka_kS_k^*)$ converge weak-$*$ on $\mathfrak S$ to
$\widehat\omega_\infty(a)=
\langle\Psi,\operatorname{ev}_\infty(a)\Psi\rangle$.
The GNS von Neumann algebra of this limiting state is
$M_\infty$. In particular, every joint $*$-moment of the
specified bounded observables converges.
\end{proposition}
\begin{proof}
The convergence follows directly from \eqref{eq:eval}. It
remains to identify the GNS representation. In the regular
representation \eqref{eq:regular}, the vector
$\eta\otimes\Omega$ is cyclic and separating. For cyclicity,
conjugate by the decomposable unitary $Z(y)=\mathsf D_{-y}$. Then
\(Z^*\pi(a)Z=a,\qquad Z^*\lambda(t)Z=\mathsf D_{-t}\lambda(t).\)
Since $\mathsf D_t\Omega=\Omega$, the orbit of $\eta\otimes\Omega$
contains products of $a\Omega$ and translates of $\eta$.
The former are total in $\F$. The translates are total in
$L^2(\R)$ because the Fourier transform of a nonzero compactly
supported smooth function is entire and is nonzero almost
everywhere on the real axis. For separation, the commutant
of \eqref{eq:regular} contains the constant algebra $\A'$
and the unitaries $\mathsf D_t\lambda(t)$. Their orbit is total by
the same argument and the cyclicity of $\Omega$ for $\A'$.
Conjugation by $G$ gives the claim for $\Psi$. The GNS
representation is therefore the concrete representation of
$M_\infty$ generated by the image of \eqref{eq:eval}.
\end{proof}

\section{Fixed windows and unbounded limiting fields}\label{sec:windows}

In a fixed rescaled energy window, the limiting field is
unbounded and the finite-$k$ norms diverge. We prove this using
many-particle vectors which stay inside that window.

Fix $I=[a,b]$, $a<b$, and write $\Delta=b-a$. Throughout this
and the following sections assume
\begin{equation}\label{eq:largerk}
 k>\max\{1,-2a\}.
\end{equation}
Then $I$ lies in the spectrum of $K_k$ and
\begin{equation}\label{eq:qk}
 q_k=\ind_I(K_k)\ne0,\qquad q=\ind_I(K_\infty),\qquad
 m_k=\tau_k(q_k)>0,
\end{equation}
where $\tau_k(x)=\tau(S_kxS_k^*)$ is the transported trace on
$M_k=S_k^*MS_k$.
This qualification is needed for arbitrary intervals: for
example $I=[-2,-1]$ and $k=1$ give $q_k=0$ and would make
normalized corner formulas undefined.

Since $K_\infty$ is unitarily equivalent to $D$, its spectral measure
has no atoms, and $q_k\to q$ strongly. Define
\begin{equation}\label{eq:bkfinite}
 b_k(f)=q_kT_k(f)q_k\in q_kM_kq_k,\qquad
 c_k(f)=q_kT_k^+(f)q_k,\qquad b_k(f)=c_k(f)+c_k(f)^*.
\end{equation}
The operators are bounded for each fixed $k$. Indeed apply the
finite-window construction of
Sections~\ref{sec:fockbound}--\ref{sec:nonhyper} to the original
energy interval $k^2/2+kI$, then divide by $k$ and integrate
the time translates. This proves boundedness also for the
separated creator. Only the self-adjoint sum is asserted to
belong to the corner algebra.

\begin{proposition}\label{prop:sharp}
On each fixed finite collection of particle sectors,
\begin{equation}\label{eq:sharpconv}
 b_k(f)v\longrightarrow qF(f)qv,\qquad
 c_k(f)v\longrightarrow qF^+(f)qv.
\end{equation}
The closure of $qF(f)q$ on $q\Hh_{\mathrm{fin}}$ is
self-adjoint on $q\Hh$ and affiliated with $qM_\infty q$.
\end{proposition}
\begin{proof}
For any fixed finite-particle $v$, set $v_k=q_kv$. Bounded
spectral calculus gives both $v_k\to qv$ and
$K_kv_k\to K_\infty qv$, since the discontinuities at $a,b$ have
zero spectral mass. Choose $\phi_j\in\D$ in the same finite
sectors, converging to $qv$ in the $K_\infty$ graph norm. For fixed
$j$, \eqref{eq:coreK} implies
\[
 \lim_{k\to\infty}\|v_k-\phi_j\|_{K_k}
        =\|qv-\phi_j\|_{K_\infty},\qquad
 \|w\|_A^2=\|w\|^2+\|Aw\|^2.
\]
Use \eqref{eq:sectorgraph}, then let $k\to\infty$ before
$j\to\infty$. The limiting field and its creator are
bounded between fixed sectors. Proposition~\ref{prop:fieldcore}
and strong convergence of the output projection now prove
\eqref{eq:sharpconv}.

The projection $q$ commutes with $N$. Consequently the
compressed symmetric field on $q\Hh_{\mathrm{fin}}$ has the
same $O(\sqrt{n+1})$ sector bounds as $F(f)$. This domain is
invariant and consists of analytic vectors, so the closure is
self-adjoint. The bounded self-adjoint corner elements
$q\sin(tF(f))q/t$ converge to it on this core as $t\to0$.
Their resolvents therefore converge strongly on $q\Hh$.
Strong closure of the corner shows that the limiting resolvent
belongs to $qM_\infty q$, proving affiliation.
\end{proof}

\begin{theorem}\label{thm:normdivergence}
Let $f\in C_c^\infty(\R,\R)$ satisfy $\int_\R f\ne0$.
For every bounded nondegenerate interval $I$,
\begin{equation}\label{eq:actualdivergence}
 \|c_k(f)\|\longrightarrow\infty,\qquad
 \|c_k(f)^*\|\longrightarrow\infty,\qquad
 \|b_k(f)\|\longrightarrow\infty.
\end{equation}
\end{theorem}

Conjugate by $G^*$ and then by $Z^*$, where $Z(y)=\mathsf D_{-y}$.
If $\mathsf D_t=e^{itP}$, direct differentiation gives
\begin{equation}\label{eq:KminusP}
 K_\infty\longmapsto D-P,\qquad F(f)\longmapsto\Phi(h_f).
\end{equation}
Use the unitary Fourier transform in $y$ with kernel
$(2\pi)^{-1/2}e^{-i\lambda y}$, so that $D$ becomes multiplication
by $\lambda$. Together with the one-particle Mellin transform, and
retaining symbols for their unitary images, this gives
\begin{equation}\label{eq:signedenergy}
 P=d\Gamma(\xi),\qquad q=\ind_I(\lambda-d\Gamma(\xi)).
\end{equation}
Unlike the positive momentum $p$, the dilation energy $\xi$
has both signs.

\begin{proofclaim}[Vectors in the fixed window]\label{clm:window-vectors}
There are normalized $n$-particle vectors in $q\mathcal H$
whose compressed-annihilation norms have the lower bound
\eqref{eq:cltlower}, with a constant $\gamma>0$.
\end{proofclaim}
\begin{proof}
By \eqref{eq:mellin} and $\widehat f(0)\ne0$,
$\widetilde h_f$ is smooth and nonzero near zero. Choose a
small $\varepsilon>0$ and a normalized one-particle vector
$v(\xi)$ supported in $(-\varepsilon,\varepsilon)$ so that
$|v|^2$ is a smooth even probability density of positive
variance and
\begin{equation}\label{eq:gamma}
 \overline{\widetilde h_f(\xi)}v(\xi)\ge0,\qquad
 \gamma=\langle\widetilde h_f,v\rangle>0.
\end{equation}
One obtains it by giving a nonnegative even smooth amplitude
the phase of $\widetilde h_f$.

Choose $r>\varepsilon$ small compared with $b-a$, and set
$J=[-r,r]$, $J_0=[-r+\varepsilon,r-\varepsilon]$. Choose a
normalized $z(\lambda)$ supported close enough to $(a+b)/2$
that
\begin{equation}\label{eq:windowchoice}
 \lambda-s\in I\quad\text{if }\lambda\in\supp z,\quad
                           s\in J+[-\varepsilon,\varepsilon].
\end{equation}
Let $X_j$ be independent variables of density $|v|^2$,
$S_n=X_1+\cdots+X_n$, and $p_n=\Pr(S_n\in J)$. The
normalized symmetric vector
\begin{equation}\label{eq:cltvector}
 \psi_n(\lambda;\xi_1,\ldots,\xi_n)
   =\frac{z(\lambda)}{\sqrt{p_n}}
       \ind_J\!\left(\sum_{j=1}^n\xi_j\right)
                       \prod_{j=1}^nv(\xi_j)
\end{equation}
satisfies $q\psi_n=\psi_n$. On the region where the remaining
sum lies in $J_0$, every annihilated momentum in
$\supp v$ makes the original sum lie in $J$. Formula
\eqref{eq:annihilation} and \eqref{eq:gamma} therefore give
\begin{equation}\label{eq:cltlower}
 \|q\,a(\widetilde h_f)q\psi_n\|^2
    \ge n\gamma^2
         \frac{\Pr(S_{n-1}\in J_0)}{\Pr(S_n\in J)}.
\end{equation}
The output restriction is contained in $q\Hh$ by
\eqref{eq:windowchoice}.
\end{proof}

\begin{proofclaim}[Positive limit of the probability ratio]\label{clm:window-ratio}
For the variables and intervals of Claim~\ref{clm:window-vectors},
the asymptotics \eqref{eq:localCLT} hold. In particular the ratio
in the lower bound stays bounded away from zero.
\end{proofclaim}
\begin{proof}
For completeness, the ratio in \eqref{eq:cltlower} has a
positive limit by an elementary local limit argument. Let
$\sigma^2>0$ be the variance and $\phi(t)$ the characteristic
function of $|v|^2$. At zero,
$\phi(t)=1-\sigma^2t^2/2+O(t^4)$ and
$|\phi(t)|\le e^{-ct^2}$ locally. Away from zero its modulus
is strictly less than one on compact sets, because the density
is absolutely continuous and nondegenerate; at infinity it
decays rapidly. Fourier inversion for the density of $S_n$,
split into these regions, and $t=u/\sqrt n$ in the first
region show uniformly for bounded $s$ that
\[
 \text{density of }S_n\text{ at }s
       \sim(2\pi n\sigma^2)^{-1/2}.
\]
For the tail, choose a large fixed $T$ with
$\sup_{|t|\ge T}|\phi(t)|<1$ and dominate
$|\phi(t)|^n$ by that supremum to the power $n-2$ times
$|\phi(t)|^2$. Thus
\begin{equation}\label{eq:localCLT}
 \Pr(S_n\in J)\sim\frac{|J|}{\sqrt{2\pi n\sigma^2}},\qquad
 \frac{\Pr(S_{n-1}\in J_0)}{\Pr(S_n\in J)}
                 \longrightarrow\frac{|J_0|}{|J|}>0.
\end{equation}
This is an analytic estimate and uses no numerical evidence.
\end{proof}

\begin{proof}[Proof of Theorem~\ref{thm:normdivergence}]
It follows that the compressed annihilator is unbounded.
Taking adjoints between adjacent particle sectors shows that
the compressed creator is unbounded as well. Their sum is
unbounded because on an $n$-particle input the two outputs
have particle numbers $n-1$ and $n+1$ and are orthogonal.
Proposition~\ref{prop:sharp} now rules out any uniformly
bounded subsequence of $c_k(f)$ or $b_k(f)$: such a subsequence
would bound the corresponding limiting operator on the dense
finite-particle domain. This proves the actual limits
\eqref{eq:actualdivergence}, and adjunction gives the remaining
norm equality.
\end{proof}

\section{Tracial second moments in the limit}\label{sec:limittrace}

Despite the norm divergence, the normalized tracial second
moments have a positive finite limit.

Set
\begin{equation}\label{eq:skZI}
 s_k(\xi)=\sqrt{k^2+2k\xi},\qquad
 Z_I=\int_a^b e^{2\pi\xi}\dd\xi.
\end{equation}
The trace formula and a change of variables give
\begin{equation}\label{eq:mkasymptotic}
 m_k=\int_a^b\frac{k}{\pi^2}\sinh(2\pi s_k(\xi))\dd\xi
       \sim\frac{ke^{2\pi k}}{2\pi^2}Z_I.
\end{equation}
Define the shifted-energy function $\xi_k(s)=(s^2-k^2)/(2k)$.
The vacuum-column calculation used in Lemma~\ref{lem:decomposition},
now with the time-smearing multiplier, gives the exact identity
\begin{equation}\label{eq:vkexact}
 v_k:=\frac{\tau_k(b_k(f)^2)}{m_k}
  =\frac1{k^2m_k}\iint_{\xi_k(r),\xi_k(s)\in I}
       G_1(r,s)|\widehat f(\xi_k(r)-\xi_k(s))|^2\dd\mu(r)\dd\mu(s),
\end{equation}

\begin{proposition}\label{prop:varianceconvergence}
For $\int f\ne0$,
\begin{equation}\label{eq:vinfty}
 v_k\longrightarrow v_\infty(f,I)
   =\frac2{Z_I}\int_I\int_I e^{\pi(\xi+\eta)}
       \frac{\xi-\eta}{\sinh\pi(\xi-\eta)}
                     |\widehat f(\xi-\eta)|^2\dd\xi\dd\eta
       \in(0,\infty).
\end{equation}
This is the normalized tracial second moment of the
self-adjoint affiliated operator $B=\overline{qF(f)q}$ in
$qM_\infty q$.
\end{proposition}
\begin{proof}
Uniformly on $I^2$, including the removable diagonal,
\begin{align}\label{eq:uniformasymptotics}
 d\mu(s_k(\xi))&\sim\frac{ke^{2\pi k}}{2\pi^2}
                              e^{2\pi\xi}\dd\xi,\notag\\
 G_1(s_k(\xi),s_k(\eta))
    &\sim4\pi^2k e^{-2\pi k-\pi(\xi+\eta)}
                  \frac{\xi-\eta}{\sinh\pi(\xi-\eta)}.
\end{align}
Here $s_k(\xi)=k+\xi+O(k^{-1})$ uniformly on $I$.
Substitution into \eqref{eq:vkexact} and
\eqref{eq:mkasymptotic} proves \eqref{eq:vinfty}. The
diagonal quotient is $1/\pi$. Continuity and
$\widehat f(0)\ne0$ give strict positivity.

We check the limiting trace independently, fixing its scalar
normalization. In the representation \eqref{eq:KminusP} after
Fourier transformation in $y$, all core operators are
decomposable in $\lambda$. Normalize the dual weight by
\begin{equation}\label{eq:dualweight}
 \widetilde\varphi(X)=\int_\R
                 \langle\Omega,X(\lambda)\Omega\rangle\dd\lambda,
                         \qquad X\ge0.
\end{equation}
This is the standard dual weight in the regular
representation, with the Haar scalar chosen as displayed
\cite{Haagerup}. Put $h_\infty=e^{-2\pi K_\infty}$, affiliated with the centralizer
of $\widetilde\varphi$. The modular group of $\widetilde\varphi$ is implemented by
$h_\infty^{it}=e^{-2\pi itK_\infty}$. Weighting by $h_\infty^{-1}$ gives a faithful
normal semifinite trace by the invariant-weight theorem
\cite{PT}. Since $P\Omega=0$, its formula is
\begin{equation}\label{eq:coretrace}
 \tau_\infty(X)=\int_\R e^{2\pi\lambda}
                 \langle\Omega,X(\lambda)\Omega\rangle\dd\lambda.
\end{equation}
The weighting is interpreted by spectral truncation of $h_\infty$;
the expression is allowed to be infinite. In particular
$\tau_\infty(H(K_\infty))=\int e^{2\pi\xi}H(\xi)\dd\xi$ and
$\tau_\infty(q)=Z_I$.

For the corner, set
$\zeta_I(\lambda)=e^{\pi\lambda}\ind_I(\lambda)\Omega$.
Its squared norm is $Z_I$. For $X\in(qM_\infty q)_+$,
the vacuum outside $I$ is killed by $q$, so
\eqref{eq:coretrace} becomes
$\tau_\infty(X)=\langle\zeta_I,X\zeta_I\rangle$.
Spectral truncation extends this identity to positive
affiliated operators. The vector $\zeta_I$ belongs to the
finite-particle domain of $B$. Only creation acts on it;
hence \eqref{eq:mellin} gives
\begin{align}\label{eq:coresecondmoment}
 \tau_\infty(B^2)=\|B\zeta_I\|^2
   &=\int_I e^{2\pi\eta}\int_I
               |\widetilde h_f(\eta-\xi)|^2\dd\xi\dd\eta\notag\\
   &=2\int_I\int_I e^{\pi(\xi+\eta)}
        \frac{\xi-\eta}{\sinh\pi(\xi-\eta)}
                     |\widehat f(\xi-\eta)|^2\dd\xi\dd\eta.
\end{align}
Dividing by $Z_I$ proves the last assertion.
\end{proof}

\section{The loss of the finite-energy obstruction}\label{sec:loss}

Set
\begin{equation}\label{eq:uk}
 u_k=e^{2\pi iK_k/\Delta}q_k,\qquad
 c_{k,j}=u_k^jc_k(f)u_k^{-j},\qquad
 b_{k,j}=c_{k,j}+c_{k,j}^*.
\end{equation}
At each fixed $k$ the finite-window Parseval argument supplies
a constant $R_k<\infty$ with
\begin{equation}\label{eq:krow}
 \sum_{j=1}^n c_{k,j}c_{k,j}^*\le R_k^2q_k\qquad(n\ge1).
\end{equation}
The proof for an arbitrary bounded energy interval is the
same as \eqref{eq:parseval}. Its length replaces $h$, and
Fourier phases on a translated interval have the same norms.
The estimates of Lemma~\ref{lem:fullfock} use only the
measure and maximal density of that interval. Time smearing
then uses Minkowski's inequality in the row Hilbert space;
the time evolution commutes with $u_k$.

Taking $n=1$ and applying Theorem~\ref{thm:normdivergence} gives
\begin{equation}\label{eq:Rnecessary}
 R_k\ge\|c_k(f)\|\longrightarrow\infty.
\end{equation}

The original $L$-energy window has length $k\Delta$.
Using \eqref{eq:d1constant} and the factor $k^{-1}$ in the
rescaled field gives the available bound, where
$\|f\|_1=\int_{\mathbb R}|f(t)|\,dt$,
\begin{equation}\label{eq:Rexplicit}
 \mathcal R_k^2:=\|f\|_1^2\frac{\Delta}{k}
       \nu(k^2/2+kb)\bigl(1+\sqrt{3m_k/2}\bigr)^2.
\end{equation}
Thus one may take $R_k=\mathcal R_k$, or bound the optimal
$R_k$ above by this number. Its asymptotic value is
\begin{equation}\label{eq:Rasymptotic}
 \mathcal R_k^2\sim
      \|f\|_1^2\frac{3\Delta e^{2\pi b}Z_I}{8\pi^4}e^{4\pi k}.
\end{equation}
At finite $k$, the tensor-norm contradiction compares
\begin{equation}\label{eq:finitekcontradiction}
 n v_k\quad\text{and}\quad
 \left\|\sum_{j=1}^n b_{k,j}\otimes b_{k,j}^{\op}\right\|_{\min}
          \le2R_k^2(\sqrt n+1).
\end{equation}
At fixed $k$, large $n$ contradicts hyperfiniteness. By
\eqref{eq:Rasymptotic} and $v_k\to v_\infty>0$, taking $n$ of
order $e^{8\pi k}$ with a sufficiently large prefactor suffices.

Theorem~\ref{thm:normdivergence} and
Proposition~\ref{prop:varianceconvergence} give
\begin{equation}\label{eq:contractionloss}
 \frac{\tau_k((b_k(f)/\|b_k(f)\|)^2)}{\tau_k(q_k)}
       =\frac{v_k}{\|b_k(f)\|^2}\longrightarrow0.
\end{equation}
Thus contraction normalization loses the second moment,
whereas the Weyl operators converge by
Theorem~\ref{thm:boundedconv}.

The mechanism is visible in \eqref{eq:signedenergy}:
the window bounds $\lambda-\sum_j\xi_j$, with $\xi_j$ of both signs.
The vectors \eqref{eq:cltvector} stay in the window while their
annihilation matrix elements grow as $\sqrt n$.

The same conclusion applies to the rescaled unsmeared point
witnesses. More explicitly, put
$e_k=\ind_{k^2/2+kI}(L)=S_kq_kS_k^*$ and, for any $t>0$, define $b_k^{\mathrm{pt}}=\frac{1}{k}S_k^*(e^{tL}e_k)A_{1,t}(e^{tL}e_k)S_k$. Let $c_k^{\mathrm{pt}}$ be its separated creation part, given by the
bounded construction of Section~\ref{sec:nonhyper} in this energy
window. Thus $b_k^{\mathrm{pt}}=c_k^{\mathrm{pt}}+
(c_k^{\mathrm{pt}})^*$, independently of $t$, and
\[
 \|b_k(f)\|\le\|f\|_1\|b_k^{\mathrm{pt}}\|,\qquad
 \|c_k(f)\|\le\|f\|_1\|c_k^{\mathrm{pt}}\|.
\]
Their norms therefore diverge. The point-field variance
formula is \eqref{eq:vinfty} with the factor
$|\widehat f(\xi-\eta)|^2$ omitted; uniform asymptotics on
$I^2$ again give a finite positive limit.

\appendix

\section{Explicit insertion blocks}\label{app:blocks}

We record the particle blocks, with the normalizations used in
the spectral estimates. All formulas are interpreted on wave packets.

We first use the one-species coordinates of \eqref{eq:4.2}. Let
$V^+$ and $V^-$ be the creation and annihilation blocks of
$V(x)=2^de^{-dx}\Phi_d(0)$ in this representation. For
$\mathbf p=(p_1,\ldots,p_n)$ put $q_n=\sum_{j=1}^np_j$, with $q_0=0$;
$\mathbf p_{\widehat j}$ means the tuple with its $j$th entry deleted.
The variables $t,s>0$ below are spectral parameters, with energies
$E(t),E(s)$. The creation block is

\begin{equation}\label{eq:4.5}
\begin{aligned}
(V^+\psi_n)_{n+1}(t;\mathbf p)
=\frac1{\sqrt{n+1}}\sum_{j=1}^{n+1}g_d(p_j)
\int_0^\infty &F_d(t,s;q_{n+1},q_{n+1}-p_j)\\[-2mm]
&\times\psi_n(s;\mathbf p_{\widehat j})d\mu(s).
\end{aligned}
\end{equation}

The annihilation block is its adjoint:

\begin{equation}\label{eq:4.6}
\begin{aligned}
(V^-\psi_{n+1})_n(t;\mathbf p)
=\sqrt{n+1}\int_0^\infty dp\,g_d(p)
\int_0^\infty &F_d(t,s;q_n,q_n+p)\\[-2mm]
&\times\psi_{n+1}(s;\mathbf p,p)d\mu(s).
\end{aligned}
\end{equation}

These are distributional/form formulas on the test space. Their bounded heat regularizations are those constructed in Section~\ref{sec:bounds}. For positive outside heat times $\epsilon_{\rm out}$ and
$\epsilon_{\rm in}$, sandwiching by the $H_R$ heat operators
multiplies either kernel by $\exp\!\left[-\frac{\epsilon_{\rm out}t^2+\epsilon_{\rm in}s^2}{4\phi_b}\right]$. For completeness, every nonconstant finite Wick monomial has an equally explicit
block formula. Let it contain \(r\) creators with wavefunctions \(f_j\)
and \(k\) annihilators with wavefunctions \(h_j\), and let its dilation
dimension be \(D>0\). For \(n\geq k\), put \(N=n-k+r\),
\(q_{\rm out}=\sum_{j=1}^Np_j\) and
\(q_{\rm in}=\sum_{j=r+1}^Np_j+\sum_{j=1}^ku_j\). If
\(\operatorname{Sym}_N\) is the orthogonal symmetrization over all
output momenta, denote the dressed monomial by $V_{r,k}$. Its block
in \eqref{eq:4.2} is

\begin{equation}\label{eq:4.6a}
\begin{aligned}
(V_{r,k}\psi_n)_N(t;\mathbf p)
={}&\frac{\sqrt{n!N!}}{(n-k)!}\operatorname{Sym}_N\Bigg[
\prod_{j=1}^rf_j(p_j)\int_{\mathbb R_+^k}d\mathbf u\,
\prod_{j=1}^k\overline{h_j(u_j)}\\
&\quad\times\int_0^\infty d\mu(v)\,
F_D(t,v;q_{\rm out},q_{\rm in})
\psi_n(v;p_{r+1},\ldots,p_N,\mathbf u)\Bigg].
\end{aligned}
\end{equation}

For \(n<k\) it is zero. The factorial follows by applying \(k\)
annihilations and then \(r\) creations; each symmetrization is
normalized as an orthogonal projection. For several species, use the
species-wise symmetrizations. Let $n_a$, $k_a$, and $r_a$ be the
input particle, annihilator, and creator counts for species $a$. On a
fixed occupation sector, put \(N_a=n_a-k_a+r_a\); the factorial in \eqref{eq:4.6a} is replaced by $\prod_a\frac{\sqrt{n_a!N_a!}}{(n_a-k_a)!}$, and the block vanishes if some \(n_a<k_a\). The total momenta
\(q_{\rm in},q_{\rm out}\) still include every species, and \(D\)
is the sum of the dimensions of all the inserted jets. The constant
monomial is treated separately as the identity, whose heat sandwich is
\(e^{-(a+b)H_R}\); it does not require a value of \(F_D\) at \(D=0\).

For a primary, the wavefunctions are the jets and polynomial
coefficients of Section~\ref{sub:primaries}. Adjointness follows
by exchanging creators and annihilators and using the symmetry
of $F_D$. Projection onto the angular basis of
Section~\ref{sub:decomposition} gives the primary-channel
coefficients.

\subsection{Comparison of energy-basis normalizations}\label{sub:normalization}

In Kolchmeyer's conventions the Hilbert-space measures are $d\mu(r)$
for the vacuum sector and $d\mu(r)d\mu(s)$ for each nonvacuum primary
sector; see \cite[equations (4.28)--(4.30)]{K}. The generalized
vacuum-to-primary block uses $\sqrt{G_{\Delta_i}(r,s)}$ as in his
(4.43). Proposition~\ref{prop:bounded-columns} gives its bounded
wave-packet interpretation. For left and right Euclidean times
$\beta_L,\beta_R>0$, the
insertion packet has coefficient $\delta_{ij}\sqrt{G_{\Delta_i}(r,s)} e^{-\beta_LE(r)-\beta_RE(s)}$. The second Boltzmann factor uses the right variable $s=s_R$; the
repetition of $s_L$ in that factor in the printed (4.38) and (4.40)
of arXiv version 2 is a typographical error.

We record a normalization issue in \cite[equation (4.46)]{K}, with equation numbers referring to arXiv version 2. Let $V_{aib}(r,s,t)$ denote his normalized three-primary coefficient
in (4.45); here $a,i,b$ are primary labels, $r,s,t>0$ are spectral
parameters, and $\Delta_j=\lambda_j$ is the weight of primary $j$.
Dividing the three-point amplitude by the two external state
normalizations gives

\begin{equation}\label{eq:4.7}
\frac{\sqrt{G_{\Delta_a}(r,s)G_{\Delta_i}(s,t)G_{\Delta_b}(t,r)}
V_{aib}(r,s,t)}
{\sqrt{G_{\Delta_a}(r,s)G_{\Delta_b}(r,t)}}
=\sqrt{G_{\Delta_i}(s,t)}V_{aib}(r,s,t).
\end{equation}

The square root therefore belongs to the inserted field and
its neighboring right energies. Our calculations use the
independently derived Bessel formula \eqref{eq:4.3}.

\section*{Acknowledgements}
This paper started from my visit to IAS and I'd like to thank IAS for
hospitality and especially Edward Witten for stimulating discussions.

The author acknowledges the use of OpenAI's ChatGPT for assistance with
literature searches, preliminary proof exploration, and the preparation and
checking of this manuscript. The author is responsible for the mathematical
content.
\section* {Conflict of Interest Statement}
The author declares no conflict of interest.

\end{document}